\documentclass[11pt,a4paper]{article}

\usepackage{siunitx}
\usepackage{amsmath}
\usepackage{amssymb}
\usepackage{bbm}
\usepackage{pennames}
\usepackage{graphicx}
\usepackage[colorlinks=true,linkcolor=blue,anchorcolor=black,citecolor=blue,filecolor=cyan,menucolor=red,runcolor=filecolor,urlcolor=blue,bookmarks=true,bookmarksnumbered=true]{hyperref}
\usepackage{algpseudocode}
\usepackage{algorithm}
\usepackage{definitions}
\usepackage{xspace}
\usepackage[dvipsnames]{xcolor}
\usepackage[T1]{fontenc}

\usepackage{authblk}
\usepackage{comment}
\usepackage{cite}

\usepackage[font=small,labelfont=bf]{caption}

\makeatletter

\DeclareMathOperator*{\argmin}{arg\,min}

\newcommand{\distas}[1]{\mathbin{\overset{#1}{\kern\z@\sim}}}%
\newsavebox{\mybox}\newsavebox{\mysim}
\newcommand{\distras}[1]{%
  \savebox{\mybox}{\hbox{\kern3pt$\scriptstyle#1$\kern3pt}}%
  \savebox{\mysim}{\hbox{$\sim$}}%
  \mathbin{\overset{#1}{\kern\z@\resizebox{\wd\mybox}{\ht\mysim}{$\sim$}}}%
}
\makeatother

\newcommand{\bn}{\ensuremath{{\boldsymbol{\nu}}}\xspace}
\newcommand{\bnn}{\ensuremath{{\bn_0}}\xspace}
\newcommand{\bno}{\ensuremath{{\bn_1}}\xspace}

\newcommand{\bt}{{\ensuremath{{\boldsymbol{\theta}}}}\xspace}
\newcommand{\btn}{\ensuremath{{\boldsymbol{\theta}_0}}\xspace}
\newcommand{\bto}{\ensuremath{{\boldsymbol{\theta}_1}}\xspace}

\newcommand{\aMCatNLO} {\textsc{MG5\_aMC@NLO}\xspace}
\newcommand{\bzero}{{\ensuremath{\mathbf{0}}}\xspace}
\newcommand{\bx}{\ensuremath{\boldsymbol{x}}\xspace}
\newcommand{\bz}{\ensuremath{\boldsymbol{z}}\xspace}
\newcommand{\dd}{\ensuremath{{\textrm{d}}}\xspace}

\title{\bf Refinable modeling for unbinned SMEFT analyses}

\date{}

\author[]{Robert Sch\"ofbeck}
\affil[]{\emph{Institute for High Energy Physics, Austrian Academy of Sciences,\\Dominikanerbastei 16, 1010 Vienna, Austria}}
\affil[]{\emph{Technische Universit\"at Wien,\\Karlsplatz 13, 1040 Vienna\\\phantom{x}\\\href{mailto:robert.schoefbeck@oeaw.ac.at}{\texttt{robert.schoefbeck@oeaw.ac.at}}
}}

\begin{document}

\baselineskip=14pt

\maketitle


\begin{abstract}

We present methods to estimate systematic uncertainties in unbinned LHC data analyses, focusing on constraining Wilson coefficients in the standard model effective field theory (SMEFT). Our approach also applies to broader parametric models of non-resonant phenomena beyond the standard model (BSM). By using machine-learned surrogates of the likelihood ratio, we extend well-established procedures from binned Poisson counting experiments to the unbinned case. This framework handles various theoretical, modeling, and experimental uncertainties, laying the foundation for future unbinned analyses at the LHC.
We also introduce a tree-boosting algorithm that learns precise parametrizations of systematic effects, providing a robust, flexible alternative to neural networks for modeling systematics. We demonstrate this approach with an SMEFT analysis of highly energetic top quark pair production in proton-proton collisions.

\end{abstract}

\vspace{\baselineskip}

\vspace{10pt}
\noindent\rule{\textwidth}{1pt}
\tableofcontents
\noindent\rule{\textwidth}{1pt}

\section{Introduction}
\label{sec:intro}


The Large Hadron Collider (LHC) generates vast amounts of data from particle decays in high-energy interactions, offering a unique opportunity to explore fundamental physics. Recent advances in machine learning (ML) provide powerful tools not only for reconstruction and object-tagging but also for novel analysis techniques. High-dimensional unbinned analyses, where dozens of features probe a large number of model parameters, are now feasible with machine-learned surrogates optimized for hypothesis testing.

In the theoretical domain, the lack of new resonance signals has led to the adoption of the standard model effective field theory (SMEFT)~\cite{Buchmuller:1985jz,Leung:1984ni,Degrande:2012wf,Brivio:2017vri, Isidori:2023pyp} as the main framework for describing phenomena below an assumed energy scale, conventionally set at $\Lambda_{\textrm{SMEFT}}=1$~\TeV. This framework extends the standard model (SM) Lagrangian with field monomials, where the Wilson coefficients serve as the parameters of interest (POIs). The SMEFT enables experimentalists to test a range of high-scale models without dealing with their fundamental parameters.

The SMEFT is organized by the mass dimension of operators, beginning with dimension six for relevant new physics scenarios at the LHC~\cite{Grzadkowski:2010es}. Since the lowest-order matrix-element (ME) modifications to Wilson coefficients are linear, cross-section deviations can be described by quadratic polynomials within the SMEFT’s validity range~\cite{Belvedere:2024nzh}. This analytic structure supports simulation-based inference~(SBI) methods that improve performance, especially when probing multiple Wilson coefficients simultaneously~\cite{GomezAmbrosio:2022mpm, Chatterjee:2022oco,Chatterjee:2021nms, Chen:2020mev,Chen:2023ind,Cranmer:2015bka,Brehmer:2018kdj,Brehmer:2018eca,Brehmer:2018hga,Brehmer:2019xox,Brehmer:2019gmn,Butter:2021rvz}. These methods offer statistically optimal observables at the detector level, with fast evaluation after an initial training stage.

Nevertheless, most current LHC measurements are straightforward Poisson counting experiments, partially because these reduce the computational demand. Large-scale computing infrastructure has become more accessible, but the application of unbinned techniques is still hampered by the absence of a comprehensive set of tools that bring decades of experience with treating systematic effects in binned Poisson measurements on par with the unbinned case. 
The available methodology for treating systematic uncertainties in unbinned simulation-based inference techniques, such as ML optimal observables, is sparse.
While optimal ML observables have a sound footing in well-developed statistical methodology,
the otherwise finely honed procedures for treating systematic uncertainties are rarely seen in this light.
The present work aims to change that situation through a comprehensive statistical interpretation of procedures for treating systematic effects in SBI. 
We explain how the factorization of individual systematic effects facilitates the training of multi-variate parametrized regressors and how to address uncertainties in the normalization of processes. The most significant advantage of this approach is its stage-wise nature. Adding new processes or systematic uncertainties does not invalidate partially available training.

\begin{figure}[t]
    \centering
    \includegraphics[width=.99\textwidth]{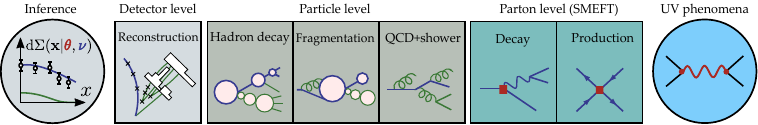}
  \caption{A hierarchy of scales governs our assumed understanding of physical phenomena and the modeling of collider physics. Staged event simulation proceeds in the ``forward mode'' from right to left, while data analyses proceed in the reverse direction, aimed at constraining models of UV phenomena or the SMEFT POIs from the detector-level measurements. The sketch is adapted from Ref.~\cite{Plehn:2022ftl}.}
  \label{fig:conditional-sequence}
\end{figure}

The conceptual cornerstone of the modeling of collider phenomena underpinning SBI relies on a hierarchical separation of processes by energy scale, starting from hypothetical UV phenomena and their SMEFT parametrization at $\Lambda_{\textrm{SMEFT}}$.
The SMEFT Wilson coefficients, our POIs, are denoted by \bt, and we aim at parameter inference through frequentist confidence intervals~\cite{Cranmer:2014lly}. 
Those parameters are, therefore, not stochastic. However, we note that there is no conceptual limitation to Bayesian SMEFT analysis.

Figure~\ref{fig:conditional-sequence} reflects this hierarchy, from right to left, by grouping unobserved (latent) variables and systematic effects at the parton, particle, and detector levels. This division balances sufficient detail with manageable notation complexity; more levels could be added but would obscure the core concepts.
Collider event simulation mirrors this staging: at the parton level, ME generators sample SMEFT ME-squared terms. Key systematic uncertainties here include unphysical effects from technically unavoidable energy scales linked to the perturbative renormalization of fixed-order predictions and the factorization of collinear and infrared radiation. Uncertainties in the parton distribution functions (PDFs) cover our lack of knowledge in the composition of the $\textrm{pp}$~initial state.
At the scale of $\Lambda_{\textrm{QCD}}$, particle-level simulations handle parton shower (PS) effects, fragmentation, hadronization, decay, and underlying event modeling. Tuning these models introduces systematic uncertainties, and modeling generator differences as a ``two-point'' systematic uncertainty can be effective. At the detector level, simulations include particle interactions with detector material, digitization, trigger logic, and event reconstruction.

Our approach provides a general method to obtain ML-based parametrizations of all these effects by grouping them into classes of systematic variations addressed individually. With these procedures in place, we can build unbinned models that can be iteratively refined. Adding new effects or background contributions does not invalidate surrogates trained on the initial model.

On the technical side, we fill a gap in the methodology by developing a tree-boosting algorithm that can learn arbitrarily accurate parametrizations of systematic effects. This is done by extending tree algorithms to produce regressors that are parametric in externally provided data; in our case, the nuisance parameters~(\bn) linked to systematic effects. The resulting ``Boosted Parametric Tree''~(BPT) offers a robust and flexible alternative to neural networks for this purpose, with the training procedure fully grounded in unbinned model building.
Therefore, it enables a full understanding of tree-boosting within the context of unbinned hypothesis tests. The terminal nodes of the trees act as measurement bins, allowing for an analytic dependence on nuisance parameters similar to the binned case. The BPT thus learns an expressive surrogate for differential cross-section ratios~(DCRs), accommodating a potentially high-dimensional set of model parameters. 

We use unbinned SMEFT analyses as our motivating case, assuming that the dependence on the POIs is learned by an algorithm from the literature~\cite{GomezAmbrosio:2022mpm, Chatterjee:2022oco,Chatterjee:2021nms, Chen:2020mev,Chen:2023ind, Cranmer:2015bka,Brehmer:2018eca, Brehmer:2018hga,Brehmer:2019xox, Brehmer:2018kdj}. 
We demonstrate our tools for modeling systematic effects through a semi-realistic SMEFT case study in top quark pair production.
However, this methodology is broadly applicable and could enhance the inference of any SM parameter with a non-resonant impact on collider data. In addition to extracting parameters like $\alpha_s$, electro-weak precision observables, or $\sin^2\mathbf{\theta}_\textrm{W}$, inclusive cross-section measurements could benefit from an unbinned treatment of the signal process.

The rest of the paper is structured as follows. We discuss the relation to existing works for unbinned SMEFT analyses in Sec.~\ref{sec:other-work}. The statistical setup for unbinned hypothesis tests is provided in Sec.~\ref{sec:unbinned-tests}, and we use it to outline the key concepts of refinable modeling.
A comprehensive review of the statistical interpretation of event simulation, suitable for developing ML tools, is given in Sec.~\ref{sec:simulation}. This part also defines the terminology for Sec.~\ref{sec:systematics-learning}, which explains how to train generic regressors for suitable parametrizations of the various model-parameter dependencies and introduces the BPT algorithm. In Sec.~\ref{sec:refinable-likleihood}, we use the enw tools as building blocks for constructing refinable unbinned models. 
Those are applied in Sec.~\ref{sec:ttbar}, where we demonstrate the application of the procedures for SMEFT analyses of top quark pair production in the two-lepton channel. We provide conclusions in Sec.~\ref{sec:conclusion}.

\section{Relation to other works}\label{sec:other-work}

Several approaches in the literature suggest ML optimal test statistics, and we incorporate aspects of the statistical methodology and ML techniques. 

The {\bf\textsc{ML4EFT}} framework~\cite{GomezAmbrosio:2022mpm} advocates unbinned SMEFT hypothesis tests using a similar statistical setting as this work and presents sensitivity studies obtained without detector simulation or systematic uncertainties. The present work takes the next step, focussing on treating systematic uncertainties, and provides the tools for capitalizing on simulated data sets, encapsulating the systematic effects from all stages of LHC event simulation. A less significant difference is that SMEFT effects in Ref.~\cite{GomezAmbrosio:2022mpm} are learned by using networks, while we use the following tree-based method for this purpose.

The {\bf Boosted Information Tree}~({\bf\textsc{BIT}})~\cite{Chatterjee:2021nms,Chatterjee:2022oco} is a tree-based algorithm for learning SMEFT effects. In this work, we adopt it for the SMEFT signal modeling and extend it toward predicting the full positive quadratic SMEFT polynomial. The BIT algorithm learns the quadratic SMEFT polynomial using the same statistical foundation as the present work. The ``Parametric Regression Tree'' in Sec.~\ref{sec:bpt} has a different goal, but the technical implementation and the statistical interpretation of the boosting are closely related.

In {\bf Parametrized classifiers for optimal EFT sensitivity}~\cite{Chen:2020mev}, the authors develop a neural-network-based approach for learning optimal SMEFT classifiers up to next-to-leading (NLO) perturbative accuracy. Ref.~\cite{Chen:2023ind} provides a reweighting-based extension. Our approach to learning the SMEFT signal dependence is a tree-based alternative, and our focus in this work is on systematic uncertainties. The idea of learning generic coefficient functions for parametric regressors, as discussed in Sec.~\ref{sec:systematics-learning}, is partly motivated by the corresponding SMEFT construction in Ref.~\cite{Chen:2020mev}. 

The authors of the {\bf \textsc{Madminer}} framework~\cite{Cranmer:2015bka, Brehmer:2018kdj, Brehmer:2018eca, Brehmer:2018hga, Brehmer:2019xox, Brehmer:2019gmn} developed the understanding of event simulation for likelihood-free inference that is also an essential basis for this work. On the technological side, {\textsc{Madminer}} provides various techniques for general parameter inference and, specifically, also for unbinned SMEFT analyses. Beyond these motivating examples, {\textsc{Madminer}} arguably established the SBI methods as a subfield in high-energy physics, to which the present work contributes. While \textsc{Madminer} can also model systematic uncertainties, we use a more general and incrementally refinable statistical model for this purpose.

The authors of {\bf Learning new physics from an imperfect machine}~\cite{dAgnolo:2021aun} use an entirely different (neural-network-based) model of the phenomena beyond the SM, which is at variance with the SMEFT case presented here. Nevertheless, the statistical setup~(Sec.~\ref{sec:unbinned-tests}), in particular, the definition of the nuisance parameters and the parametrization of systematic effects, are similar to this work. 

In the {\bf\textsc{Inferno}} approach~\cite{DeCastro:2018psv}, a non-linear summary statistic is constructed by minimizing inference-motivated losses via stochastic gradient descent.
The algorithm uses Fisher's information on the parameters of interest and accounts for nuisance parameters, but it is not specific to SMEFT.
In Ref.~\cite{Layer:2023lwi}, the method is used to reduce the systematic uncertainties in the measurement of the top quark pair production in the $\tau$+jets channel.

\section{Unbinned likelihood ratio tests}\label{sec:unbinned-tests}

Given a data set $\mathcal{D}$, confidence level~(CL) intervals for the POIs $\bt$ are determined from the profiled likelihood ratio test statistic.
In this section, we relate it to quantities that a machine can learn. Splitting the data in a primary set $\mathcal{D}$ and an auxiliary set $\mathcal{A}$, we have
\begin{align}    
q_{\bt}(\mathcal{D})&=-2\log\frac{\max_{\bn\phantom{,\bn}}L(\mathcal{D},\mathcal{A}|\bt,\bn)}{\max_{\bn,\bt}L(\mathcal{D},\mathcal{A}|\bt,\bn)}=-2\log\frac{L(\mathcal{D},\mathcal{A}|\bt,{{\boldsymbol{\hat\nu}}}_{\bt})}{L(\mathcal{D},\mathcal{A}|\boldsymbol{\hat\theta},\boldsymbol{\hat\nu})\;\;\,}.\label{eq:test-statistic}
\end{align}
The auxiliary data set has components to constrain systematic uncertainties, such as in the integrated luminosity or the jet energy scale calibration. The maximum-likelihood estimate~(MLE) of the nuisance parameters for a given $\bt$ is $\boldsymbol{\hat{\nu}}_\bt$, while $(\boldsymbol{\hat\theta},\boldsymbol{\hat\nu})$ represents the MLE or all model parameters simultaenously.

By design, $q_{\bt}(\mathcal{D})$ is always non-negative for any $\bt$, with larger values indicating greater incompatibility between the model defined by $\bt$ and the data $\mathcal{D}$. Without nuisance parameters, the Neyman-Pearson lemma states that a hypothesis test of fixed size $\alpha$ based on $q_{\boldsymbol{\theta}}(\mathcal{D})$ has maximum power, meaning it is most likely to correctly reject the null hypothesis when the alternative is true~\cite{Neyman:1933wgr}.

Technically, $\mathcal{A}$ should be an argument of $q_{\bt}(\mathcal{D})$, but we omit it in the notation, as an analytic approximation of the corresponding likelihood factor, described below, will capture all its effects. We assume that $\mathcal{A}$ and $\mathcal{D}$ do not overlap and that SMEFT effects, parametrized by \bt, are negligible in $\mathcal{A}$. Under this assumption, the auxiliary data set produces a multiplicative term $L(\mathcal{A}|\bn)$ in the likelihood,
\begin{align}
L(\mathcal{D},\mathcal{A}|\bn,\bt)=L(\mathcal{D}|\bn, \bt)L(\mathcal{A}|\bn).
\end{align}

According to Wilks' theorem~\cite{Wilks:1938dza}, if $\mathcal{D}$ is distributed under $\boldsymbol{\theta}$, then $q_{\bt}(\mathcal{D})$ is asymptotically distributed as a $\chi^2_{N_{\bt}}$ distribution, where $N_{\bt}$ is the number of independent degrees of freedom in $\bt$. This asymptotic distribution is independent of the nuisance parameters, which simplifies the limit-setting procedure and is a primary reason why LHC data analyses commonly use the profiled likelihood ratio test statistic.
The confidence intervals at a confidence level of, e.g., $1-\alpha=95\%$, are given by solving
\begin{align}
q_{\boldsymbol{\theta}}=F^{-1}_{\chi^2_{N_{\boldsymbol{\theta}}}}(1-\alpha),
\end{align}
where $F_{\chi^2_{N_{\bt}}}$ is the cumulative distribution function of the $\chi^2_{N_{\bt}}$ distribution. 

To obtain confidence intervals with some POIs profiled, these parameters are treated as nuisance parameters in $q_{\boldsymbol{\theta}}$, reducing $N_\bt$ accordingly. However, large quadratic terms in the SMEFT expansion can invalidate Wilks' theorem~\cite{Bernlochner:2022oiw}. In such cases, the distribution must be determined by other means, such as toy experiments, to ensure the desired confidence level.

The likelihood functions in Eq.~\ref{eq:test-statistic} are ``extended'' likelihoods: a Poisson-distributed counting variable describes the random fluctuation in the total number of observed events. The remaining discriminating information is encoded in the fiducial detector-level probability density function (pdf) $p(\bx|\bt,\bn)$, which relates to the fiducial differential cross-section as
\begin{align}
\dd\Sigma(\bx|\bt,\bn)=\sigma(\bt,\bn)\,p(\bx|\bt,\bn)\,\dd\bx.\label{eq:fiducial-pdf}
\end{align}
We denote the inclusive fiducial cross-section by $\sigma(\bt,\bn)$. In general, $\dd\Sigma(\bx|\bt,\bn)$ includes contributions from multiple processes.
With the integrated luminosity $\mathcal{L}(\bn)$, subject to systematic uncertainties whose effects we parametrize by nuisance parameters, the likelihood to observe a data set $\mathcal{D}$ of size $N$ can, in general, be written in terms of the differential cross-section as
\begin{align}
L(\mathcal{D}|\bt,\bn)&=\textrm{P}_{\mathcal{L}(\bn)\sigma(\bt,\bn)}(N)\prod_{i=1}^{N} p(\boldsymbol{x}_i|\bt,\bn){}=\textrm{P}_{\mathcal{L}(\bn)\sigma(\bt, \bn)}(N)\prod_{i=1}^{N} \frac{1}{\sigma(\bt,\bn)}\frac{\dd\Sigma(\bx_i|\bt,\bn)}{\dd\bx},\label{eq:general-likleihood}
\end{align}
where $\textrm{P}_\lambda$ denotes the Poisson distribution with mean $\lambda$. 
The extended log-likelihood ratio for two sets of model parameters becomes
\begin{align}
\log\frac{L(\mathcal{D}|\bto,\bno)}{L(\mathcal{D}|\btn,\bnn)}&=-\mathcal{L}(\bno)\sigma(\bto,\bno)+\mathcal{L}(\bnn)\sigma(\btn,\bnn)+\sum_{i=1}^{N}\log\left(\frac{\mathcal{L}(\bno)}{\mathcal{L}(\bnn)}\frac{\dd\Sigma(\bx_i|\bto,\bno)}{\dd\Sigma(\bx_i|\btn,\bnn)}\right).\label{eq:likleihoodratio}
\end{align}

Each of the $K$ sources of systematic uncertainty is associated with a nuisance parameter~$\nu_k$, collectively denoted by \bn. Systematic effects related to detector calibration, the measurement of the integrated luminosity, theoretical calculations, and more are modeled with these nuisance parameters. Measurements can constrain some of these uncertainties through the observed auxiliary data set~$\mathcal{A}_0$, which is the specific instance of $\mathcal{A}$ found in real data and, therefore, not a random quantity~\cite{dAgnolo:2021aun}. 
We set $\bn=\bzero$ to correspond to the maximum of $L(\mathcal{A}_0|\bn)$, so that by definition,
\begin{align}
\max_\bn L(\mathcal{A}_0|\bn)=L(\mathcal{A}_0|\bzero).
\end{align}
The central value $\bn=\bzero$ represents the best available calibrations across all modeling aspects before considering $\mathcal{D}$. The nuisance parameters then parametrize deviations from this hypothesis. Combined with $\bt=\bzero$, this choice defines an SM reference hypothesis with likelihood
\begin{align}
L(\mathcal{D},\mathcal{A}|\textrm{SM})\equiv L(\mathcal{D}|\bzero,\bzero)L(\mathcal{A}|\bzero),\label{eq:SM-ref-likleihood}
\end{align}
which describes the SM without any SMEFT effects and includes the best available calibrations.
We can parametrize the systematic effects to make the $\nu_k$ as uncorrelated as possible and scale them so that the auxiliary log-likelihood ratio becomes a simple analytic expression in terms of $\bn$. In the Gaussian approximation, this is achieved by diagonalizing the Hessian of the auxiliary likelihood function at $\bn=\bzero$, resulting in a penalty of the form
\begin{align}
-2\log\frac{L(\mathcal{A}|\bn)}{L(\mathcal{A}|\bzero)}=\sum_{k=1}^K\nu_k^2,\label{eq:penalty}
\end{align}
though generalizations to other probability distributions are possible.

If a specific nuisance parameter is only constrained by the primary data set and not by $\mathcal{A}_0$, it is excluded from the penalty term in Eq.~\ref{eq:penalty} and referred to as ``floating''. While some uncertainties, such as those related to PDFs, are clearly interpretable in terms of SM parameters, this is not always the case. For example, uncertainties from renormalization or factorization scales address limitations in perturbative accuracy but do not guarantee statistical coverage when these scales vary in the simulation.
With this caveat in mind, we treat all systematic uncertainties heuristically in the same way.

We normalize the likelihoods in Eq.~\ref{eq:test-statistic} by dividing both the numerator and denominator by the SM reference likelihood in Eq.~\ref{eq:SM-ref-likleihood}. This yields
\begin{align}
    q_{\bt}(\mathcal{D}) = \min_{\bn} u(\mathcal{D},\mathcal{A}|\bn,\bt) - \min_{\bn,\bt} u(\mathcal{D},\mathcal{A}|\bn,\bt),\label{eq:test-stat}
\end{align}
where
\begin{align}
-\frac{1}{2}u(\mathcal{D},\mathcal{A}|\bn,\bt) &= -\mathcal{L}(\bn)\,\sigma(\bt,\bn) + \mathcal{L}_0\,\sigma(\textrm{SM}) + \sum_{i=1}^{N(\mathcal{D})}\log\left(\frac{\mathcal{L}(\bn)}{\mathcal{L}_0}\frac{\dd\Sigma(\bx_i|\bt,\bn)}{\dd\Sigma(\bx_i|\textrm{SM})}\right) - \frac{1}{2}\sum_{k=1}^K\nu_k^2,\label{eq:likelihoodratio-normalized}
\end{align}
and $\mathcal{L}_0$ denotes the central value of the auxiliary luminosity measurement.

The main drawback of the unbinned likelihood ratio test statistic in Eq.~\ref{eq:likelihoodratio-normalized} is the need to evaluate the DCR, inclusive cross-section, and integrated luminosity as functions of the model parameters. While log-normal (multiplicative) nuisances effectively model the integrated luminosity dependence around the central value~\cite{CMS:2021xjt, ATLAS:2022hro}, we also require estimates for the DCR and inclusive cross-section.
Event generators cannot provide these estimates parametrically in terms of model parameters, as they operate in ``forward'' mode through sequential stochastic processes. However, for inference, the minimization in Eq.~\ref{eq:test-statistic} requires to evaluate the DCR for externally provided simulated or real events.

\begin{figure}[t]
    \centering
    \includegraphics[width=.99\textwidth]{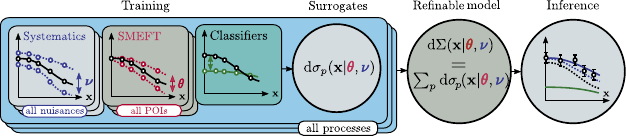}
  \caption{Overview of refinable modeling in SBI, proceeding in the backward mode, from left to right. For each process, we machine-learn the dependence on the SMEFT POIs and on the nuisance parameters and combine these elements into a refinable surrogate model for the global DCR in the measurement region, suitable for inference.}
  \label{fig:sketch}
\end{figure}

\section{Simulation for inference}\label{sec:simulation}
Although computationally intensive, robust predictions from Monte Carlo~(MC) simulations are invaluable for modeling highly energetic scattering processes~\cite{Campbell:2022qmc}. 
Our strategy is to provide the necessary parametric estimates by breaking down the primary task of modeling the DCR into smaller, manageable machine-learning tasks. The tasks are based on simulation and capitalize on the high quality of the MC methods. As illustrated in Figure~\ref{fig:sketch}, we proceed in the backward mode, from left to right, and separate the training into distinct parton-level processes that are simulated separately. 
For each process, we learn parametrizations of systematic effects and POI dependencies with the help of efficiently generated ``synthetic'' data sets that correspond to the systematic variations in the binned approach. 
Inheriting these procedures, we leverage the extensive experience from over a decade of binned LHC data analyses. 

Training tasks for systematics can be divided into uncorrelated groups of nuisance parameters which then separately estimate these effects. A high granularity allows a gradual refinement of each aspect of the final model without invalidating unrelated tasks.
In the following sections, we describe how the results of ML training tasks combine into a refinable surrogate model $\dd\Sigma(\bx|\bt,\bn)$ that provides the needed DCR in Eq.~\ref{eq:likelihoodratio-normalized}. 

\subsection{Hierarchical data representations and staged event simulation}\label{sec:staged-simulation}

Systematic effects can modify predictions at any energy scale. As shown in Fig.~\ref{fig:conditional-sequence}, we broadly categorize these into parton level~(p), particle level~(ptl), detector reconstruction level~(reco), and observed features derived from reconstruction, denoted by $\bx$. By definition, the observables~$\bx$ are the quantities used in Eqs.~\ref{eq:fiducial-pdf}~-~\ref{eq:likelihoodratio-normalized}. All other quantities, including $\bz_{\textrm{reco}}$ (e.g., low-level hit patterns in sub-detectors accessible in real data), are denoted by $\bz$.

This grouping strikes a balance between detail and notational simplicity. It also reflects the approximate separation of systematic effects from UV and SMEFT energy scales to those relevant to QCD and detector signals. This staging is flexible; additional stages can be introduced as long as an event representation can be defined.

Event simulation for LHC proton-proton~(pp) collisions is divided among several computer codes, each addressing modeling at a specific energy scale with specialized techniques. At the parton level, ME generators like \aMCatNLO~\cite{Alwall:2014hca, Frederix:2012ps}, \SHERPA~\cite{Sherpa:2019gpd}, and \POWHEG~\cite{Nason:2004rx, Frixione:2007vw, Frixione:2007nw, Alioli:2010xd, Campbell:2014kua} provide a sampling of purely perturbative ME-squared predictions for the hard-scatter interaction. The dynamics of fundamental particles from the hard scatter and subsequent decays of heavy SM particles are represented by $\bz_\textrm{p}$.
The PS evolves $\bz_\textrm{p}$ down to energies where perturbative methods are no longer valid. Together with color reconnection, hadronization, decays of unstable hadrons, underlying event, particles from multiple-parton interactions, and both initial- and final-state radiation, it defines the particle level. This is simulated with general-purpose tools like \PYTHIA~\cite{Sjostrand:2014zea} or \HERWIG\cite{Bellm:2015jjp}, which require tuning of phenomenological parameters to reliably describe data. The resulting particle-level event is represented by $\bz_\textrm{ptl}$.
The parton and particle levels are latent and cannot be directly observed in real data.

Particle-level events are processed with detector-specific simulation tools like \GEANT~\cite{GEANT4:2002zbu}, using conditions for each data-taking period and mixed with simulations of separate hard-scatter events to model pile-up. After simulating detector hits, event reconstruction proceeds to the particle candidate level, using, e.g.,  a version of the \textsc{ParticleFlow} algorithm~\cite{ATLAS:2017ghe, CMS:2017yfk}. 
Together with jet clustering, lepton identification, and disambiguation, this process is similar in both real and simulated data. The result is the reconstruction-level event, with features denoted by $\bz_{\textrm{reco}}$.
Simplified event reconstruction is available from, e.g., \textsc{Delphes}~\cite{deFavereau:2013fsa}.

Most data analyses derive high-level observables \bx from $\bz_{\textrm{reco}}$. These observables capture all event features included in the hypothesis test. In principle, \bx can represent the entire reconstruction level, including the variable-length list of all reconstructed particle candidates in an event~\cite{Komiske:2018cqr}. Such approaches have been used to constrain SMEFT effects~\cite{Chatterjee:2024pbp}; here, we focus on high-level event features.
Any real-data event features not already included in \bx belong to $\bz_{\textrm{reco}}$ and are latent in the hypothesis test.

For bookkeeping, we group the nuisance parameters into $\bn_{\textrm{p}}$, $\bn_{\textrm{ptl}}$, and $\bn_{\textrm{reco}}$, collectively denoted by $\bn$ when simplifying notation. The differential cross-section in the fiducial phase space from Eq.~\ref{eq:fiducial-pdf} then becomes
\begin{align}
&\dd\sigma(\bx|\bt,\bn_{\textrm{reco}},\bn_{\textrm{ptl}},\bn_{\textrm{p}})=\sigma(\bt,\bn_{\textrm{reco}},\bn_{\textrm{ptl}},\bn_{\textrm{p}})\nonumber\\&\qquad\qquad\times\int\dd\bz_{\textrm{reco}}\int\dd\bz_{\textrm{ptl}}\int\dd\bz_{\textrm{p}}\,p(\bx,\bz_{\textrm{reco}},\bz_{\textrm{ptl}},\bz_{\textrm{p}}|\bt,\bn_{\textrm{reco}},\bn_{\textrm{ptl}},\bn_{\textrm{p}})\,\dd\bx.\label{eq:fiducial-pdf-int}
\end{align}

The hierarchical event representation enables a natural factorization of the pdf as
\begin{align}
&p(\bx,\bz_{\textrm{reco}},\bz_{\textrm{ptl}},\bz_{\textrm{p}}|\bt,\bn_{\textrm{reco}},\bn_{\textrm{ptl}},\bn_{\textrm{p}})\nonumber\\&\qquad\qquad=p(\bx|\bz_{\textrm{reco}})\,p(\bz_{\textrm{reco}}|\bz_{\textrm{ptl}},\bn_{\textrm{reco}}) p(\bz_{\textrm{ptl}}|\bz_{\textrm{p}},\bn_{\textrm{ptl}})\,p(\bz_{\textrm{p}}|\bt,\bn_{\textrm{p}}).\label{eq:pdf-factorized}
\end{align}
This pdf depends on both latent and observable features~\cite{Brehmer:2018kdj}. In ratios, the conditional factors\footnote{The factor $p(\bx|\bz_{\textrm{reco}})$ could also be conditional on nuisance parameters related to uncertainties in analysis-dependent parameters that may be involved when computing $\bx$ from $\bz_{\textrm{reco}}$. This extension is straightforward. } can partially or entirely cancel, enabling efficient generation of synthetic data sets for trainng ML surrogates~\cite{Brehmer:2019xox}.

\subsection{Semi-analytic modeling at the parton level}\label{sec:parton-level-modeling}
At the parton level, the ME generators provide a (possibly weighted) sampling of the ME-squared SMEFT terms.
The generic parton-level DCR for a single process is
\begin{align}
&\dd\sigma_{\textrm{SMEFT}}(\bz_p|\bt,\nu_R,\nu_F,\bn_{\textrm{PDF}})\propto\sum_{f_1,f_2,h}\left|\mathcal{M}_{\textrm{SMEFT}}(\boldsymbol{z}_p,h|\bt,\mu_R(\nu_R),\mu_F(\nu_F))\right|^2\nonumber\\&\qquad\qquad\times\textrm{PDF}(f_1,x_{\textrm{Bjorken},1},\mu_F(\nu_F),\bn_{\textrm{PDF}})\textrm{PDF}(f_2,x_{\textrm{Bjorken},2},\mu_F(\nu_F),\bn_{\textrm{PDF}})\dd \bz_p,\label{eq:parton-level-hel-ignorant}
\end{align}
where $\mu_R$ and $\mu_F$ are the renormalization and factorization scales, respectively. 
For single-operator insertions, the dependence on the SMEFT Wilson coefficients \bt is accurately described by a quadratic polynomial.
The flavors of the incoming partons are denoted by $f_{1/2}$ and take values in $\{\Pqu, \overline{\Pqu}, \Pqd, \overline{\Pqd}, \Pqc, \overline{\Pqc}, \Pqs, \overline{\Pqs}, \Pqb, \overline{\Pqb}, \Pg\}$. Furthermore, we denote the Bjorken scaling variables by $x_{\textrm{Bjorken},1/2}$. 
The relevant latent parton-level configuration, sufficient for  evaluating the parton distribution functions~(PDFs)~\cite{Buckley:2014ana}, is then given by $
\{f_{1},f_{2},x_{\textrm{Bjorken},1},x_{\textrm{Bjorken},2}\}$. 
Equation~\ref{eq:parton-level-hel-ignorant} also sums over the initial and final-state helicity configurations, denoted by $h$. This choice removes the helicity configuration from the parton-level latent-space event representation $\bz_{\textrm{p}}$ and is called ``helicity-ignorant''\footnote{Helicity-aware and helicity-ignorant SMEFT predictions are compared in Ref.~\cite{Belvedere:2024nzh}.} in the context of reweighted predictions~\cite{Mattelaer:2016gcx}. A list of four-momenta then represents the final state parton-level dynamics, and we arrive at the parton-level representation 
\begin{align}
\bz_{\textrm{p}}=\{f_{1},f_{2},x_{\textrm{Bjorken},1},x_{\textrm{Bjorken},2},p^\mu_1,p^\mu_2,\ldots\}.\label{eq:parton-level-repr}
\end{align}
If the helicity configuration is kept~(``helicity-aware''~\cite{Mattelaer:2016gcx}) and enters the particle-level simulation, $h$ is also included. The only required change in Eq.~\ref{eq:parton-level-hel-ignorant} in that case is to use a separate differential $\dd\bz_{\textrm{p}}^{(h)}$ for each helicity configuration~\cite{Belvedere:2024nzh}. 

Uncertainties in the PDFs can be expressed as variations along Hessian eigen-directions of an underlying parametrization~\cite{Buckley:2014ana}, with the corresponding nuisance parameters denoted by $\bn_{\textrm{PDF}}$. These PDF variations represent different hypotheses about proton parton dynamics and have a clear physical interpretation. In contrast, the dependence on $\mu_R$ and $\mu_F$ arises from technical artifacts, namely the finite perturbative order and the separation of collinear radiation between the ME and the PDF.
Despite this, we treat scale uncertainties with standard nuisance parameters $\nu_R$ and $\nu_F$, corresponding to up and down variations around the central values $\mu_{R,0}$ and $\mu_{F,0}$. If values of $\pm 1$ correspond to variation factors of 2, then
\begin{align}
\mu_R(\nu_R)=2^{\nu_R}\mu_{R,0}\quad\textrm{and}\quad \mu_F(\nu_F)=2^{\nu_F}\mu_{F,0}.\label{eq:scale-nuisances}
\end{align}

We account for uncertainties in the overall normalization of the process with log-normal nuisances in Sec.~\ref{sec:refinable-likleihood} and will not discuss them here. However, we include a general parton-level ad-hoc reweighting function $\alpha_{\textrm{rw}}(\bz_{\textrm{p}},\bn_{\textrm{rw}})$, useful when higher-order perturbative corrections significantly depend on the parton-level configuration and we want to adjust the parton-level distribution with an ad-hoc modification.
The total parton-level prediction is then written as
\begin{align}
\dd\sigma(\bz_p|\bt,\bn_{\textrm{p}})&=\alpha_{\textrm{rw}}(\bz_{\textrm{p}},\bn_{\textrm{rw}})\,\dd\sigma_{\textrm{SMEFT}}(\bz_p|\bt,\nu_R,\nu_F,\bn_{\textrm{PDF}})
\label{eq:prediction-tot}
\end{align}
An example of this type is the modeling of the transverse top quark momenta in the $\textrm{t}\bar{\textrm{t}}$ process, which is well understood from higher-order pertubation theory~\cite{Serkin:2021bbn}, but is not yet available from SMEFT ME generators. The nuisances $\bn_{\textrm{rw}}$ modify parameters in the reweighting function $\alpha_{\textrm{rw}}(\bz_{p},\bn_{\textrm{rw}})$ within their uncertainties. Since such procedures are highly application-dependent, we do not elaborate further, except to note that $\alpha_{\textrm{rw}}(\bz_{p},\bn_{\textrm{rw}})$ must always be positive.
The parton-level nuisance parameters considered so far are
\begin{align}
\bn_{\textrm{p}}=\{\nu_R,\nu_F,\bn_{\textrm{PDF}},\bn_{\textrm{rw}}\}
\end{align}
and are associated with systematic effects that can be modeled semi-analytically, allowing for efficient computation.

We use Eq.~\ref{eq:prediction-tot} to define a parton-level pdf and the inclusive parton-level cross-section generically as 
\begin{align}
\dd\sigma(\bz_{\textrm{p}}|\bt,\bn_\textrm{p})=\bar\sigma(\bt,\bn_{\textrm{p}})\,p(\bz_{\textrm{p}}|\bt,\bn_{\textrm{p}})\,\dd\bz_{\textrm{p}}.\label{eq:parton-pdf}
\end{align}
The bar on $\bar\sigma(\bt,\bn_{\textrm{p}})$ indicates that the inclusive cross-section pertains to the entire kinematic phase-space, unaffected, e.g., by the finite detector acceptance. By definition, we have
\begin{align}
\int d\sigma(\bz_{\textrm{p}}|\bt,\bn_{\textrm{p}})=\bar\sigma(\bt,\bn_{\textrm{p}})\quad\textrm{and}\quad\int\dd\bz_{\textrm{p}}\,p(\bz_{\textrm{p}}|\bt,\bn_{\textrm{p}})=1,
\end{align}
which differs from the fiducial cross-section $\sigma(\bt,\bn)$ in Eq.~\ref{eq:fiducial-pdf} by the acceptance effects and the event selection from the subsequent modeling stages, in particular at the detector level.
With an ME generator, we obtain a possibly weighted sample of identically and independently distributed events from Eq.~\ref{eq:prediction-tot} as 
\begin{align}
    \{w_i,\bz_{p,i}\}\stackrel{\textrm{i.i.d.}} \sim \bar\sigma(\bt,\bn_{\textrm{p}})\,p(\bz_{\textrm{p}}|\bn_{\textrm{p}}).\label{eq:parton-level-sample}
\end{align}
The overall normalization of weights $w_i$ can be set to $\sum w_i=\bar\sigma(\bt,\bn)$.

These parton-level systematic effects enable tractable simulation, allowing an inexpensive way to modify an existing sample, such as by reweighting, to approximate model parameters beyond the nominal values. Many other effects, such as the choice of different ME generators, do not allow for tractable simulation. If we model differences between such ``two point alternatives'' with nuisance parameters, it becomes impossible to compute the likelihood ratio for one generator choice while sampling from another.
However, in Sec.~\ref{sec:systematics-learning}, we show how to use machine learning to create parametrizations that interpolate between two point modeling alternatives using a nuisance parameter. This approach is crucial in practice, as uncertainties, especially those in modeling nonperturbative aspects of QCD, are often of this type. 

\subsection{Forward-mode event generation at particle and reconstruction level}

Particle-level simulation, staged as described in Sec.~\ref{sec:staged-simulation}, includes the PS, initial- and final-state radiation, multiple-parton interactions, color reconnection, hadronization, the underlying event, and hadron decay. The subsequent reconstruction-level simulation models the detector interaction and event reconstruction.
For each parton-level event $\{w_i,\bz_{\textrm{p},i}\}$, the particle and detector-level simulations, along with observable reconstruction, produce an event representation of the form
\begin{align}
\{w_i,\bx_i,\bz_{\textrm{reco},i},\bz_{\textrm{ptl},i},\bz_{\textrm{p},i}\}.
\end{align}

Due to finite detector acceptance, some events will not pass the event selection. Poor reconstruction performance near reconstruction thresholds motivates defining a fiducial region, denoted by $\mathcal{X}$. Analysis-specific selections, including reliable object-level calibrations, background reduction, and prior knowledge of SMEFT sensitivity, are incorporated into the definition of $\mathcal{X}$. We only require that for a real event, it is possible to determine if it belongs in $\mathcal{X}$.
In particular, $\mathcal{X}$ can include requirements on latent variables like $\bz_{\textrm{reco}}$: online selection acceptance, thresholds on reconstructed object properties, and various data-cleaning event vetoes are part of the definition of $\mathcal{X}$, even if these variables are typically not in $\bx$.

In contrast to the parton level, the particle and reconstruction-level simulation is computationally expensive, and most effects at these stages can not be simulated tractably. Event samples are, therefore, only available for a limited set of model parameters for different $\bn_{\textrm{ptl}}$ and $\bn_{\textrm{reco}}$. 
For each such configuration, the event sample is written as 
\begin{align}
    \mathcal{D}(\bt,\bn)=\{w_i,\bx_i,\bz_{\textrm{reco},i},\bz_{\textrm{ptl},i},\bz_{\textrm{p},i}\}\stackrel{\textrm{i.i.d.}} \sim \sigma(\bt,\bn)\,p(\bx,\bz|\bt,\bn)\quad\textrm{if}\quad\{\bx_i,\bz_{\textrm{reco},i}\}\in\mathcal{X},\label{eq:det-level-sample}
\end{align}
where the total fiducial cross-section in Eq.~\ref{eq:det-level-sample} is
\begin{align}
\sigma(\bt,\bn)=\sum_{\bx_i,\bz_{\textrm{reco},i}\in\mathcal{X}}w_i.\label{eq:fiducial-normalization}
\end{align}
There is a conceptual difference between Eq.~\ref{eq:parton-level-sample} and Eq.~\ref{eq:det-level-sample}. While the parton-level distribution in Eq.~\ref{eq:parton-level-sample} on the r.h.s. is analytically known and used in the MC sampling, Eq.~\ref{eq:det-level-sample} should be understood in the reverse direction. The simulated sample approximates the joint pdf in the fiducial region on the r.h.s., which is unavailable otherwise. 
Concretely, the formal approximation of the joint pdf is
\begin{align}
\sigma(\bt,\bn)\,p(\bx,\bz|\bt,\bn)\approx\sum_{\mathcal{D}(\bt,\bn)}w_i\delta(\bx-\bx_i)\delta(\bz-\bz_i),\label{eq:simu-pdf}
\end{align}
and we can interpret the event weights as
\begin{align}
w_i=\sigma(\bt,\bn) p(\bx_i,\bz_i|\bt,\bn).\label{eq:weight-as-joint-likleihood}
\end{align}
At NLO, the generated samples are necessarily weighted at the ME stage, and the weights can partly be negative~\cite{Frixione:2002ik}. Negative weights, in principle, invalidate the interpretation in Eq.~\ref{eq:weight-as-joint-likleihood}, but Eq.~\ref{eq:simu-pdf} still holds, provided the large sample limit is respected.

To connect to the binned analyses, the expected yield in a given bin $\Delta\bx\subset\mathcal{X}$ is given by $\lambda_{\Delta\bx}(\bt,\bn)=\mathcal{L}(\bn)\sigma_{\Delta\bx}(\bt,\bn)$ with
\begin{align}
\sigma_{\Delta\bx}(\bt,\bn)&=\int_{\Delta\bx}\dd\bx\frac{\dd\sigma(\bx|\bt,\bn)}{\dd\bx}=\sigma(\bt,\bn)\int_{\Delta\bx}
\dd\bx\,p(\bx|\bt,\bn)\nonumber\\
&=\sigma(\bt,\bn)\int_{\Delta\bx}
\dd\bx\int\dd\bz\,p(\bx,\bz|\bt,\bn)\approx\sum_{\bx_i\in\mathcal{D}(\bt,\bn)\,\cap\,\Delta\bx}w_i,\label{eq:binned-prediction}
\end{align}
where the sum extends over all events within $\mathcal{D}(\bn,\bt)$ that fall in the volume $\Delta\bx$.
We denote this selection by $\mathcal{D}(\bt,\bn)\cap\Delta\bx$. Cross-section weighted expectation values, used to define the loss functions in Sec.~\ref{sec:systematics-learning}, are approximated as
\begin{align}
    \langle\mathcal{O}(\bx,\bz)\rangle_{\bx,\bz|\bt,\bn}\equiv\int\dd\bx\,\dd\bz\, \sigma(\bt,\bn)\,p(\bx,\bz|\bt,\bn)\,\mathcal{O}(\bx,\bz)\approx\sum_{\mathcal{D}(\bt,\bn)}\,w_i \mathcal{O}(\bx_i,\bz_i).\label{eq:op-exp}
\end{align}
We emphasize that the per-event weights $w_i$ in Eq.~\ref{eq:op-exp} are typically only known for a small set of model parameter configurations. 

\subsection{Synthetic data sets and tractable simulation}\label{sec:synthetic-data}

\begin{figure}[t]
    \centering
    \includegraphics[width=.99\textwidth]{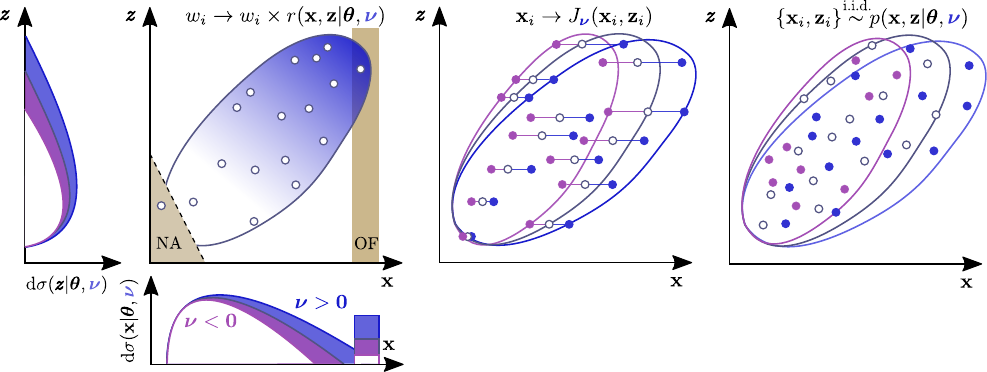}
  \caption{Illustration of the modeling of systematic effects. Left: A reweighting function, depending on $\bx$ and $\bz$, modifies the event weight and, thereby, tractably changes the distributions of both the latent and observed event features. Events in the region ``NA'' fail acceptance, and the region ``OF'' depicts an overflow bin. Middle: A function $J_{\bn}(\bx,\bz)$ provides variations of observed features, for example, related to uncertainties in the calibration of the underlying reconstructed objects. The white, blue, and magenta markers correspond to the nominal simulation and the up and down variations, respectively. The simulation is tractable in this case. Right: Non-tractable systematic effect can only be modeled by independent sampling at different parameter points.}
  \label{fig:joint-space-systematics}
\end{figure}

Simulated samples are computationally expensive. While it is always possible to obtain a simulation based on Eq.~\ref{eq:det-level-sample} for a specific model parameter configuration $(\btn,\bnn)$, it is practically important to know whether we can efficiently generate a new simulated (synthetic) data set from an existing one when $\bt\neq\btn$ or $\bn\neq\bnn$ for some model parameters. In such cases, the simulation is called ``tractable'' for these parameters.

We discuss two tractable cases: likelihood-based reweighting and variations in the calibration of the reconstructed objects. Since SMEFT effects can also be modeled tractably, we address this separately. If a simulation is non-tractable for a model parameter, we must use Eq.~\ref{eq:det-level-sample} to obtain systematically varied data sets. A visualization of these approaches is shown in Fig.~\ref{fig:joint-space-systematics}.

\subsubsection{Uncertainties in the calibration of reconstructed objects}\label{sec:uncertainty-calibration}
An important type of tractable simulation addresses uncertainties in the calibration of underlying object properties, such as jet and lepton momenta or the discriminator value of a $\Pqb$-tagging algorithm. Variations of $\bx$ from these uncertainties are obtained by recomputing it based on modified event properties, defining a function $\bx_\bn=J_{\bn}(\bx,\bz)$. This function provides adjusted values of $\bx$ that depend on latent object-level and event properties, so $J_{\bn}$ also depends on $\bz$.

Evaluating $J_\bn$ provides information on how the observation changes as a function of the model parameters, but it does not give the likelihood or cross-section ratio as a function of the model parameters for a fixed observation, which is needed for Eq.~\ref{eq:likelihoodratio-normalized}. 
For simulated data at a reference parameter point $\{w_i,\bx_i,\bz_i\}\distas{\textrm{iid}}p(\bx,\bz|\btn,\bnn)$, applying $J_\bn$ instead implies
\begin{align}
p(\bx_i,\bz_i|\btn,\bnn)=p\left(J_\bn(\bx_{i},\bz_i),\bz_i|\btn,\bn\right),\label{eq:ll-identity-2}
\end{align}
meaning the joint likelihood remains unchanged with $\bn$ when we simultaneously modify the observation to $\bx_{\bn,i}= J_{\bn}(\bx_i,\bz_i)$.
Synthetic data samples can be efficiently generated as
\begin{align}
\mathcal{D}(\btn,\bn)=\{w_i,\bx_{\bn,i}=J_{\bn}(\bx_i,\bz_i),\bz_i\}_{i=1}^{N_{\textrm{sim}}}\quad\textrm{for all}\quad\{w_i,\bx_i,\bz_i\}\in \mathcal{D}(\btn,\bnn),\label{eq:synthetic-calibration}
\end{align}
and are enough to learn a DCR surrogate as a function of $\bn$, as discussed in Sec.~\ref{sec:systematics-learning}. We assume that calibration-type uncertainties are independent of the POIs, so $\btn$ appears on both sides of Eq.~\ref{eq:ll-identity-2}.

\subsubsection{Synthetic data from event-reweighting}\label{sec:synthetic-reweighting}

When the change in the likelihood of observing an event as a function of a specific model parameter can be computed without resampling the pdf, we can generate reweighted synthetic data sets as 
\begin{align}
\mathcal{D}(\bt,\bn)=\{w_i \times r(\bx_i,\bz_i|\bt,\bn,\btn,\bnn),\bx_i,\bz_i\}_{i=1}^{N_{\textrm{sim}}}\quad\textrm{for all}\quad\{w_i,\bx_i,\bz_i\}\in \mathcal{D}(\btn,\bnn).\label{eq:synthetic-reweighting}
\end{align}
A reweighting function $r(\bx_i,\bz_i|\bt,\bn,\btn,\bnn)$ must be available for the corresponding model parameter and can depend on both observables and latent features. Important tractable cases occur at the parton level, where access to the analytic form of ME-squared terms allows computation of the joint likelihood ratio.
In addition to parton-level parametrizations of the form  
\begin{align}
r(\bx_i,\bz_i|\btn,\bn)=\alpha_{\textrm{rw}}(\bz_{i,\textrm{p}},\bn_{\textrm{rw}})
\end{align}
designed to approximate higher-order perturbative corrections as in Eq.~\ref{eq:prediction-tot}, it is useful to describe uncertainties related to other theoretical inaccuracies with nuisance parameters. For example, the predicted rates of events with high particle-level jet multiplicity ($N_{\textrm{gen-jet}}$) depend on the specifics of matching between the ME calculation and the PS~\cite{Hoeche:2005vzu}, often with significant uncertainties. To address this, we can introduce ad-hoc uncertainties for events with different particle-level jet multiplicities through the variation
\begin{align}
r(\bx_i,\bz_i|\btn,\bn)=1+\alpha_1^{\nu_{\textrm{gen-jet}}}\delta_{1,N_{\textrm{gen-jet}}}+\alpha_2^{\nu_{\textrm{gen-jet}}}\delta_{2,N_{\textrm{gen-jet}}}+\cdots\label{eq:gen-jet-reweighting}
\end{align}
where $\alpha_1$, $\alpha_2$, etc., are constants, and $\nu_{\textrm{gen-jet}}$ is the associated nuisance parameter.

A special case is reweighting functions that depend solely on $\bx$. In this case, we can access the DCR as a function of the observed features without requiring any learning. 
For example, consider a selection with a fixed lepton multiplicity $N_{\ell}$, where the reconstruction efficiency has a relative uncertainty $\Delta \textrm{SF}(\ell)/\textrm{SF}(\ell)$, and a corrective scale factor $\textrm{SF}(\ell)$ is already included in the nominal simulation. We treat the uncertainty $\Delta\textrm{SF}(\ell)$ with a nuisance parameter $\nu_\ell$. If \bx includes the properties of the leptons, allowing $\Delta\textrm{SF}(\ell)$ to be computed solely from $\bx$, we have
\begin{align}
r(\bx_i|\nu_\ell)=\prod_{i=1}^{N_{\ell}}\left(1+\frac{\Delta\textrm{SF}(\ell_i)}{\textrm{SF}(\ell_i)}\right)^{\nu_\ell},\label{eq:lepton-sf}
\end{align}
which provides the DCR as a function of $\nu_\ell$ without needing a surrogate. However, if $\bx$ alone is insufficient to compute the scale factors, the right-hand side of Eq.~\ref{eq:lepton-sf} represents $r(\bx_i,\bz_i|\nu)$ rather than $r(\bx_i|\nu)$, and a surrogate is required.

\subsubsection{SMEFT modeling}\label{sec:synthetic-SMEFT}

SMEFT effects can be modeled with synthetic data, which initially motivated the development of optimal SMEFT observables~\cite{Cranmer:2015bka, Brehmer:2018kdj, Brehmer:2018eca, Brehmer:2018hga, Brehmer:2019xox, Brehmer:2019gmn}. The ME in Eq.~\ref{eq:parton-level-hel-ignorant} can be efficiently recomputed as a function of the POIs \bt. The SMEFT dependence of the DCR is
\begin{align}
&r(\bx_i,\bz_i|\bt,\btn)\\\nonumber
&=\frac{\sigma(\bt)\;\;\;\,}{\sigma(\textrm{SM})}\frac{p(\bx_i,\bz_{\textrm{reco},i}, \bz_{\textrm{ptl},i},\bz_{\textrm{p},i}|\bt)\;\;\;\,}{p(\bx_i,\bz_{\textrm{reco},i},\bz_{\textrm{ptl},i},\bz_{\textrm{p},i}|\textrm{SM})}=\frac{\sigma(\bt)\;\;\;\,}{\sigma(\textrm{SM})}\frac{p(\bx|\bz_{\textrm{reco}})}{p(\bx|\bz_{\textrm{reco}})}\frac{p(\bz_{\textrm{reco}}|\bz_{\textrm{ptl}})}{p(\bz_{\textrm{reco}}|\bz_{\textrm{ptl}})}\frac{p(\bz_{\textrm{ptl}}|\bz_{\textrm{p}})}{p(\bz_{\textrm{ptl}}|\bz_{\textrm{p}})}\frac{p(\bz_{\textrm{p}}|\bt)\;\;\;\,}{p(\bz_{\textrm{p}}|\textrm{SM})}\nonumber\\
&=\frac{\sigma(\bt)\;\;\;\,}{\sigma(\textrm{SM})}\frac{p(\bz_{\textrm{p},i}|\bt)\;\;\;\,}{p(\bz_{\textrm{p},i}|\textrm{SM})}=\frac{\left|\mathcal{M}(\bz_{\textrm{p},i},\bt)\right|^2\;\;\;\,}{\left|\mathcal{M}(\bz_{\textrm{p},i},\textrm{SM})\right|^2}=1+\theta_m r^{(m)}(\bz_{\textrm{p},i})+\theta_m \theta_n r^{(mn)}(\bz_{\textrm{p},i}),\label{eq:r-SMEFT}
\end{align}
where we have omitted the nuisance parameter dependence, as the numerator and denominator are evaluated for the same $\bnn$.
The conditional probabilities in the third term, which are not tractable, cancel in the ratio with excellent accuracy. The remainder is the parton-level DCR, which is available at the level of the ME generator.
Thus, $r(\bx_i,\bz_i|\bt,\btn)$ does not depend on $\bx_i$. By calculating the per-event polynomial coefficients $r^{(m)}(\bz_{\textrm{p},i})$ and $r^{(mn)}(\bz_{\textrm{p},i})$ with $m,n=1,\ldots,N_{\bt}$ from the ME-squared terms at various \bt values, we can construct a parametrization of $r(\bx_i,\bz_i|\bt,\btn)$ valid across the entire SMEFT parameter space, making synthetic data sets $\mathcal{D}(\bt,\bnn)$ readily available. We provide further details in Appendix~\ref{sec:EFT-weights}. Note that the SM point in the denominator is not unique; Eq.~\ref{eq:r-SMEFT} can be applied for any $\btn$, allowing for simulation at EFT parameter points other than the SM.

\subsection{Large sample limit and overflow bins} 

Learning surrogates of the likelihood ratio requires sufficient simulated data, and in the following sections, we assume the large sample limit when minimizing loss functions. For any finite simulated data set, observables with energy units often imply a threshold beyond which simulation becomes too sparse. 
Since SMEFT effects can increase with energy, removing events in the tails of energetic variable distributions is generally counterproductive. Instead, for each such variable in $\bx$, we can define a threshold beyond which we do not fully trust the modeling of the differential cross-section but still find acceptable uncertainties in the cumulative yield.

We accumulate all events with features above certain thresholds in several ``overflow bins''~(OF) and treat these bins using standard Poisson likelihoods. Suppose $x_n$ is a feature in the vector of observables $\bx'$ representing the event before introducing overflow. In that case, we must address the fact that synthetic data insufficiently samples $p(\bx'|x_n > x_{n,0}, \bt, \bn)$. 
The simplest solution is to drop all event features for $x_n > x_{n,0}$ and represent the event only by its presence in the overflow of $x_n$, i.e.,
\begin{align}
  \bx'=\begin{cases} 
      \bx & x_{n}<x_{n,0} \\
      \textrm{OF}_{n}&\textrm{otherwise}.
   \end{cases}
\end{align}
For a number of overflow bins~($N_{\textrm{OF}}$), one for each required observable, we reduce the observation to the counting variable $N(\textrm{OF}_n)$ instead of using fully unbinned information. The number $N_{\textrm{OF}}$ characterizes the measurement, while the observed number of events in the $n$-th overflow bin $N(\textrm{OF}_n)$ characterizes the observation. Explicitly, $\mathcal{D}'=\mathcal{D} \cup \{N(\textrm{OF}_n)\}_{n=1}^{N_{\textrm{OF}}}$.
The likelihood for the overflow bin observation follows from Eq.~\ref{eq:general-likleihood} as a Poisson factor for each overflow bin as
\begin{align}
L(\mathcal{D}'|\bt,\bn)=L(\mathcal{D}|\bt,\bn)\times\prod_{n=1}^{N_{\textrm{OF}}}\,\textrm{P}\left(N(\textrm{OF}_n)|\lambda(\textrm{OF}_n|\bt,\bn)\right).\label{eq:OF-LL}
\end{align}
As assumed, the predicted yield in each overflow bin as a function of the POIs and nuisances, denoted by $\lambda(\textrm{OF}_n|\bt,\bn)$, is available from simulation with sufficient precision. This additional factor is a standard binned likelihood term and should be handled as in traditional binned SMEFT analyses. To simplify the formulas in the following sections, we assume that this factor is included in Eq.~\ref{eq:general-likleihood} and that $\bx\in\mathcal{X}$ implies the event is not in any overflow bin.

\section{Learning from simulation}\label{sec:systematics-learning}
With procedures in place for obtaining simulated and synthetic data sets, we now outline how parametrizations can be learned using common loss functions, such as cross-entropy loss. We consider a general expressive function $\hat f(\bx)$, without specifying its implementation; it could be a neural network, the BPT from Sec.~\ref{sec:bpt}, or any other trainable multivariate predictive function.

\subsection{Likelihood-ratio trick and cross-entropy loss}\label{sec:likelihood-ratio-trick-and-CE}
We begin with the well-known ``likelihood-ratio trick,'' which underpins the learning tasks in this work: a sufficiently expressive machine trained on a classification task learns a (monotonic function of) the likelihood ratio. If the training data is normalized by the differential cross-section, the classifier learns the DCR.
This fact is at the heart of learning techniques for parametric surrogates.
Let us take two fixed hypotheses (\btn, \bnn) and (\bto, \bno), obtained either by independent simulation or produced synthetically as described in Sec.~\ref{sec:synthetic-data}, and minimize the cross-entropy loss function
\begin{align}
    L_{\textrm{CE}}[\hat f]&=-\langle\log\hat f(\boldsymbol{\bx})\rangle_{\bx,\bz|\btn,\bnn}-\langle\log(1-\hat f(\bx))\rangle_{\bx,\bz|\bto,\bno}.
\label{eq:CE-2samp}
\end{align}
The minimum of Eq.~\ref{eq:CE-2samp} for $\hat f(\bx)$ is attained where the function derivative vanishes,
\begin{align}
\frac{\delta L_{\textrm{CE}}[\hat f]}{\delta\hat f(\bx)}=0.
\end{align}
Because $\hat f(\bx)$ does not depend on $\bz$ by construction, we can formally integrate over the latent configuration using  $\int\dd\bz\,p(\bx,\bz|\bt,\bn)= p(\bx|\bt,\bn)$, separately for the summands in Eq.~\ref{eq:CE-2samp}.  The solution is then expressed in terms of latent-space integrals as
\begin{align}
    f^{\ast}_{\textrm{CE}}(\boldsymbol{\bx})&\equiv\argmin_{\hat f}L_{\textrm{CE}}[\hat f]=\left(1+\frac{\sigma(\bto,\bno)\,\int\dd\bz p(\bx,\bz|\bto,\bno)}{\sigma(\btn,\bnn)\,\int\dd\bz p(\bx,\bz|\btn,\bnn)}\right)^{-1}=\left(1+\frac{\dd\sigma(\bx|\bto,\bno)}{\dd\sigma(\bx|\btn,\bnn)}\right)^{-1}.\label{eq:CE-analytic-sol}
\end{align}
This expression is a monotonous function of the DCR for two fixed choices of the model parameters and can be rearranged to
\begin{align}
 \frac{\dd\sigma(\bx|\bto,\bno)}{\dd\sigma(\bx|\btn,\bnn)}=\frac{1}{f^{\ast}_{\textrm{CE}}(\bx)}-1.\label{eq:generic-minimum}
\end{align}
Alternative loss functions and their minima are discussed in Appendix~\ref{sec:alternate-losses}.
The simulation-based approximation of Eq.~\ref{eq:CE-2samp}, suitable for a concrete implementation in computer code, is
\begin{align}
        L_{\textrm{CE}}[\hat f]\approx-\sum_{\mathcal{D}(\btn,\bnn)}w_{i}\log \hat f(\bx_i)-\sum_{\mathcal{D}(\bto,\bno)}w_{i}\log (1-\hat f(\bx_i)),\label{eq:CE-2samp-empirical}
\end{align}
and explicitly uses two different samples in the two terms.
For the reweighting-based tractable effects in Sec.~\ref{sec:synthetic-reweighting} and Sec.~\ref{sec:synthetic-SMEFT}, the per-event joint likelihood is available, and we can use Eq.~\ref{eq:synthetic-reweighting} to rewrite the cross-entropy loss with a single pdf as 
\begin{align}
    L_{\textrm{CE}}[\hat f]&=-\int\dd\bx\,\dd\bz\,\sigma(\btn,\bnn) \,p(\bx,\bz|\btn,\bnn)\left( \log \hat f(\bx)+ r(\bx,\bz|\bto,\bno,\btn,\bnn)\log (1-\hat f(\bx))\right).\label{eq:CE-1samp}
\end{align}
The two terms in Eq.~\ref{eq:CE-1samp} separately agree with the two expectations in Eq.~\ref{eq:CE-2samp}.
The approximation of Eq.~\ref{eq:CE-1samp} for a synthetic data set is
\begin{align}
    L_{\textrm{CE}}[\hat f]&\approx-\sum_{\mathcal{D}(\btn,\bnn)}w_{i}\left(\log \hat f(\bx_i)+r(\bx_i,\bz_i|\bto,\bno,\btn,\bnn)\log (1-\hat f(\bx_i))\right).\label{eq:CE-1samp-empircal}
\end{align}
The main difference to Eq.~\ref{eq:CE-2samp-empirical} is the absence of independent stochastic fluctuations in the two terms. In both cases, $\hat f(\bx)$ is implemented as a finitely but sufficiently expressive ML algorithm, approximating the exact solution. We denote this approximation by 
\begin{align}
\hat f (\boldsymbol{\bx})&\simeq\left(1+\frac{\dd\sigma(\bx|\bto,\bno)}{\dd\sigma(\bx|\btn,\bnn)}\right)^{-1}.
\end{align}

\subsection{Machine-learning systematic parametrizations}\label{sec:ml-parametrizations}

To learn parametrizations, we replace $\hat f$ with a suitable parametric ansatz that captures the $\bn$ dependence. The loss function is summed over a set of model parameter points (base points), denoted by $\mathcal{V}$. The parametrization can be determined using synthetic data from a sufficient number of base points. We omit \bt in the formulas, as we will factorize this dependence in Sec.~\ref{sec:refinable-likleihood}. The fully calibrated SM parameter point serves as the reference, $\bnn=\bzero$.
The ansatz
\begin{align}
\hat f(\bx)&=\frac{1}{1+\exp(\hat T(\bx|\bn))}\label{eq:parametric-ansatz}
\end{align} 
eliminates the monotonous dependence from Eq.~\ref{eq:CE-analytic-sol}. 
The ML estimate of the DCR is then $\hat S(\bx|\bn)=\exp(\hat T(\bx|\bn))$.
The exponential function removes the necessity of ensuring that $\hat S(\bx|\bn)$ must be positive.
Inserting Eq.~\ref{eq:parametric-ansatz} into Eq.~\ref{eq:CE-2samp}, leads to
\begin{align}
L_\textrm{CE}[\hat T(\bx|\bn)]&=\left\langle\textrm{Soft}^+(\hat T(\bx|\bn))\right\rangle_{\bx,\bz|\bzero}+\left\langle\textrm{Soft}^+(-\hat T(\bx|\bn))\right\rangle_{\bx,\bz|\bn}\label{eq:loss-analytic-SoftMax}
\end{align}
where $\textrm{Soft}^+(x)=\log(1+\exp(x))$.
Next, we approximate the logarithm of the DCR with a polynomial ansatz in \bn in terms of coefficient functions as
\begin{align}
\hat T(\bx|\bn)=\nu_a\hat\Delta_{a}(\bx)+\nu_a\nu_b\hat\Delta_{a,b}(\bx)+\ldots. \label{eq:gnu-expansion}
\end{align}
The ellipsis indicates that cubic or higher terms can be added as needed, allowing to parametrize the systematic effects, in principle, with arbitrary precision. The functional form is equivalent to the ansatz in Refs.~\cite{dAgnolo:2021aun, Brehmer:2019xox}. Notably, we include the possibility that some of the coefficient functions are chosen to be absent.

To determine $\hat\Delta_a(\bx)$, $\hat\Delta_{ab}(\bx)$, etc., via a suitable loss function, we note that Eq.~\ref{eq:gnu-expansion} is a \textit{linear} equation; only the coefficients in this system are polynomial in \bn.  Without loss of generality, we assume triangular coefficient functions $\hat \Delta_{abc\cdots}(\bx)$, i.e., $\hat \Delta_{abc}(\bx)=0$ unless $a\leq b\leq c$, etc., and we denote their total number by $N_{\Delta}$.
To reduce the notational clutter, we next introduce a multi-index\footnote{We use the Einstein sum convention for the nuisance parameter index, labeled by $a$, $b$, $\ldots$,  as well as for the multi-index labeled by $A$, $B$, $\ldots$.} $A=1,\ldots,N_{\Delta}$. In the most general case, $A$ labels the set $\{a,(ab),(abc),\ldots\}$ where $a$ labels the $N_\bn$ linear terms, $(ab)$ the $N_\bn(N_\bn+1)/2$ quadratic terms, etc. 
For a given nuisance-parameter point~$\bn$, we similarly write $\nu_A=\{\nu_a,\nu_{a}\nu_b,\nu_{a}\nu_b\nu_c,\ldots\}_A$, where each element corresponds to one of the coefficient functions.
This notation simplifies Eq.~\ref{eq:gnu-expansion} to
\begin{align}
    \hat T(\boldsymbol{x}|\bn)= \nu_A\hat\Delta_{A}(\bx).\label{eq:gnu-expansion-multi}
\end{align}

With a sufficiently large number of base points and the corresponding, possibly synthetic, data sets, we add up copies of the loss function in Eq.~\ref{eq:loss-analytic-SoftMax}, one for each $\bn\in\mathcal{V}$, 
\begin{align}
L[\hat\Delta_A(\bx)]&=\sum_{\bn\in\mathcal{V}}L_\textrm{CE}[\nu_A\hat\Delta_A(\bx)]\label{eq:sum-loss-generic}\\
&=\sum_{\bn\in\mathcal{V}}\Bigg(\left\langle\textrm{Soft}^+(\nu_A\hat\Delta_A(\bx))\right\rangle_{\bx,\bz|\bzero}+\left\langle\textrm{Soft}^+(-\nu_A\hat\Delta_A(\bx))\right\rangle_{\bx,\bz|\bn}\Bigg)\label{eq:softmax-loss}
\end{align}
It is straightforward to show that the minimum approximates the DCR as
\begin{align}
 \hat S(\bx|\bn)=\exp\left(\nu_A\hat\Delta_A(\bx)\right)&\simeq\frac{\dd\sigma(\bx|\bn)}{\dd\sigma(\bx|\bzero)},\label{eq:S-estimate}
\end{align}
if the base points $\mathcal{V}$ span the space of the nuisance parameters, i.e., the $N_\bn\times N_{\Delta}$-dimensional matrix of the base-point coordinates $\{\nu_A\}_{\bn\in\mathcal{V}}$ has at least rank $N_\Delta$. This implies, in particular, that we need at least as many base points and synthetic training data sets as there are independent coefficient functions.

Each term in the sum in Eq.~\ref{eq:softmax-loss} contains two expectations; note that the first expectation in each term corresponds to the SM. The coefficient functions $\hat\Delta_A(\bx)$ can be implemented as neural networks or any other trainable regressor. In most cases, the factorization of systematic effects is a reliable simplification, so the number of coefficient functions that need to be learned simultaneously remains small.
For a single nuisance, one or two coefficient functions are enough to achieve linear or quadratic accuracy in Eq.~\ref{eq:S-estimate}, which is usually sufficient. Higher-order terms can be added as needed.

\subsection{Two-point alternatives}\label{sec:2-point-alternatives}
For some systematic effects, like binary generator choices, no tractable simulation exists--only two alternate simulations. In this case, the sum over $\mathcal{V}$ in Eq.~\ref{eq:softmax-loss} becomes trivial, simplifying the learning task. A single nuisance parameter $\nu_{2P}$ interpolates between the nominal choice ($\nu_{\textrm{2P}}=0$) and the alternative ($\nu_{\textrm{2P}}=1$). Without predictions for more values, we can only learn a linear approximation. With $\nu_\textrm{2P}=1$ as the sole value in $\mathcal{V}$, we get
\begin{align}
L[\hat\Delta]=&\langle\textrm{Soft}^+(\hat\Delta(\bx))\rangle_{\bx,\bz|0}+\left\langle\textrm{Soft}^+(-\hat\Delta(\bx))\right\rangle_{\bx,\bz|1}\label{eq:softmax-loss-two-point}
\end{align}
where $\hat\Delta(\bx)$ is a single-valued coefficient function.
Minimization provides an estimate
\begin{align}
\hat S_{\textrm{2P}}(\bx|\nu_{\textrm{2P}})=\exp\left(\nu_{\textrm{2P}}\hat\Delta(\bx)\right)\simeq\left(\frac{\dd\sigma_1(\bx)}{\dd\sigma_0(\bx)}\right)^{\nu_{\textrm{2P}}},\label{eq:S-estimate-2-point}
\end{align}
which is an \bx-dependent linear interpolation of the logarithm of the DCR. 

When profiling the nuisance parameter $\nu_{2\textrm{P}}$, it takes values other than 0 and 1, even though predictions are only well-defined at these points. Is the interpolation meaningful during profiling? This is a modeling question and cannot be resolved by statistical or ML methodology.
As in the binned case, we generally recommend avoiding two-point alternatives and instead using a single model with meaningful and flexible parameters. Two-point alternatives, such as using alternative generators, are beneficial as cross-checks. When profiling the effects of $\hat S_{\textrm{2P}}(\bx|\nu_{2\textrm{P}})$, it should be ensured that its impact is not substantial or dominant; otherwise, the validity of the measurement could be in doubt. Regardless of the modeling decision, two-point alternatives are accounted for by Eq.~\ref{eq:S-estimate-2-point}.

\subsection{The Boosted Parametric Tree algorithm}\label{sec:bpt}

The coefficient functions in Eq.~\ref{eq:gnu-expansion-multi} could be implemented using standard neural networks. However, with hundreds of nuisance parameters in a realistic analysis, it is advantageous to develop a low-maintenance, flexible algorithm to separately learn the numerous systematic dependencies. In the following, we describe a tree-boosting regressor, the ``Boosted Parametric Tree''~(BPT), designed for learning systematic effects. Its complete derivation is provided in Appendix~\ref{sec:parametric-regression-trees-details}.

Boosted tree algorithms have a strong track record in classification and regression tasks and were recently applied to novel searches for resonant phenomena~\cite{Finke:2023ltw}. They use an additive sequence of weak learners, each generating a coarsely binned prediction based on hierarchical phase-space partitioning, which is computationally efficient. 
Each tree’s terminal nodes are linked to a predictive function that varies non-linearly across the phase-space boundaries of these nodes. Here, we extend standard tree-based regression algorithms, such as those in {\textsc{TMVA}}~\cite{Speckmayer:2010zz}, by introducing a more flexible terminal-node predictor that provides the DCR to arbitrary order in the expansion in \bn. The summed prediction from these weak learners, trained iteratively through boosting, is both smooth and arbitrarily expressive. 

The simplicity of the weak learner leads to relatively mild failure modes. For instance, if a tree is trained with insufficient data, it cannot extrapolate incorrectly to phase-space regions beyond the training data; it will simply predict the value of the highest populated bin it finds, respecting regularization requirements, such as a minimum number of events per terminal node. 

Additionally, a tree algorithm with axis-linear node splits, like the ``Classification and Regression Tree''~(CART)~\cite{breiman1984classification} algorithm, does not interpolate stepwise features in the training data. If features take only discrete values, the algorithm will partition phase space based on selections at these discrete values. Therefore, features related to object multiplicity need no special handling. Replacing nominal training data with digitized values according to a chosen binning can serve as validation, allowing sensitivity comparisons between unbinned and binned reference results.

The trees' terminal nodes can be associated with more complex quantities than simple class probabilities or regression values, as shown in applications learning polynomial SMEFT dependence~\cite{Chatterjee:2021nms,Chatterjee:2022oco}. 
We exploit this flexibility for developing a boosted tree algorithm where terminal nodes of the weak learner are linked to parametrizations of systematic effects of the training data in the nodes.
With the tools from Sec.~\ref{sec:ml-parametrizations}, the BPT provides a tree-based estimate $\hat T(\bx|\bn)$ of the logarithm of the DCR in the polynomial expansion
\begin{align}
     \hat T(\bx|\bn)=\nu_A\hat\Delta_A(\bx)\simeq\log\frac{\dd\sigma(\bx|\bn)\;\;\;}{\dd\sigma(\bx|\textrm{SM})}
\end{align}
so that $\hat S(\bx|\bn)=\exp(\hat T(\bx|\bn))$. 
The free parameters in $\hat\Delta_A(\bx)$ are trained with the CE loss function in Eq.~\ref{eq:softmax-loss} by an iterative boosting algorithm.
It fits the weak learners of an additive expansion, one at a time, to the pseudo-residuals of the preceding boosting iteration. The complete construction is discussed in Appendix~\ref{sec:parametric-regression-trees-details}. Here, we describe the resulting algorithm.

During training, each weak learner captures only part of the total parameter dependence in each terminal node, with this fraction controlled by the algorithm’s learning rate. We set a number $B$ of boosting iterations and choose learning rates $0 < \eta^{(b)} < 1$ for $b = 1, \ldots, B$ to form an additive expansion of $\hat T(\boldsymbol{x})$ in terms of the weak learners.
The $\eta^{(b)}$ can be chosen constant, and values between $10^{-3}$ and $3\cdot10^{-1}$ for this universal learning rate have proven efficient. 
At each iteration $b$, the weak learner is a tree with terminal nodes corresponding to phase-space partitioning—a set of non-overlapping regions $\mathcal{J}^{(b)}$ that together cover $\mathcal{X}$.
This leads to
\begin{align}
     \hat \Delta_A(\bx)=\sum_{b=1}^B\eta^{(b)}\sum_{j\in\mathcal{J}^{(b)}} \mathbbm{1}_j(\bx)\hat\Delta^{(b)}_{A,j}\label{eq:delta-hat-expansion}
\end{align}
where the indicator function $\mathbbm{1}_j(\bx)$ equals one if \bx is in the phase-space region of terminal node $j$ and zero otherwise. Training iteration $b$ involves finding the partitioning $\mathcal{J}^{(b)}$ whose terminal-node predictions minimize the loss. 
Each terminal node prediction is based on constants $\hat\Delta^{(b)}_{A,j}$, which are best-fit polynomial coefficients approximating the nuisance parameter dependence in terminal node $j$. These coefficients, labeled by the multi-index $A$, are determined from the training sets $\mathcal{D}_\bzero$ and $\mathcal{D}^{(b)}_\bn$, where we have one $\mathcal{D}^{(b)}_\bn$ for each $\bn \in \mathcal{V}$. We initialize with $\hat \Delta^{(0)}_{A,j}=0$. 

To proceed from iteration $b-1$ to $b$, we remove a fraction $\eta^{(b)}$ of the previous iteration’s fit result from the training data. Since we estimate the logarithm of the DCR, the reweighting
\begin{align}
    \mathcal{D}_{\bn}^{(b)}= \left\{\exp(-\eta^{(b-1)} t^{(b-1)\ast}(\bx_i|\bn))\,w^{(b-1)}_i, \bx_i,\bz_i\right\} \;\; \textrm{for all} \;\; \{w^{(b-1)},\bx_i,\bz_i\}\in\mathcal{D}_{\bn}^{(b-1)}\;\;\textrm{for all}\;\;\bn\in\mathcal{V},
\end{align}
produces the corresponding $\mathcal{D}_{\bn}^{(b)}$.
\begin{algorithm}[tbh]
\caption{Boosted Parametric Tree~(BPT) for learning of systematic uncertainties}\label{alg:boosting}
\begin{algorithmic}
\Require base points $\bn\in\mathcal{V}$, sample $\mathcal{D}_\bzero$ and $\mathcal{D}_{\bn}\;\textrm{for all} \;\mathbf{\nu}\in\mathcal{V}$,\\\quad\quad\quad\;\,
boosting iterations B, learning rates $0\leq\eta^{(b)}\leq1$ for $b=1,...,B$.
\Ensure $\sum_{\bn\in\mathcal{V}}\nu_A\nu_B$ has full rank
\State $t^{(0)\ast}_{\bn}(\bx)\gets 0$
\State $\hat T^{(0)}_{\bn}(\bx) \gets 0$
\State $\mathcal{D}^{(0)}_{\bn} \gets  \mathcal{D}_{\bn}$ for all $\bn\in\mathcal{V}$
\For{$b=1,\dots,B$} 
\State  $\mathcal{D}_{\bn}^{(b)}\gets \left\{w^{(b)}_i \gets \exp(-\eta^{(b-1)} t^{(b-1)\ast}_{\bn}(\mathbf{x_i}))\,w^{(b-1)}_i, \bx_i,\bz_i\right\}$ for all $\{w^{(b-1)},\bx_i,\bz_i\}\in\mathcal{D}_{\bn}^{(b-1)}$
\State $\mathcal{J}^{(b)} \gets \argmin_{\mathcal{J}}L[\mathcal{J}]$ with $\mathcal{D}^{(b)}_{\bzero}$ and $\mathcal{D}^{(b)}_{\mathbf{\bn}}$ using CART or TAO
\ForAll{$j\in \mathcal{J}^{(b)}$}
\State $\sigma_{j,\bzero}\gets\sum_{(\bx_i, 
w_i)\in\mathcal{D}_{\bzero}\cap j}w_{i}$
\State $\sigma_{j,\bn}\gets\sum_{(\bx_i, 
w_i)\in\mathcal{D}_{\bn}^{(b)}\cap j}w_{i}\quad\textrm{for all}\quad\bn\in\mathcal{V}$
\State $\hat\Delta^{(b)}_{A,j}\gets\left[\;\sum_{\nu\in\mathcal{V}}\bn\bn^{\textrm{T}}\right]_{AB}^{-1} \left[\sum_{\;\bn\in\mathcal{V}}\bn\log\frac{\sigma_{j,\bn}}{\sigma_{j,\bzero}}\right]_B$
\EndFor
\State $t^{(b)\ast}_{\bn}(\bx)\gets \sum_{j\in\mathcal{J}^{(b)}}\mathbbm{1}_j(\bx)\left(\nu_A\hat\Delta^{(b)}_{A,j}\right)$
\State $\hat T^{(b)}_{\bn}(\bx)\gets\hat T^{(b-1)}(\bx)+\eta^{(b)}t^{(b)\ast}(\bx)$
\EndFor
\State \Return $\hat T(\bx|\bn)=\sum_{b=1}^B\eta^{(b)}\sum_{j\in\mathcal{J}^{(b)}}\mathbbm{1}_j(\bx)\nu_A\hat\Delta^{(b)}_{A,j}$
\end{algorithmic}
\end{algorithm}
The nominal SM training sample $\mathcal{D}_\bzero$ is unchanged. 

The quantity in the exponent is the best fit at iteration $b-1$,
\begin{align}
t^{(b-1)\ast}(\bx|\bn)=\nu_A\sum_{j\in\mathcal{J}^{(b-1)}} \mathbbm{1}_j(\bx)\hat\Delta^{(b-1)}_{A,j},
\end{align} 
whose polynomial coefficients also appear on the r.h.s. of Eq.~\ref{eq:delta-hat-expansion}. 
To obtain $\hat\Delta_{A,j}^{(b)}$, we use the new training data to predict per-node cross-section values 
\begin{align}
\sigma^{(b)}_{j,0}=\sum_{(\bx_i,w_i)\,\in\,\mathcal{D}^{(b)}_\bzero\cap\Delta\bx_j}w_i\qquad\textrm{and}\qquad\sigma^{(b)}_{j,\bn}=\sum_{(\bx_i,w_i)\,\in\,\mathcal{D}^{(b)}_\bn\cap\Delta\bx_j}w_i\label{eq:sigma-at-each-nu}
\end{align}
for the nominal $\bzero$ and each $\bn\in\mathcal{V}$. The notation $\mathcal{D}^{(b)}_\bn\cap\Delta\bx_j$ indicates summing over events from $\mathcal{D}^{(b)}_\bn$ that fall within the phase-space region $\Delta\bx_j$ of terminal node $j$.
These estimates, valid for $\bn=\bzero$ and $\bn\in\mathcal{V}$, yield the new polynomial interpolation at iteration $b$, with coefficients
\begin{align}
\hat\Delta_{A,j}^{(b)}=\left[\;\sum_{\nu\in\mathcal{V}}\bn\bn^{\textrm{T}}\right]_{AB}^{-1} \left[\sum_{\;\bn\in\mathcal{V}}\bn\log\frac{\sigma^{(b)}_{j,\bn}}{\sigma^{(b)}_{j,\bzero}}\right]_B.\label{eq:weak-learner-prediction}
\end{align}
The inverse matrix in the first factor exists if $\mathcal{V}$ has a full-rank coordinate matrix, so we need at least as many training samples $\bn\in\mathcal{V}$ as there are coefficient functions $\hat\Delta_A(\bx)$. This linear relation of log-ratios makes Eq.~\ref{eq:weak-learner-prediction} highly efficient to evaluate. We can then use the CART or ``Tree Alternate Optimization''~(TAO)~\cite{TAO-1,TAO-2,TAO-3,TAO-4} algorithms to determine the optimal phase-space partitioning $\mathcal{J}^{(b)}$, completing iteration $b$.
After $B$ boosting iterations, all constants in Eq.~\ref{eq:delta-hat-expansion} are determined, giving the final DCR estimate 
\begin{align}
     \hat S(\bx|\bn)=\exp(\nu_A\hat\Delta_A(\bx))\simeq\frac{\dd\sigma(\bx|\bn)\;\;\;}{\dd\sigma(\bx|\textrm{SM})}
\end{align}
It is fast to evaluate, parametric in $\bn$, satisfies $\hat S(\bx|\bzero)=1$, and is continuous in both $\bx$ and $\bn$. 
Algorithm~\ref{alg:boosting} provides a pseudo-code summary of the steps, defining the BPT algorithm. Appendix~\ref{sec:parametric-regression-trees-details} contains a detailed derivation and a simple analytic toy example.

\section{Gradually refinable modeling}\label{sec:refinable-likleihood}

With the setup for the training of parametric regressors in place, we discuss gradually refinable modeling as a flexible approach to unbinned analyses, incorporating incremental improvements while preserving existing results. 
Based on procedures commonly employed in binned analyses, we discuss the factorization of POI and nuisance parameter dependencies in the unbinned case. Uncorrelated groups of systematic effect are isolated and subsequently learned independently. With the help of an additive model, summing over the various contributing processes, we can incrementally extend and refine an existing model without invalidating existing components.
In this way, gradually refinable modeling aligns with established practices for systematic uncertainty management in binned analyses, extending these strategies to unbinned data while allowing the analysis to evolve with growing data and improved modeling techniques.

\subsection{The binned Poisson likelihood}

We begin with the binned Poisson likelihood for several observations ($N_{\textrm{bin}}$) in disjoint phase-space regions (bins), a setup described in detail in Refs.~\cite{CMS:2024onh,Cranmer:2012sba}. Multiple processes, labeled by $p=1,\ldots,N_p$, contribute to the cross-section component $\sigma_{n,p}(\textrm{SM})$ in bin $n$. 
The dependence on SMEFT POIs is assumed to factorize from the systematic effects. Since SMEFT ME-squared terms are polynomial or can be truncated to polynomial form, a small set of non-zero values of $\bt$ suffices to determine the coefficients in the SMEFT parametrization $\sigma_{n,p}(\theta)=R_{n,p}(\bt)\sigma_{n,p}(\textrm{SM})$~\cite{Belvedere:2024nzh}. Here, $R_{n,p}(\bt)$ satisfies $R_{n,p}(\textrm{SM})=1$ and fully encodes the SMEFT dependence in each bin.
If $R_{n,p}(\bt)= 1$ holds to a good approximation for all $n$, we call the process $p$ a background. 

The Poisson expectation of the yield in bin $n$ can then be expressed as
\begin{align}
\lambda_n(\bt,\bn)=\mathcal{L}(\bn)\sum_{p=1}^{N_p}R_{n,p}(\bt)\,\exp\left(\bn^\intercal\Delta_{n,p,1}+\nu^\intercal\Delta_{n,p,2}\bn\right)\sigma_{n,p}(\textrm{SM}),\label{eq:binned-expansion}
\end{align}
where the exponential is a second-order interpolation\footnote{A detailed account of the options for interpolating binned yields is provided in Ref.~\cite{Cranmer:2014lly}.}  of the systematic effects in terms of $K$-dimensional vectors $\Delta_{n,p,1}$ and  $K\times K$ matrices $\Delta_{n,p,2}$. The nuisances \bn are conventionally chosen to minimize their linear correlation. In the uncorrelated case, off-diagonal entries in $\Delta_{n,p,2}$ vanish. Small linear nuisance parameter correlations can be accounted for in the penalty~\cite{CMS:2024onh}. 

The binned ansatz in Eq.~\ref{eq:binned-expansion} reflects inductive bias. First, systematic effects are modeled in a factorized form, meaning the constants $\Delta_{n,p,1/2}$ are assumed to be accurately determined by individually varied simulations, with all other model parameters held fixed. Second, the model is additive. 
This inconspicuous fact, combined with the factorization of systematic effects, is key to enabling gradual model refinement, an implicit feature in the binned case. Once the per-process expectations and systematic parametrizations in Eq.~\ref{eq:binned-expansion} are established, most of these values can remain unchanged even if the model is refined to include a new process or nuisance parameter. Although this computational saving is modest in the binned case, the unbinned analysis replaces the bin-by-bin constants in Eq.~\ref{eq:binned-expansion} with $\bx$-dependent ML parametrizations. Selecting a flexible additive model minimizes the need for re-training when refining the model, offering potentially significant gains in computational efficiency.

The additivity in Eq.~\ref{eq:binned-expansion} naturally accommodates per-process nuisances related to normalization uncertainties with two key applications. First, higher-order perturbative corrections (``k-factors''), derived from theoretical predictions, enhance the accuracy of inclusive parton-level predictions. These $k$-factors generally apply to a single process, with reduced uncertainties best captured by nuisances that scale only this component. Second, normalization nuisances are useful for small backgrounds where the pdf in $\mathcal{D}$ can be estimated from $\mathcal{A}$, but normalization uncertainties remain significant. In this case, a normalization nuisance allows for in-situ constraints from $\mathcal{D}$.
Specifically, setting $\Delta_{n,p,1}=\log\alpha_{\textrm{norm},\,p}$ and $\Delta_{n,p,2}=0$ for all $n$ results in a scaling of process $p$, where $\alpha_{\textrm{norm},\,p}$ is a positive constant that normalizes the impact of the nuisance $\nu_{\textrm{norm},\,p}$. Additionally, we can omit $\nu_{\textrm{norm},\,p}$ from the penalty, allowing the process's normalization to float during profiling.

\subsection{Approximate factorization of systematic effects}\label{sec:unbinned}

We now substitute the DCR in Eq.~\ref{eq:likleihoodratio} with an ML surrogate model. ``Likelihood-free'' inference refers to techniques that rely on parametrically evaluating ratios of the extended likelihood, and thus ratios of the differential cross-section $\textrm{d}\Sigma(\bt,\bn)$. These alone are enough to evaluate Eq.~\ref{eq:test-stat}.

Constructing a generic ML surrogate starts by expressing the unbinned model $\textrm{d}\Sigma(\bt,\bn)$ as a sum over weighted sub-processes, with normalization uncertainties treated separately\footnote{The reason for the separate treatment of normalization nuisances is best seen in comparing the Taylor expansion in $\bn$ with the corresponding expansion of the purely multiplicative model in Ref.~\cite{Brehmer:2019xox} where nuisances are modeled relative to the total differential cross-section instead of per-process. For arbitrary values of normalization nuisances, a polynomial expansion of the logarithm of the {\textit{total}} differential cross-section requires arbitrarily many terms that would have to be learned individually. The ansatz in Eq.~\ref{eq:norm-nuisance-ansatz} will reduce the ensuing ML task to a straightforward classification problem, one for each process.}. Nuisance parameters $\nu_{p,\textrm{norm}}$ are introduced for this purpose.
We have
\begin{align}
\dd\Sigma(\bx|\bt,\bn)&=\sum_p\alpha_{\textrm{norm},\,p}^{\nu_{\textrm{norm},\,p}}\;\dd\sigma_p(\bx|\bt,\bn),\label{eq:norm-nuisance-ansatz}
\end{align}
where event samples for each component $\dd\sigma_p(\bx|\bt,\bn)$ can be obtained from Eq.~\ref{eq:det-level-sample}, Eq.~\ref{eq:synthetic-calibration}, or Eq.~\ref{eq:synthetic-reweighting}.
Next, we factorize systematic effects and POI dependence. The SM point is at \bt=\bn=\bzero, and for each $\dd\sigma_p(\bx|\bt,\bn)$ we have
\begin{align}
\frac{\dd\sigma_p(\bx|\bt,\bn)}{\dd\sigma_p(\bx|\bzero,\bzero)\,}&=\frac{\dd\sigma_p(\bx|\bt,\bn)\,}{\dd\sigma_p(\bx|\bzero,\bn)\;}\frac{\dd\sigma_p(\bx|\bzero,\bn)}{\dd\sigma_p(\bx|\bzero,\bzero)}\nonumber\\
&\approx\frac{\dd\sigma_p(\bx|\bt,\bzero)}{\dd\sigma_p(\bx|\bzero,\bzero)\,}\frac{\dd\sigma_p(\bx|\bzero,\bn)}{\dd\sigma_p(\bx|\bzero,\bzero)}\equiv\underbrace{\frac{\dd\sigma_p(\bx|\bt,\bzero)\,}{\dd\sigma_p(\bx|\textrm{SM})\;\,}}_{\hat R_p(\bx|\bt)}\underbrace{\frac{\dd\sigma_p(\bx|\bzero,\bn)}{\dd\sigma_p(\bx|\textrm{SM})\,}}_{\hat S_p(\bx|\bn)}.
\end{align}
This factorization works if SMEFT effects are independent of systematic effects, i.e.,
\begin{align}
\frac{\dd\sigma_p(\bx|\bt,\bn)}{\dd\sigma_p(\bx|\bzero,\bn)}\approx\frac{\dd\sigma_p(\bx|\bt,\bzero)}{\dd\sigma_p(\bx|\bzero,\bzero)}\label{eq:factorization}.
\end{align}
The factor
\begin{align}
\hat R_p(\bx|\bt)\simeq\frac{\dd\sigma_p(\bx|\bt,\bzero)}{\dd\sigma_p(\bx|\textrm{SM})\;}
\end{align}
approximates SMEFT variations as a polynomial in \bt and can be obtained from methods in Refs.~\cite{GomezAmbrosio:2022mpm,Chatterjee:2022oco,Chatterjee:2021nms,Chen:2020mev,Chen:2023ind,Cranmer:2015bka,Brehmer:2018eca,Brehmer:2018hga,Brehmer:2019xox,Brehmer:2018kdj}. Systematic effects are parametrized by
\begin{align}
\hat S_p(\bx|\bn)\simeq\frac{\dd\sigma_p(\bx|\bzero,\bn)\,}{\dd\sigma_p(\bx|\textrm{SM})\;\,}.\label{eq:general-sys-param}
\end{align}
We can learn this parametric dependence, one effect at a time, using the strategy in Sec.~\ref{sec:systematics-learning}. The ML model can be a neural network or the tree-based algorithm from Sec.~\ref{sec:bpt}.

The validity of Eq.~\ref{eq:factorization} must be established on a case-by-case basis and can be verified through simulation. The separation of particle and detector-level effects from SMEFT effects is generally accurate due to the different energy scales involved; the POIs typically do not influence low-energy detector interactions. At the parton level, we need to verify the independence of POIs from systematic effects. PDFs, for example, may depend on SMEFT POIs~\cite{Kassabov:2023hbm}, so this correlation should not be neglected without careful consideration.
Similarly, the linear and quadratic SMEFT terms may have scale uncertainties differing from the SM prediction. In this case, $\hat S_p(\bx|\nu_R,\nu_F)$ should be trained with synthetic scale variations that cover scale variations for non-zero POIs. A suitably flexible model should accommodate these subtle analysis-dependent effects, which we leave to future treatment.
From now on, we assume the factorization
\begin{align}
\hat R_p(\bx|\bt)\,\hat S_p(\bx|\bn)\simeq\frac{\dd\sigma_p(\bx|\bt,\bn)}{\dd\sigma_p(\bx|\textrm{SM})\,}
\end{align}
holds accurately. Following the same steps, we factorize $\hat S_{p}(\bx|\bn)$ into uncorrelated groups of systematic uncertainties and train each factor with Eq.~\ref{eq:S-estimate}. 
For instance, uncorrelated one-parameter systematic uncertainties with quadratic accuracy simplify the surrogate to
\begin{align}
\hat S_p(\bx|\bn)=\prod_{k=1}^K\exp\left(\nu_k\hat\Delta_{p,k,1}(\bx)+\nu_k^2\hat\Delta_{p,k,2}(\bx)\right)\label{eq:S-fac-2}
\end{align}
with $2K$ real-valued functions $\hat\Delta_{p,k,1}(\bx)$ and $\hat\Delta_{p,k,2}(\bx)$ for each $p$. In most cases, first or second-degree polynomials are sufficient, though the method allows higher degrees.

\subsection{A general unbinned surrogate model}
In analogy to Eq.~\ref{eq:binned-expansion}, we define a general model for the fiducial differential cross-section,
\begin{align}
\dd\Sigma(\bx|\bt,\bn)=\sum_{p=1}^{N_p}\,\hat R_p(\bx|\bt)\,\alpha_{\textrm{norm},\,p}^{\nu_{\textrm{norm},\,p}}\,
\hat S_p(\bx|\bn)
\,\dd\sigma_p(\bx|\textrm{SM}).\label{eq:general-diff-xsec-ansatz}
\end{align}
Next, we create a likelihood-free ML surrogate, relying only on differential cross-section ratios. Dividing by
\begin{align}
\dd\Sigma(\bx|\textrm{SM})=\sum_p\dd\sigma_p(\bx|\textrm{SM}).\label{eq:general-diff-xsec-divisor}
\end{align}
gives per-process DCRs that we replace with surrogates.
The simplest approach divides each $\dd\sigma_p(\bx|\textrm{SM})$ in Eq.~\ref{eq:general-diff-xsec-ansatz} by Eq.~\ref{eq:general-diff-xsec-divisor},
\begin{align}
\frac{\dd\Sigma(\bx|\bt,\bn)}{\dd\Sigma(\bx|\textrm{SM})\;}&=\sum_{p=1}^{N_p}\,\hat R_p(\bx|\bt)\,\alpha_{\textrm{norm},\,p}^{\nu_{\textrm{norm},\,p}}\,
\hat S_p(\bx|\bn)
\,\hat g_p(\bx)\quad\textrm{where}\quad\hat g_p(\bx)\simeq\frac{\dd\sigma_p(\bx|\textrm{SM})}{\sum_{q}\dd\sigma_{q}(\bx|\textrm{SM})}\label{eq:generic-ratio}
\end{align}
estimates the DCR of each process relative to the SM total.

A classifier trained to distinguish process $p$ from the total SM simulation can learn $\hat g_p(\bx)$ using the likelihood ratio trick. The DCR for arbitrary denominators, as required for profiling in Eq.~\ref{eq:test-statistic}, can then be obtained from double ratios, allowing $\dd\Sigma_p(\bx|\textrm{SM})$ to cancel out.

Equation~\ref{eq:generic-ratio} provides significant modeling flexibility, as the quotient of Eq.~\ref{eq:general-diff-xsec-ansatz} and Eq.~\ref{eq:general-diff-xsec-divisor} can be represented in various ways using the surrogates $\hat g_p(\bx)$. We present two examples demonstrating how this flexibility can solve common challenges in analysis development.

\subsection{Refining an existing model}\label{sec:refining-existing-model}
New systematic effects can be gradually incorporated, as expanding the dimension of the nuisance vector \bn in, for example, Eq.~\ref{eq:S-fac-2} does not invalidate existing surrogates $\hat\Delta_{p,1/2}$. Only the new components require training.

Similarly, additional background sources can be seamlessly included. With the additive structure of Eq.~\ref{eq:general-diff-xsec-ansatz}, a new process can be added by adjusting the steps leading to Eq.~\ref{eq:generic-ratio}. For instance, if a missing background $\dd\sigma_{\textrm{BKG}}(\bx)$, with $R_p(\bx|\bt)=1$, is identified, we can extend $\dd\Sigma(\bx|\bt,\bn)$ by adding a term,
\begin{align}
\dd\Sigma'(\bx|\bn,\bt)=\dd\Sigma(\bx|\bn,\bt)+\dd\sigma_{\textrm{BKG}}(\bx).\label{eq:extended-model}
\end{align}
We can express the DCR in terms of the existing model as 
\begin{align}
\frac{\dd\Sigma'(\bx|\bt,\bn)}{\dd\Sigma'(\bx|\textrm{SM})\;}&=\frac{\dd\Sigma(\bx|\bt,\bn)+\dd\sigma_{\textrm{BKG}}(\bx)}{\dd\Sigma(\bx|\textrm{SM})+\dd\sigma_{\textrm{BKG}}(\bx)\;}=\frac{\frac{\dd\Sigma(\bx|\bt,\bn)}{\dd\Sigma(\bx|\textrm{SM})}+\frac{\dd\sigma_{\textrm{BKG}}(\bx)}{\dd\Sigma(\bx|\textrm{SM})}}{1+\frac{\dd\sigma_{\textrm{BKG}}(\bx)}{\dd\Sigma(\bx|\textrm{SM})}\;}\nonumber\\
&\simeq\frac{\sum_{p=1}^{N_p}\,\hat R_p(\bx|\bt)\,\alpha_{\textrm{norm},\,p}^{\nu_{\textrm{norm},\,p}}\,
\hat S_p(\bx|\bn)
\,\hat g_p(\bx)+\hat g'(\bx)}{1+\hat g'(\bx)}
\end{align}
where
\begin{align}
\hat g'_p(\bx)\simeq\frac{\dd\sigma_{\textrm{BKG}}(\bx)}{\sum_{q}\dd\sigma_{q}(\bx|\textrm{SM})}.\label{eq:gprime}
\end{align}
The only new component is $\hat g'(\bx)$, a classifier that gives the DCR for the new process relative to the previous total. The new process adds to both the numerator and denominator, with no change to the rest of the model. If the new background has uncertainties, we replace $\dd\sigma_{\textrm{BKG}}(\bx)$ with $\dd\sigma_{\textrm{BKG}}(\bx|\bn)$ in Eq.~\ref{eq:extended-model} and repeat the derivation, yielding
\begin{align}
\frac{\dd\Sigma'(\bx|\bt,\bn)}{\dd\Sigma'(\bx|\textrm{SM})\;}
&\simeq\frac{\sum_{p=1}^{N_p}\,\hat R_p(\bx|\bt)\,\alpha_{\textrm{norm},\,p}^{\nu_{\textrm{norm},\,p}}\,
\hat S_p(\bx|\bn)
\,\hat g_p(\bx)+\hat S_{\textrm{BKG}}(\bx|\bn)\hat g'(\bx)}{1+\hat g'(\bx)},
\end{align}
where $g'(\bx)$ from Eq.~\ref{eq:gprime} remains, and we only need to learn one extra factor, $\hat S_{\textrm{BKG}}(\bx|\bn)$, to model the background’s systematic effects,
\begin{align}
\hat S_\textrm{BKG}(\bx|\bn)\simeq\frac{\dd\sigma_\textrm{BKG}(\bx|\bzero,\bn)\,}{\dd\sigma_\textrm{BKG}(\bx|\textrm{SM})\;\,},
\end{align}
similar to Eq.~\ref{eq:general-sys-param}. Refinable modeling thus avoids retraining existing regressors and enables incremental analysis development. Because an event sample for $\dd\sigma_{\textrm{BKG}}$ is the only ingredient for obtaining $\hat g'(\bx)$, it could alternatively be measured in real-data side bands, supporting the development of data-driven unbinned estimation strategies. 

\subsection{Refinement for a high-purity process}\label{sec:refining-high-purity}

Now, consider a single SMEFT-dependent signal process $\dd\sigma_{\textrm{SMEFT}}(\bx|\bt,\bn)$ and multiple \bt-independent backgrounds, $\dd\sigma_{p}(\bx|\bn)$. To form the DCR, we divide each term in the sums of both sides of Eq.~\ref{eq:generic-ratio} by $\dd\sigma_{\textrm{SMEFT}}(\bx|\textrm{SM})$, which leaves the result unchanged but modifies the parametrization to
\begin{align}
\frac{\dd\Sigma(\bx|\bt,\bn)}{\dd\Sigma(\bx|\textrm{SM})\;}&\simeq\frac{\alpha_{\textrm{norm}}^{\nu_\textrm{norm}}\;\hat R(\bx|\bt)\;\hat S(\bx|\bn)+\sum_{p=1}^{N_p}\,\alpha_{\textrm{norm},\,p}^{\nu_{\textrm{norm},\,p}}\,
\hat S_p(\bx|\bn)
\,\hat g''_p(\bx)}{1+\sum_p\hat g''_p(\bx)}
\label{eq:1samp-ratio}
\end{align}
where
\begin{align}
    g''_p(\bx)\simeq\frac{\dd\sigma_p(\bx|\textrm{SM})}{\dd\sigma_{\textrm{SMEFT}}(\bx|\textrm{SM})},
\end{align}
and signal quantities have no process index. The classifier $\hat g''(\bx)$ is trained to distinguish each background from the SMEFT signal at the SM point. Adding a new background process only requires training one additional classifier, as expanding the sum over $p$ leaves the rest of the model unaffected.

The steps in this and the previous section can be combined and repeated as modeling is refined. Equation~\ref{eq:general-diff-xsec-ansatz} thus supports incremental refinements, similar to the binned model in Eq.~\ref{eq:binned-expansion}. Once a new effect prediction is available, it can be incorporated. Unlike in the binned case, expectations here come from separately trained surrogates rather than binned yields.

\section{Top quark pair production in the \texorpdfstring{$2\ell$}{2l} channel}\label{sec:ttbar}

As an example, we study dileptonic top quark pair production in pp collisions at $\sqrt{s}=13~\TeV$, $\textrm{pp}\rightarrow\ttbar\rightarrow\textrm{b}\ell^+\nu_\ell\overline{\textrm{b}}\ell^-\overline{\nu}_\ell$, or $\ttbar(2\ell)$ for short. The event simulators provide all necessary quantities at the parton and particle levels. For calibration of reconstructed objects (jets, missing transverse momentum, and leptons), ATLAS and CMS open data projects~\cite{CMS-Open-Data,ATLAS-Open-Data} supply uncertainty information. This is enough to demonstrate the tools’ application. A fully realistic detector simulation with all data-dependent systematic effects is neither feasible nor needed.
We focus on a heuristic treatment of the main uncertainties in the differential cross-section. A detailed binned measurement of $\ttbar(2\ell)$, including a full account of systematic uncertainties, is available from ATLAS~\cite{ATLAS:2023gsl} and CMS~\cite{CMS:2024ybg}.
We focus on a heuristic treatment of the main uncertainties in the differential cross-section. A detailed binned measurement of $\ttbar(2\ell)$, including a full account of systematic uncertainties, is available from ATLAS~\cite{ATLAS:2023gsl} and CMS~\cite{CMS:2024ybg}.

\subsection{Event simulation}\label{sec:tt2l-event-simulation}

\begin{figure}[pth]
\centering
\includegraphics[width=.305\textwidth]{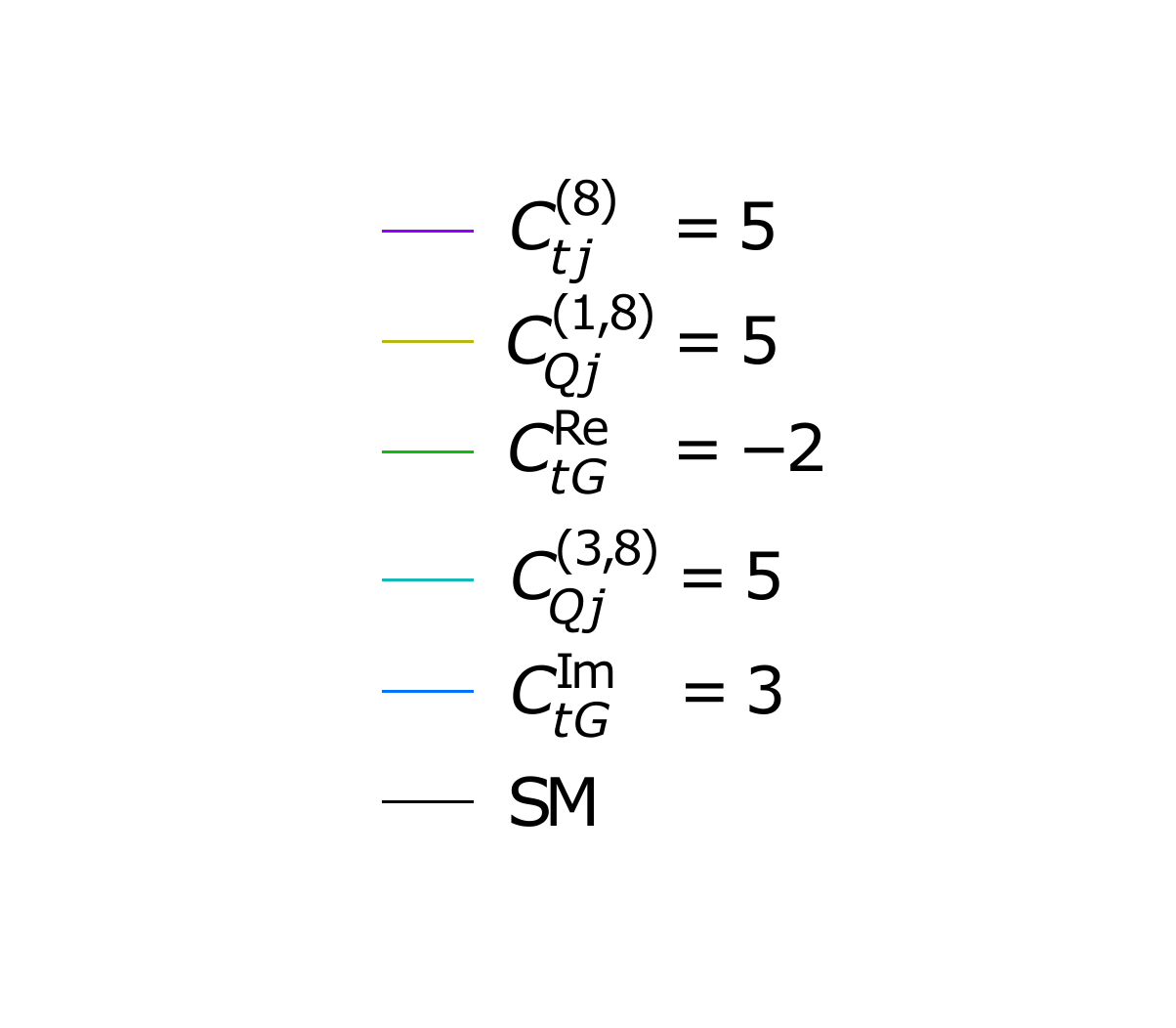}\hfill
\includegraphics[width=.305\textwidth]{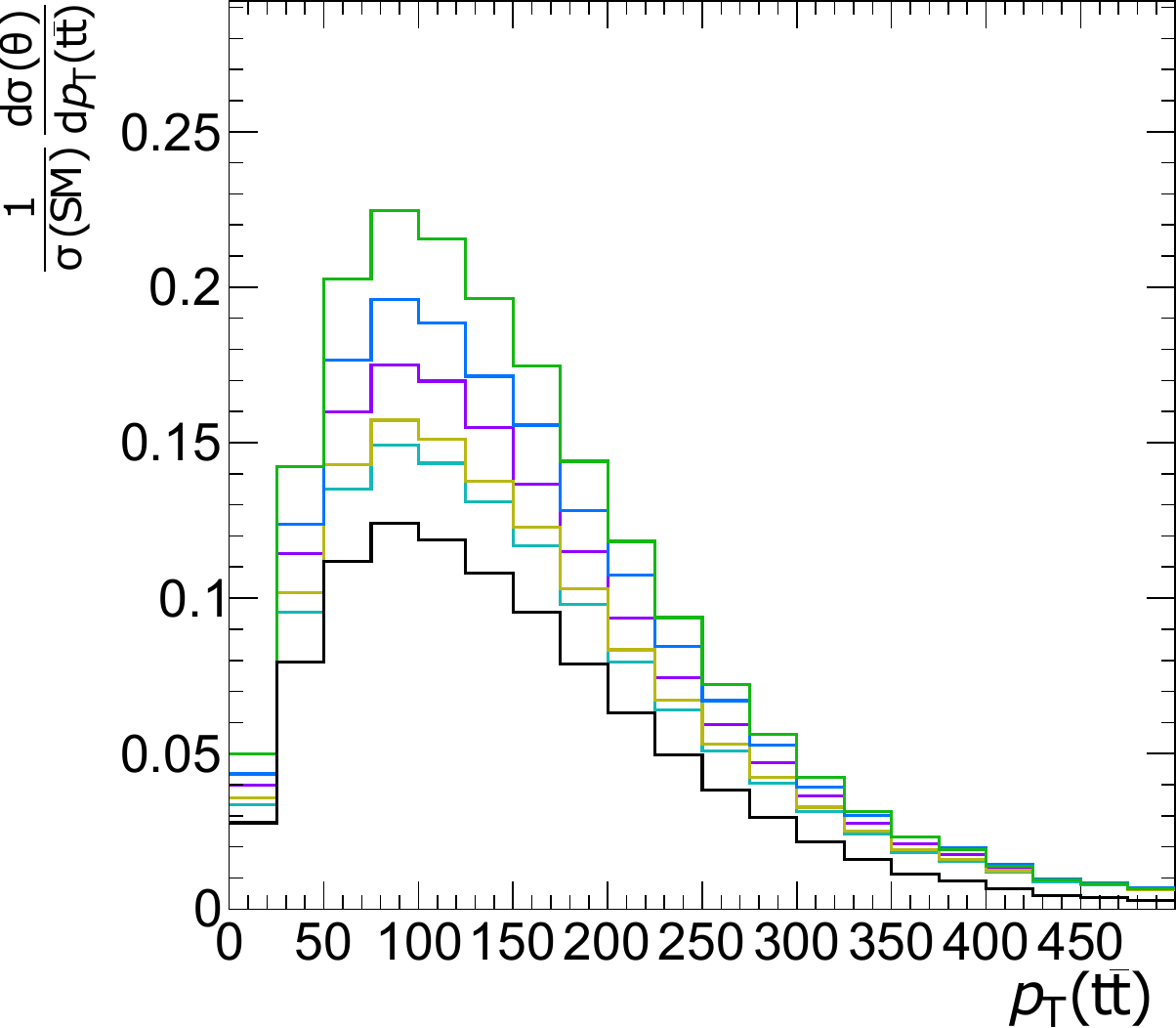}\hfill
\includegraphics[width=.305\textwidth]{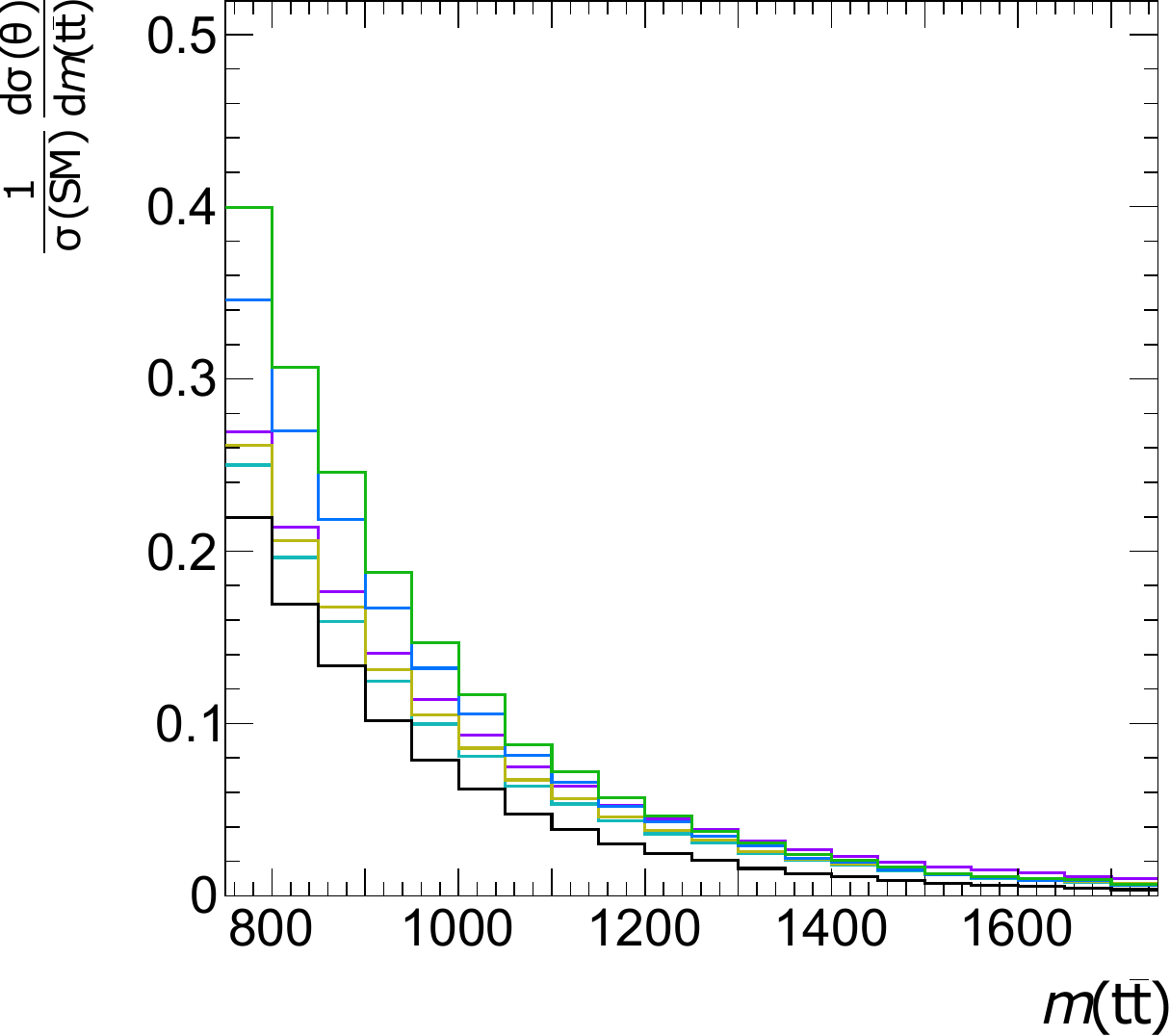}\\\vspace{.2cm}
\includegraphics[width=.305\textwidth]{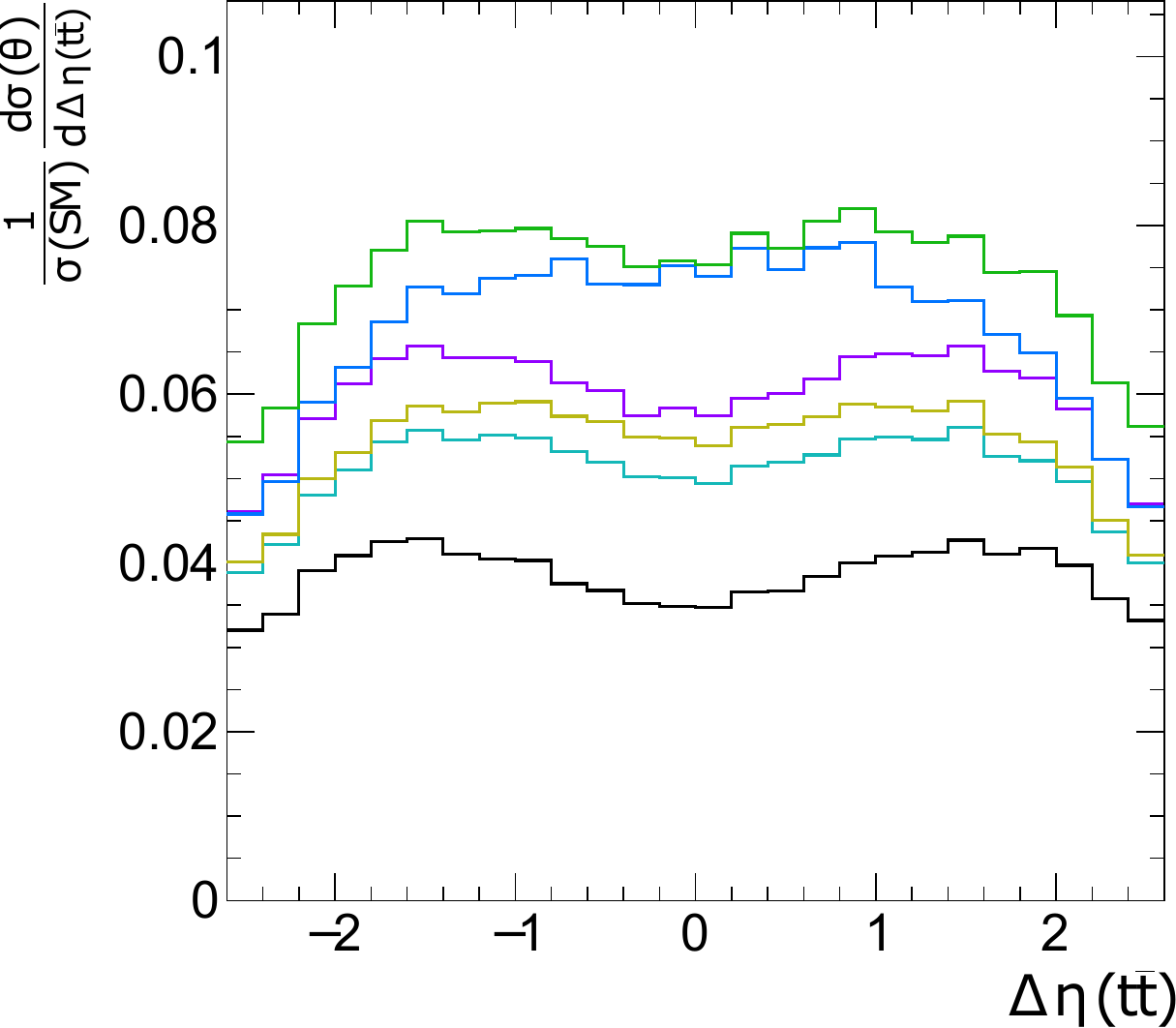}\hfill
\includegraphics[width=.305\textwidth]{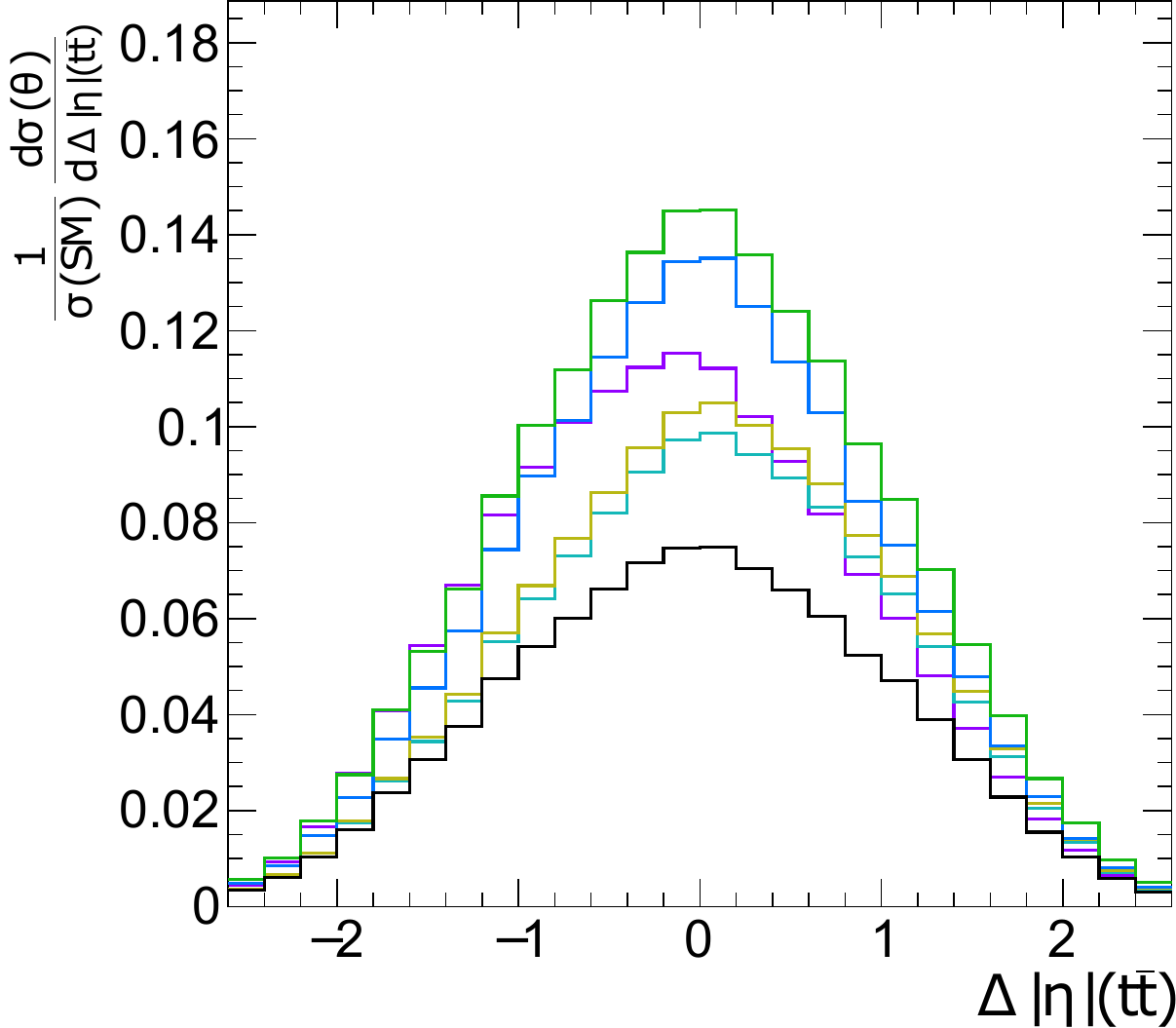}\hfill
\includegraphics[width=.305\textwidth]{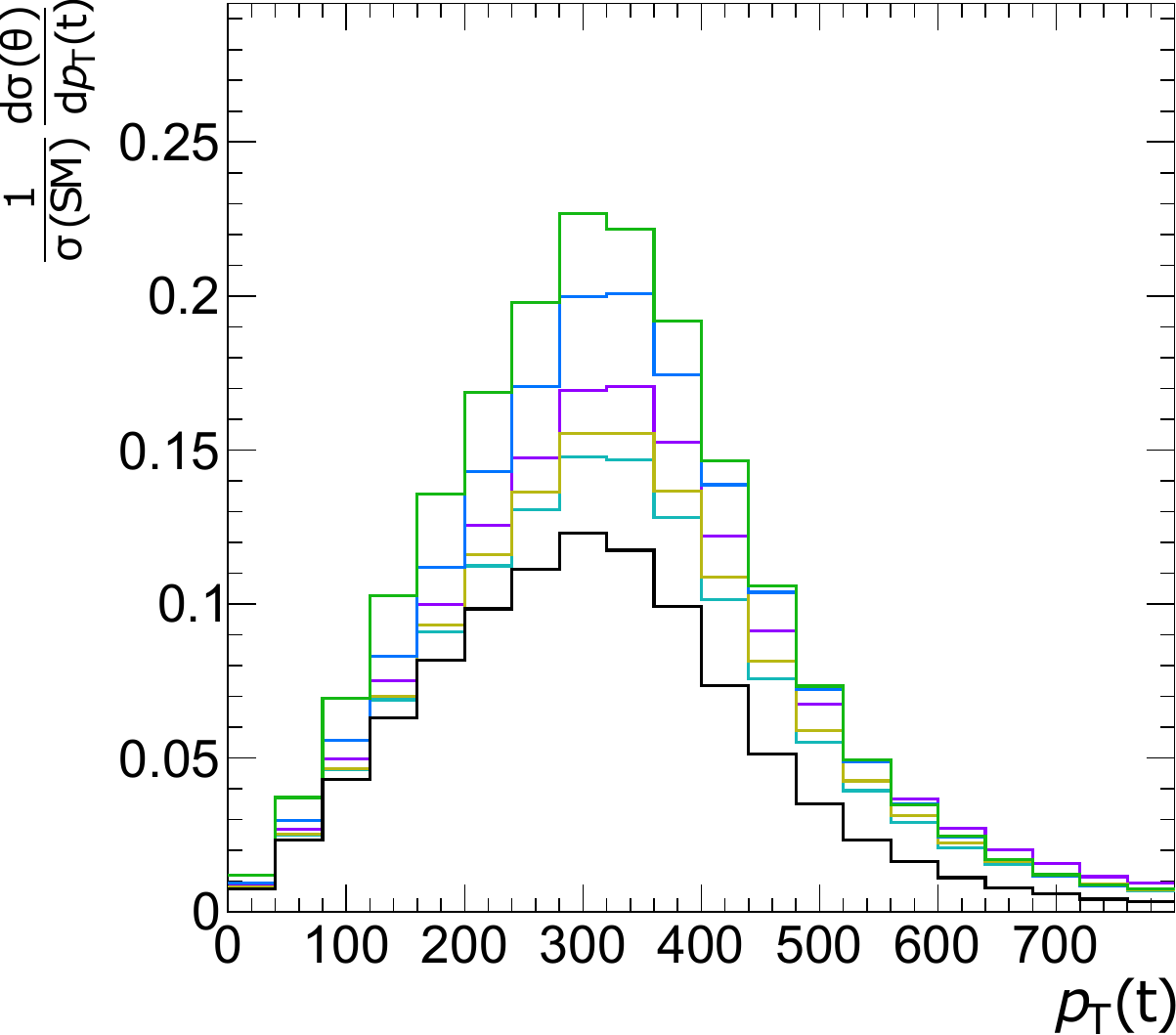}\\\vspace{.2cm}
\includegraphics[width=.305\textwidth]{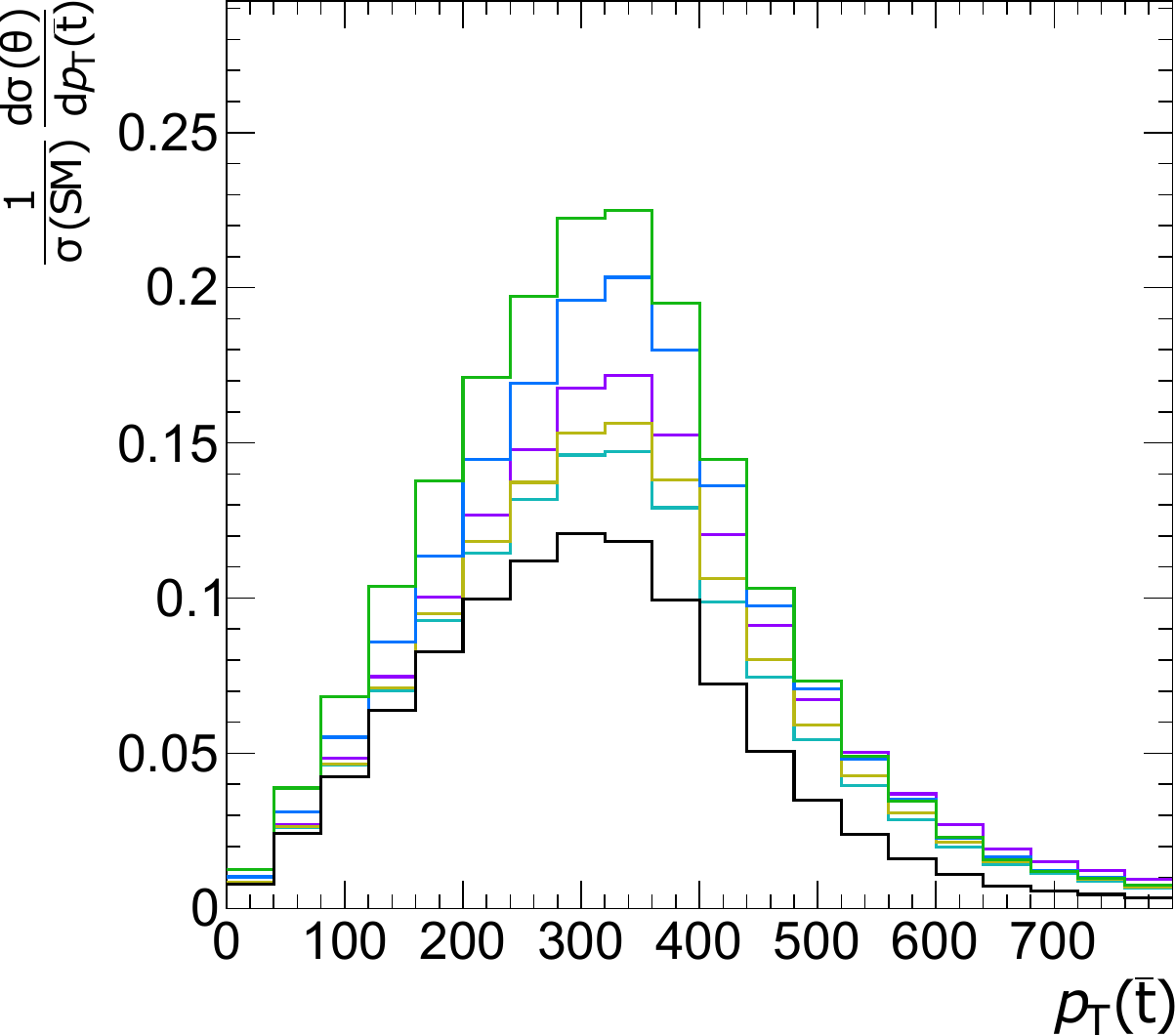}\hfill
\includegraphics[width=.305\textwidth]{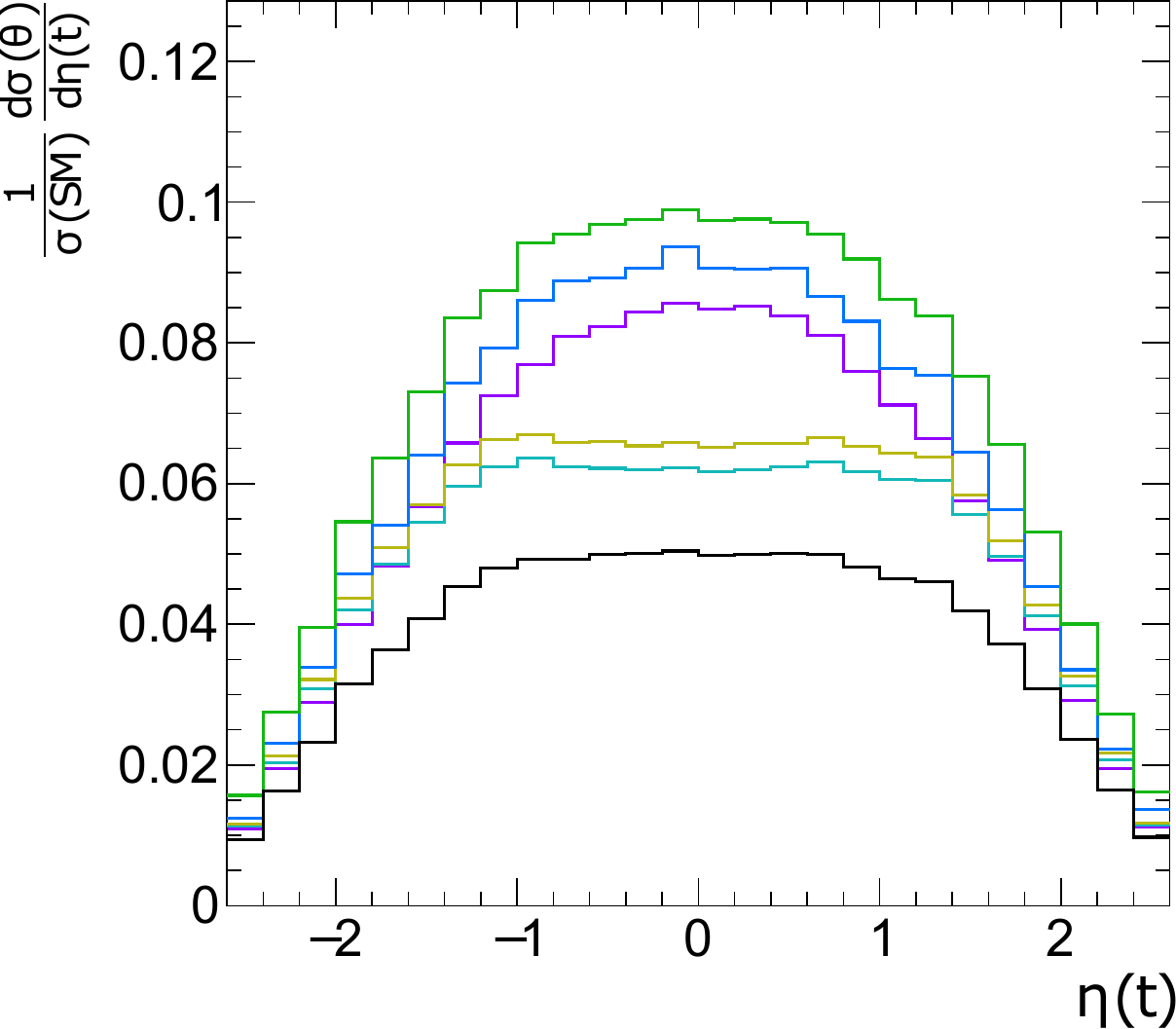}\hfill
\includegraphics[width=.305\textwidth]{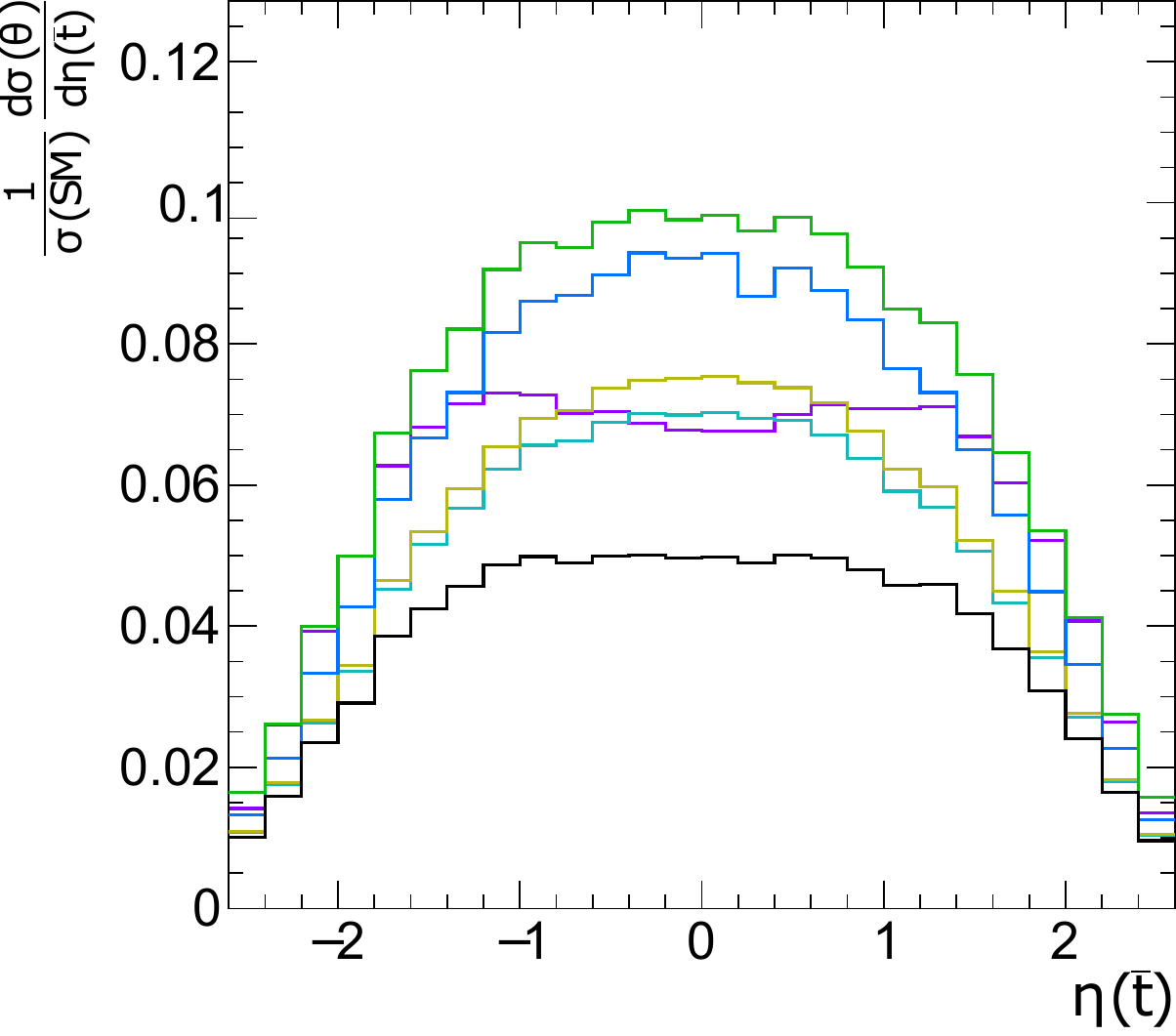}\\\vspace{.2cm}
\includegraphics[width=.305\textwidth]{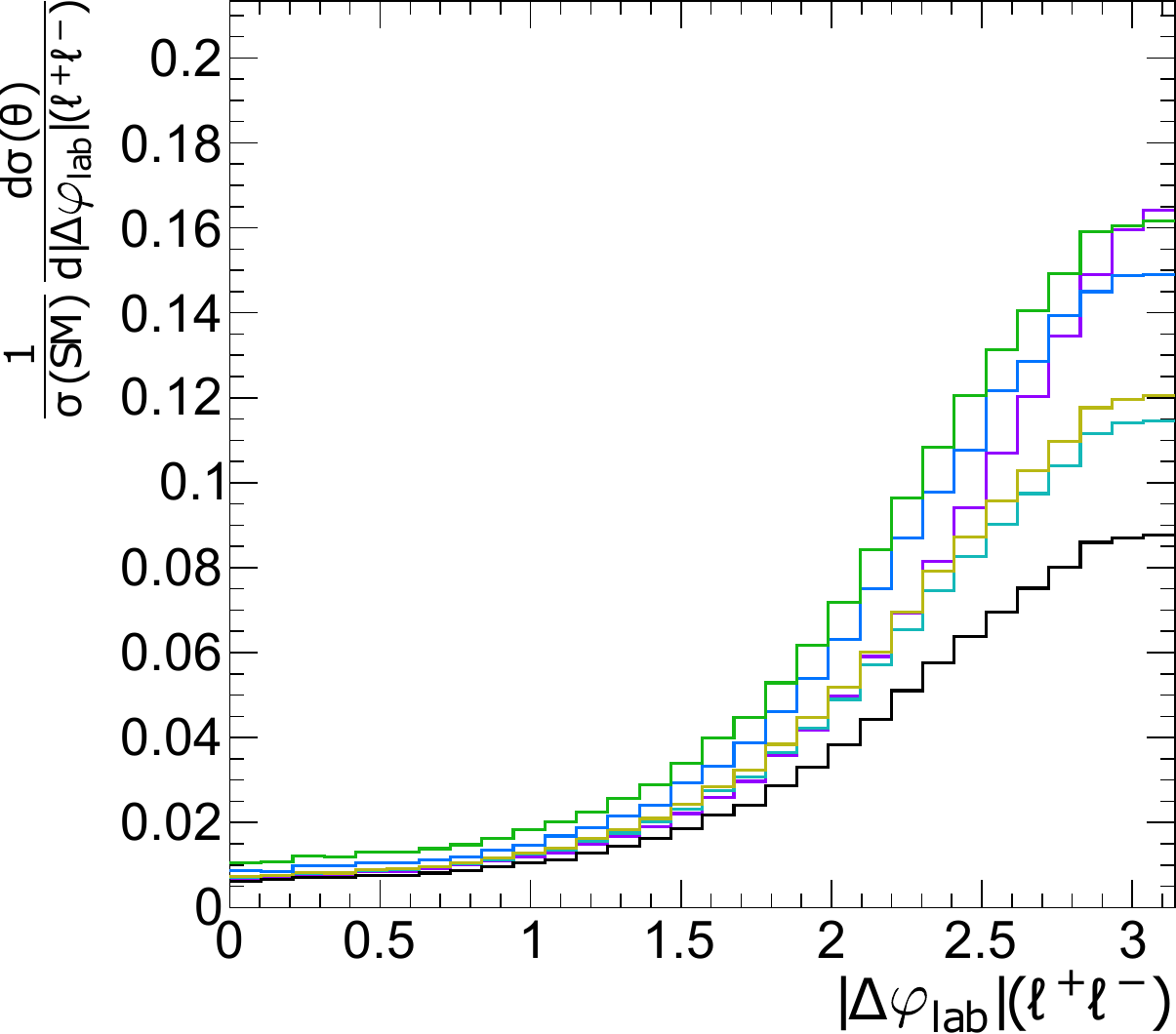}\hfill
\includegraphics[width=.305\textwidth]{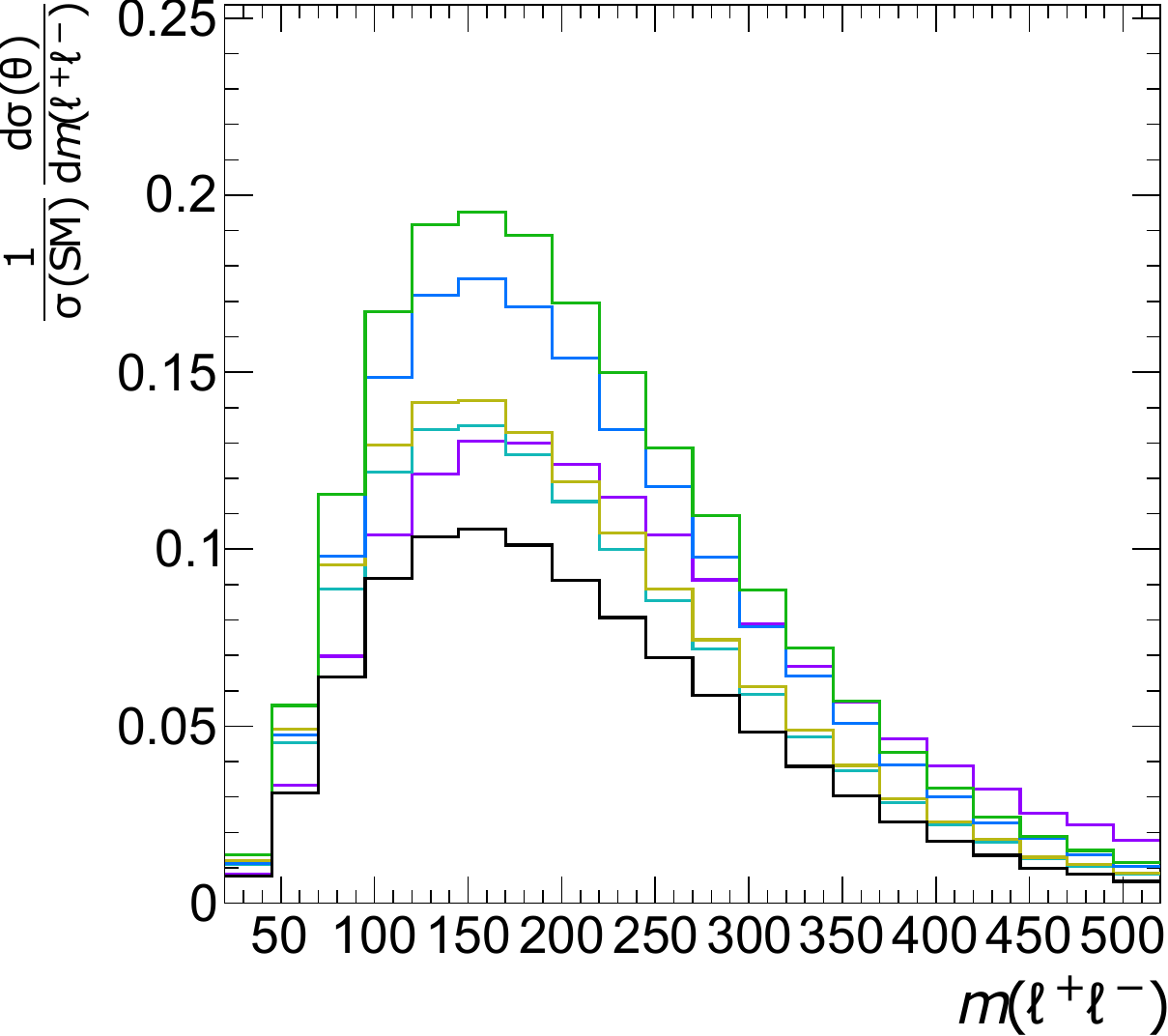}\hfill
\includegraphics[width=.305\textwidth]{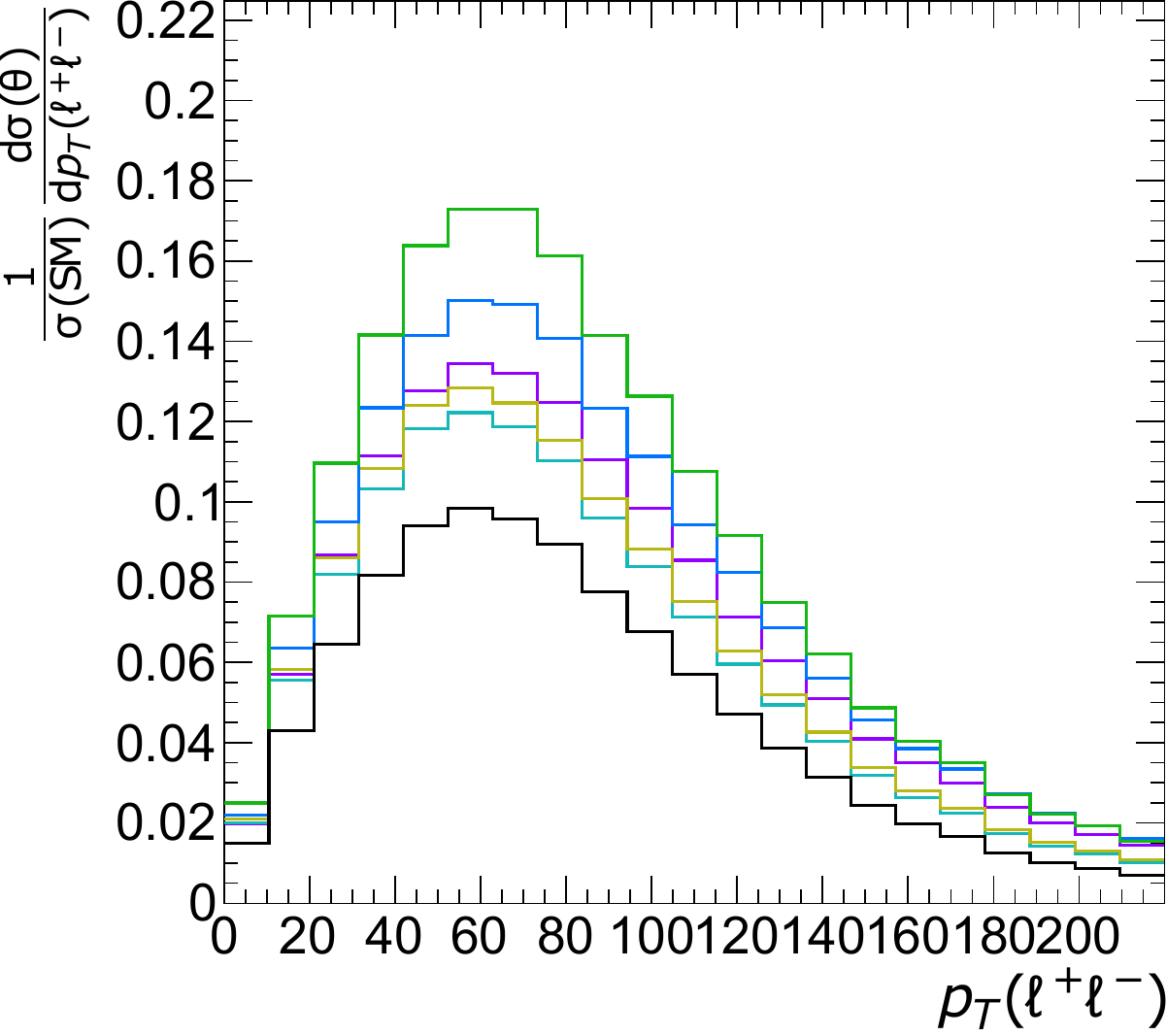}\\\vspace{.2cm}
\includegraphics[width=.305\textwidth]{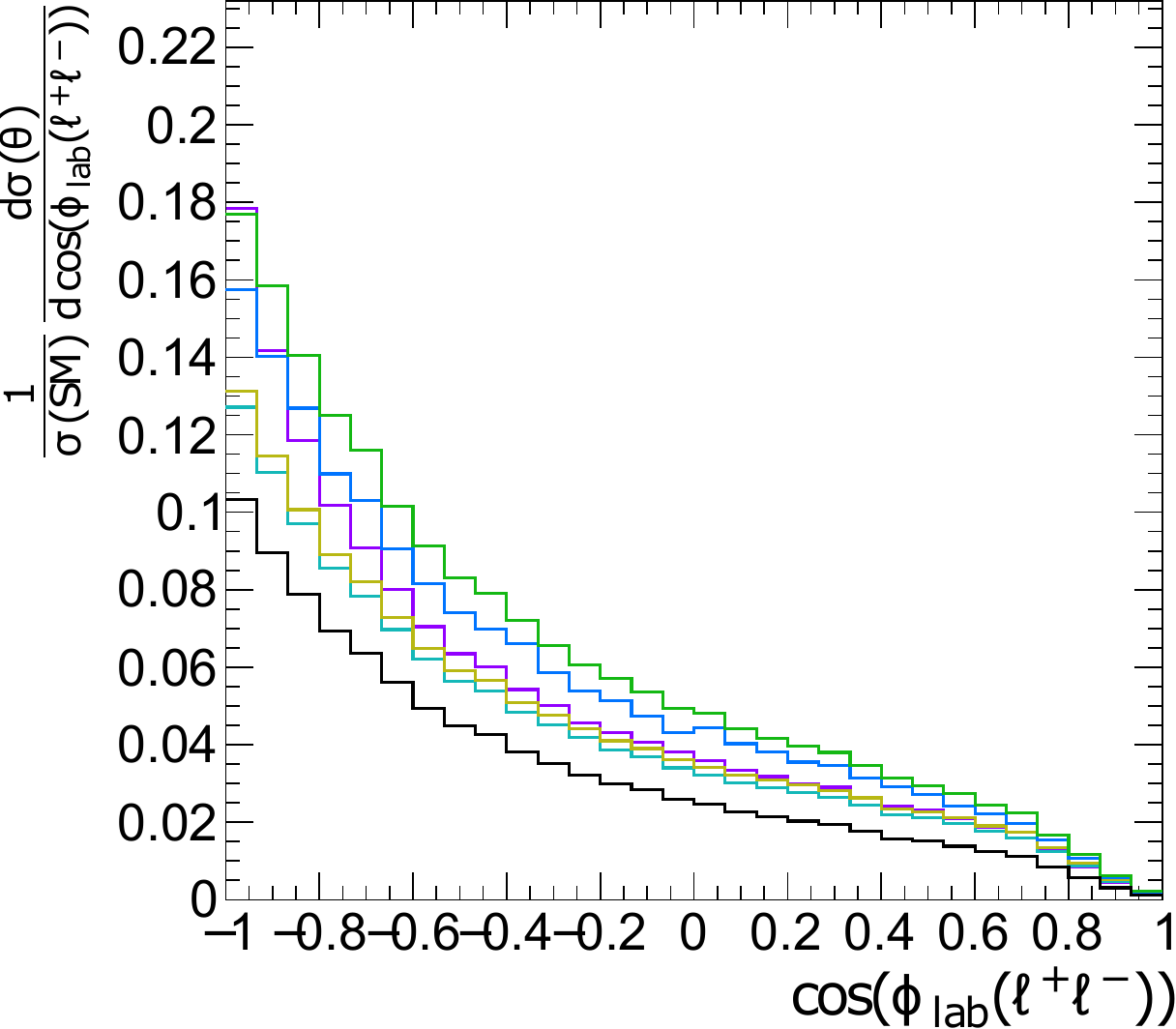}\hfill
\includegraphics[width=.305\textwidth]{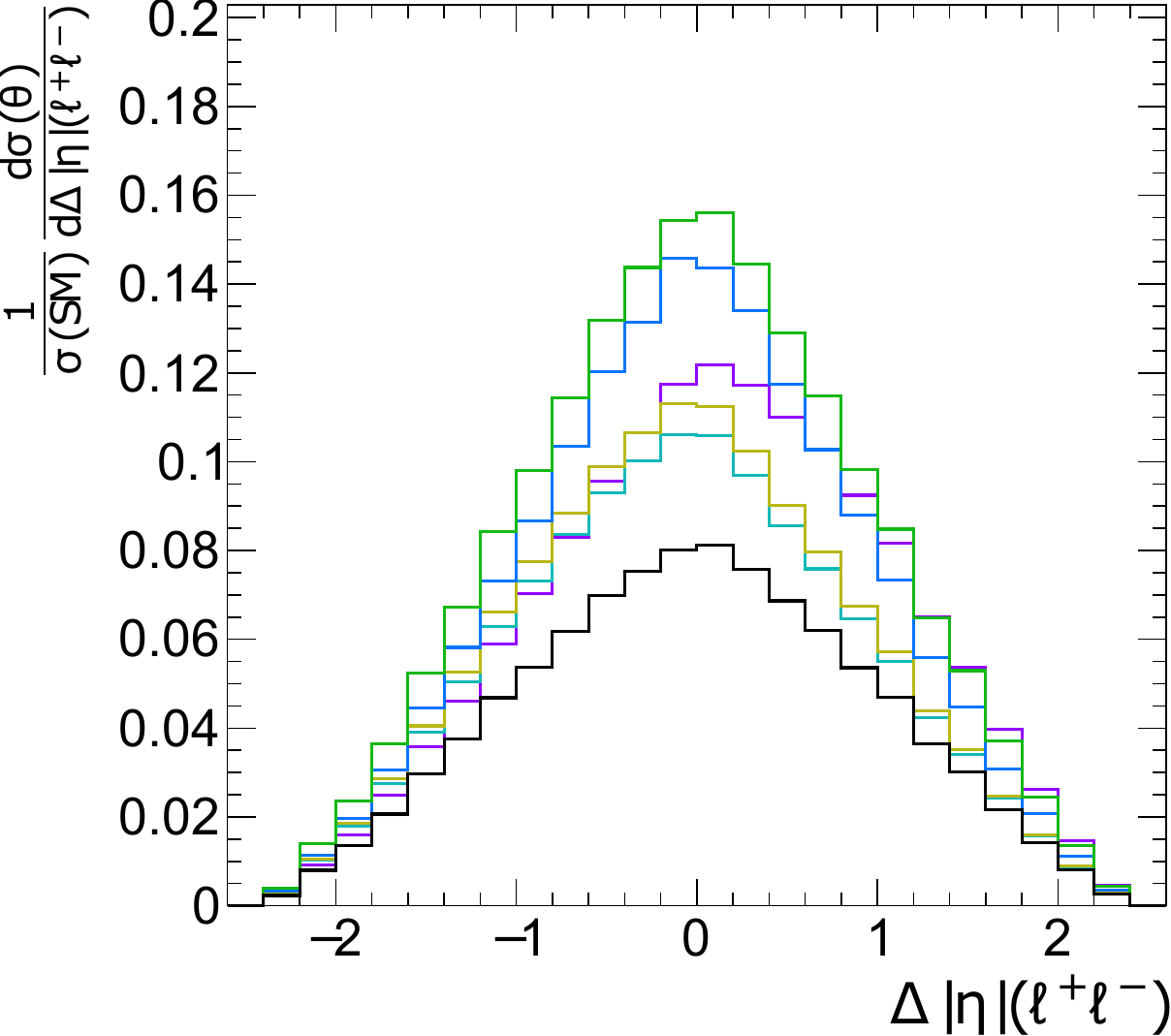}\hfill
\includegraphics[width=.305\textwidth]{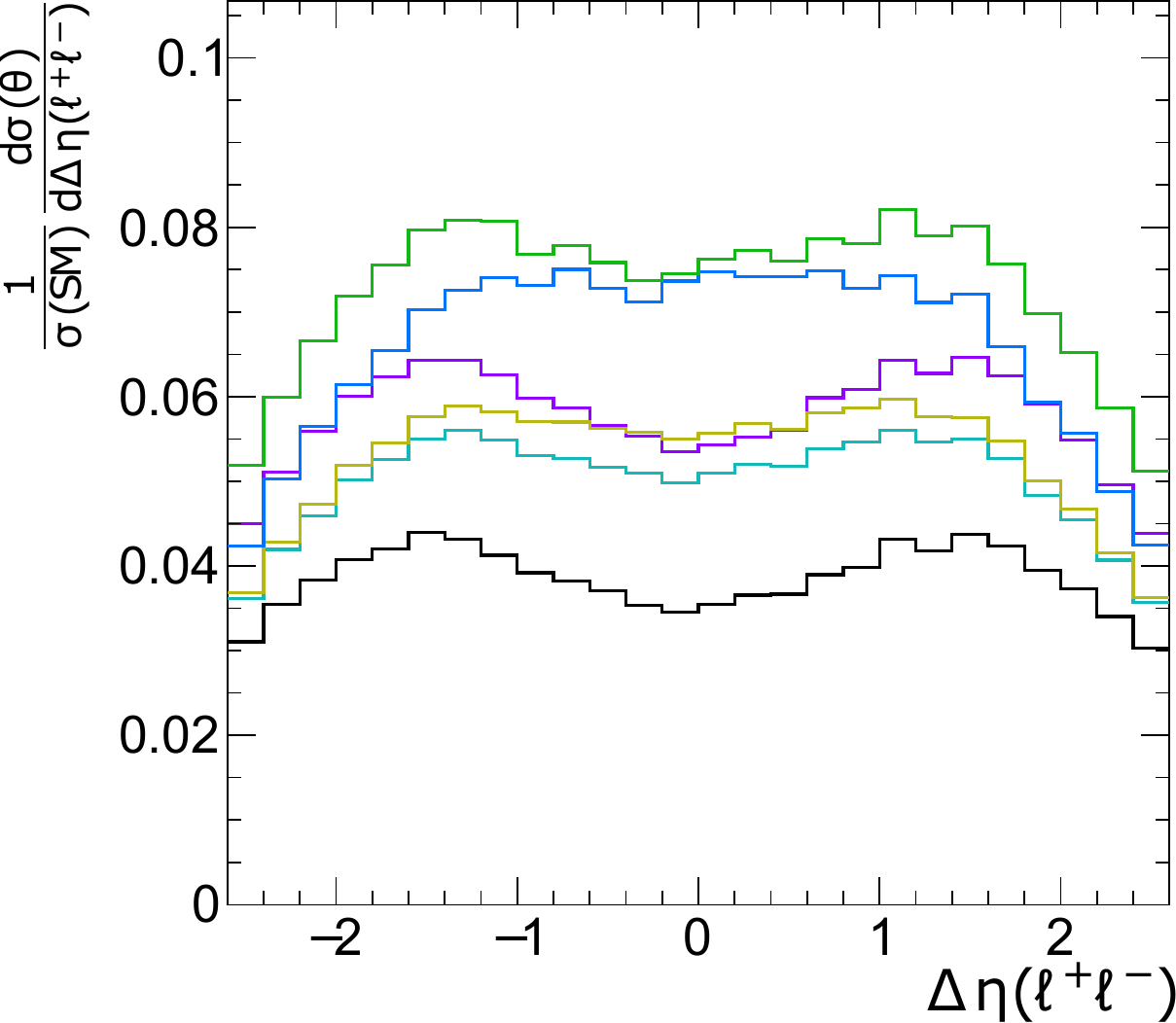}
  \caption{Kinematic features of the $\ttbar(2\ell)$ SMEFT simulation for different values of the Wilson coefficients as described in the text.}
  \label{fig:ttbar-kinematics}
\end{figure}

We generate the $\ttbar(2\ell)$ signal process with \MGvATNLO v2.6.5~\cite{Alwall:2014hca} at leading order and use the NNPDF parton distribution functions~v3.1~\cite{Ball_2017}.
We simulate the top quark pairs at $\sqrt{s}=13\TeV$, followed by leptonic decays of the W bosons ($\ell=e,\mu,\tau$), and employ the \textsc{SMEFTsim} v3.0 model~\cite{Brivio:2020onw} for simulating the parton-level SMEFT effects.
The ME simulation is interfaced to {\textsc{PYTHIA}}~v8.226~\cite{Skands:2014pea} using the CP5 tune~\cite{CMS:2015wcf,CMS:2019csb} for fragmentation, PS, and hadronization of partons in the initial and final states, along with the underlying event and multiparton interactions.
The ME for the \ttbar signal includes up to one extra parton. 
Double counting of the partons generated with \MGvATNLO and \PYTHIA is removed using the MLM~\cite{Alwall:2007fs} scheme. 
The events are subsequently processed with a \Delphes-based simulation model of the CMS detector~\cite{deFavereau:2013fsa}. Kinematic requirements are placed on jets, electrons, and muons.
Jets are reconstructed with anti-$k_T$ algorithm\cite{Cacciari:2008gp} using a distance parameter of 0.4 in the \textsc{FastJet} software package\cite{Cacciari:2011ma}.
The nominal \cPqb~tagging of jets in \Delphes is based on parton-matching and a parametrization of the nominal CMS \cPqb-tagging efficiency.
Electrons and muons must be isolated from jets, satisfy $\pt>20$\GeV, and be reconstructed within absolute pseudorapidity $|\eta|<2.5$.  
If there are two same-flavor lepton candidates of opposite electric charge within a 10\GeV window around the \PZ~boson mass, ${|m(\ell^+\ell^-)-m_\PZ|<10\GeV}$, the event is rejected.
According to Ref.~\cite{CMS:2024ybg}, the purity after the \PZ~boson mass veto is 95\%, with a small background from the Drell-Yan process. We ignore the contribution from Drell-Yan in the following.
Jets must satisfy $\pt>30$\GeV and $|\eta|<2.4$, and there must be more than two jets, among which at least two must be \cPqb~tagged.  

Using the \Delphes objects, we reconstruct the top quark kinematic quantities described in Ref.~\cite{CMS:2024ybg}. This provides access to SMEFT-sensitive observables, including the top quarks' invariant masses, angles, and transverse momenta. To reduce the computational demand while keeping sensitivity to SMEFT effects, we require $m(\ttbar)>750$~\GeV, corresponding to an inclusive fiducial cross-section of 0.31~\textrm{pb}~\cite{CMS:2024ybg}. We normalize the \Delphes simulation of a total of $1.2\times10^6$ events to this value and use a central value for the integrated luminosity $\mathcal{L}_0=137\,$fb$^{-1}$ with a conservative 5\% log-normal uncertainty, 
\begin{align}
\mathcal{L}(\bn)=\mathcal{L}_0\,\alpha_{\textrm{lumi}}^{\nu_{\textrm{lumi}}}.\label{eq:lumi-example}
\end{align}

We simulate the effects from the real and imaginary part of the Wilson coefficient $C_{tG}$, and the four-fermion operators $C_{Qq}^{(1,8)}$, $C_{Qq}^{(3,8)}$, and $C_{qt}^{(8)}$. Our five POIs are, therefore, the Wilson coefficients
\texttt{ctGRe}, \texttt{ctGIm}, \texttt{cQj18}, \texttt{cQj38}, and  \texttt{ctj8} in the conventions of Ref.~\cite{Brivio:2020onw}, multiplying the operators 
\begin{align}
\begin{aligned}
O_{tG}&=(\bar{Q}\sigma^{\mu\nu}T^a t)\tilde HG^a_{\mu\nu},\nonumber\\
O_{Qq}^{(1,8)}&=(\bar{Q}T^a\gamma_\mu  Q)(\bar{q}T^a\gamma^\mu  q),
\end{aligned}\qquad\qquad
\begin{aligned}
O_{Qq}^{(3,8)}&=(\bar{Q}\sigma^i T^a\gamma_\mu  Q)(\bar{q}\sigma^i T^a\gamma^\mu  q),\nonumber\\
O_{tj}^{(8)}&=(\bar{t}T^a\gamma_\mu  t)(\bar{u}T^a\gamma^\mu  u),
\end{aligned}
\end{align}
where the left and right-chiral lower-case quark fields $q$ and $u$ belong to the first and second generation. The third-generation left-chiral quark doublet is denoted by $Q$. 
A non-zero value of $C_{tG}$ provides CP-even and CP-odd modifications to the top-gluon interaction. The four fermion operators add contact interactions of the first and second with the third-generation quark currents. 
For obtaining the SMEFT predictions, we use the reweighting technique discussed in Sec.~\ref{sec:synthetic-SMEFT} and  Ref.~\cite{Belvedere:2024nzh}.
The dominant effect of these operators on the $\ttbar(2\ell)$ cross-section is linear with the Wilson coefficients~\cite{Elmer:2023wtr} so that complications from dominantly quadratic predictions that violate Wilk's theorem when computing the distribution of the profile likelihood test statistic can be neglected~\cite{Bernlochner:2022oiw}. 

The following event-level features define the observation $\bx$. From the reconstructed top quark momenta, we compute the invariant mass $m(\ttbar)$, the transverse momentum $p_\textrm{T}({\textrm{\ttbar}})$, the rapidity difference $\Delta\eta(\ttbar)=\eta(\textrm{t})-\eta(\overline{\textrm{t}})$ and the difference of absolute rapidities of the top and anti-top quark, $\Delta|\eta|(\ttbar)=|\eta(\textrm{t})|-|\eta(\overline{\textrm{t}})|$. The quantities $m(\ttbar)$ and $p_\textrm{T}({\textrm{\ttbar}})$ are sensitive to SMEFT effects with energy-growth while  $\Delta|\eta|(\ttbar)$ is sensitive to the effects of the charge-asymmetry~\cite{CMS:2022ged,ATLAS:2022waa}, modified, e.g., by $C_{tj}^{(8)}$.
Furthermore, we include the transverse momentum and the pseudo-rapidity of the top and anti-top quark. Because leptons are clean probes of the possible SMEFT effects, independent of the hadronic activity, we also include the invariant mass $m(\ell^+\ell^-)$, the transverse momentum  $p_{\textrm{T}}(\ell^+\ell^-)$, the rapidity difference $\Delta\eta(\ell^+\ell^-)$, and the difference of absolute rapidities $\Delta|\eta|(\ell^+\ell^-)$ of the dilepton system. Finally, we include the absolute value of the difference of the azimuthal lepton angles $|\Delta\varphi_{\textrm{lab}}|(\ell^+\ell^-)$ and the cosine of the spatial angle between the leptons $\cos(\phi_{\textrm{lab}}(\ell^+\ell^-))$ as measured in the lab frame. The distributions of these observables for the SM and non-zero values for the Wilson coefficients are shown in Fig.~\ref{fig:ttbar-kinematics}.
We find good agreement with the study in Ref.~\cite{GomezAmbrosio:2022mpm}. 
The CMS measurement of the spin-correlation in \ttbar~\cite{CMS:2019nrx} found constraining power for $C_{tG}$ in the distribution of products of angular observables of the leptons, measured in a specific reference frame spanned by the top quark momentum and the beam plane~\cite{Bernreuther:2015yna}. These variables characterize the spin-density matrix of the $\ttbar(2\ell)$ system, and we present a brief description and their distributions in Appendix~\ref{sec:ttbar-xi}.
The resulting distributions for non-zero values of the Wilson coefficients are shown in Fig.~\ref{fig:ttbar-kinematics}.

The estimate of the detector-level SMEFT dependence of the signal process $\hat R(\bx|\bt)$ can be learned by one of the tools in Refs.~\cite{GomezAmbrosio:2022mpm,Chatterjee:2022oco,Chatterjee:2021nms,Chen:2020mev,Chen:2023ind,Cranmer:2015bka,Brehmer:2018eca,Brehmer:2018hga,Brehmer:2019xox,Brehmer:2018kdj}.
We use the Boosted Information Tree technique~\cite{Chatterjee:2022oco,Chatterjee:2021nms} to learn the polynomial dependence up to the quadratic order. Concretely, we train trees with a maximum depth of four in $B=300$ boosting iterations with a learning rate of $\eta=0.2$. We regularize each tree by requiring at least 50 events in each node. 

We use the parametric tree from Sec.~\ref{sec:bpt} to estimate the systematic effects discussed in the following sections. Similar settings turn out to be almost universally applicable. We keep the maximum tree depth at four in all cases and use $B=300$ boosting iterations and a learning rate of $\eta=0.2$. When we obtain the synthetic data from reweighting, such that there are no independent stochastic fluctuations in the various terms in the loss function, a minimum node size requirement of 50 events proves sufficient to regularize the trees. For systematic variations where \bx changes with \bn, we raise this regulator requirement to 500. 

In the subsequent sections, we discuss uncertainties in the renormalization and factorization scales~(scale), the difference between the \MGvATNLO and the \POWHEG event generator~(POW), the normalization of the signal process~(norm), the jet momentum calibration~(JES), the b-tagging efficiency~(HF) and light-quark mis-tagging probability~(LF), and the lepton efficiency calibration~($\ell$). 
The model, therefore, is given in terms of the DCR 
\begin{align}
\mathcal{R}(\bx|\bt,\bn)\equiv\frac{\dd\Sigma(\bx|\bt,\bn)}{\dd\Sigma(\bx|\textrm{SM})\,}=&\;\alpha_{\textrm{norm}}^{\nu_{\textrm{norm}}}\,\hat R(\bx|\bt)\;\hat S_{\textrm{scale}}(\bx|\nu_{R},\nu_F)\;\hat S_{\textrm{POW}}(\bx|\nu_{\textrm{POW}})\,\hat S_{\textrm{JES}}(\bx|\nu_{\textrm{JES}})\nonumber\\&
\times\,\hat S_{\textrm{LF}}(\bx|\nu_{\textrm{LF}})\,\hat S_{\textrm{HF}}(\bx|\nu_{\textrm{HF}})\,\hat S_{\ell}(\bx|\nu_{\ell})\label{eq:tt2l-diff-xsec-model}
\end{align}
with the individual factors defined in the following. 

\subsection{Parton-level uncertainties}

\begin{figure}[tp]
\centering
\includegraphics[width=.32\textwidth]{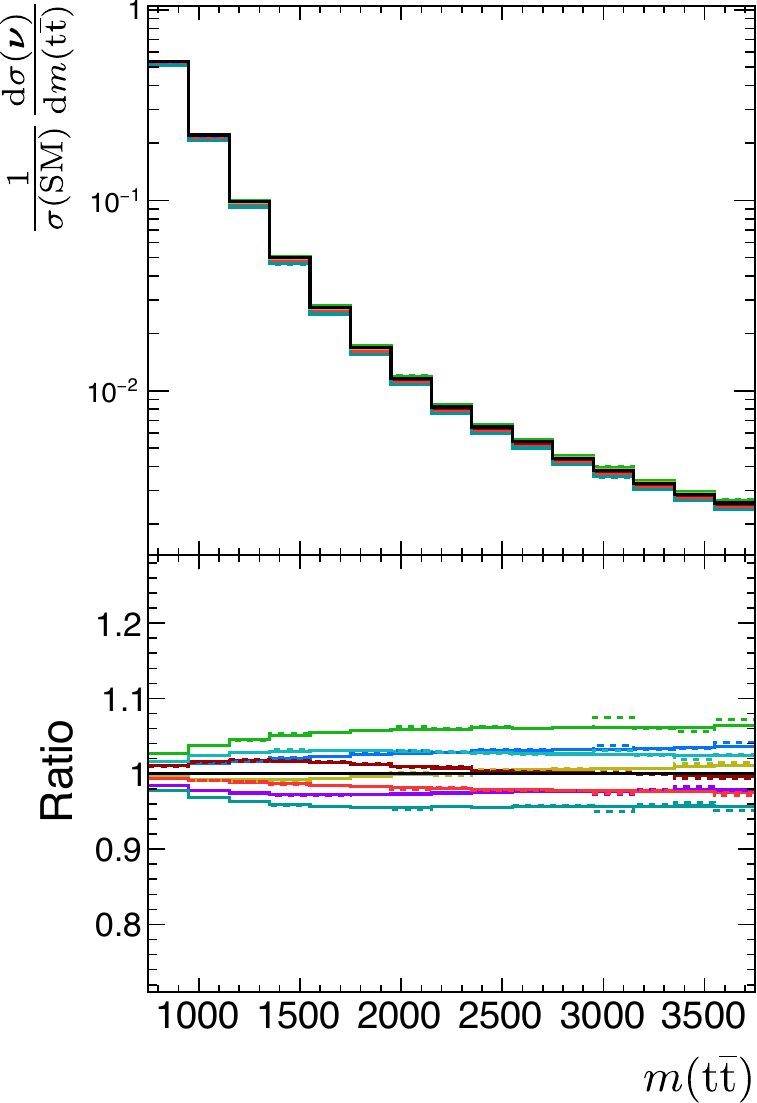}\hfill
\includegraphics[width=.32\textwidth]{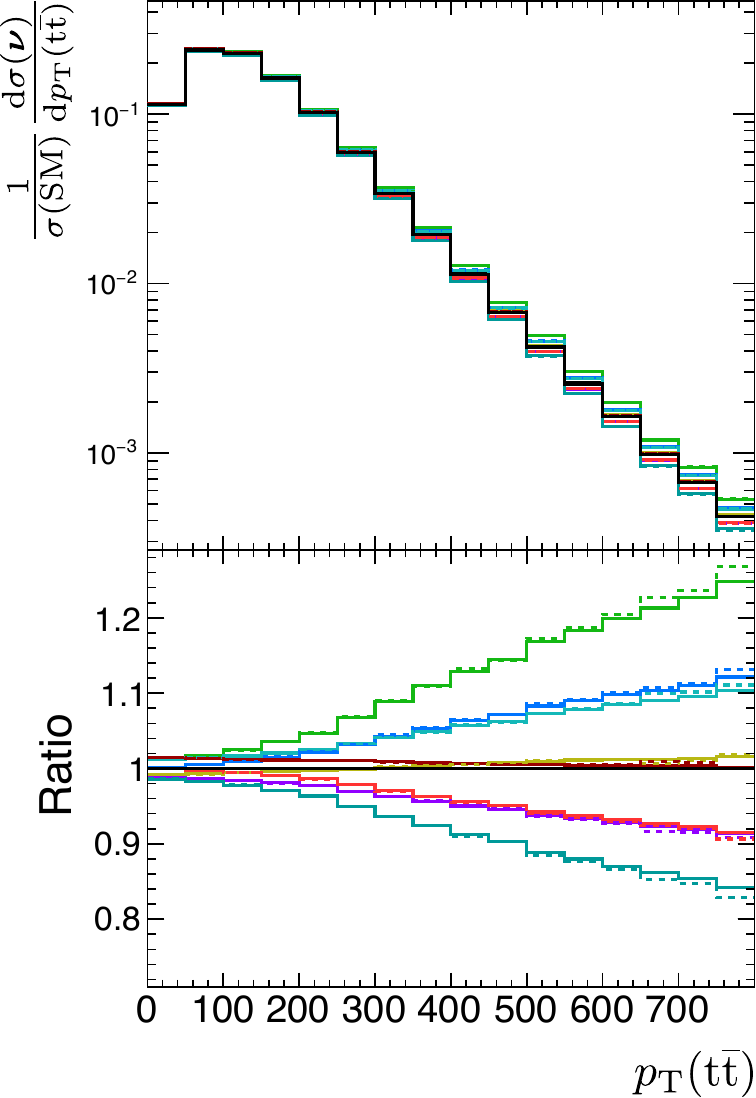}\hfill
\includegraphics[width=.32\textwidth]{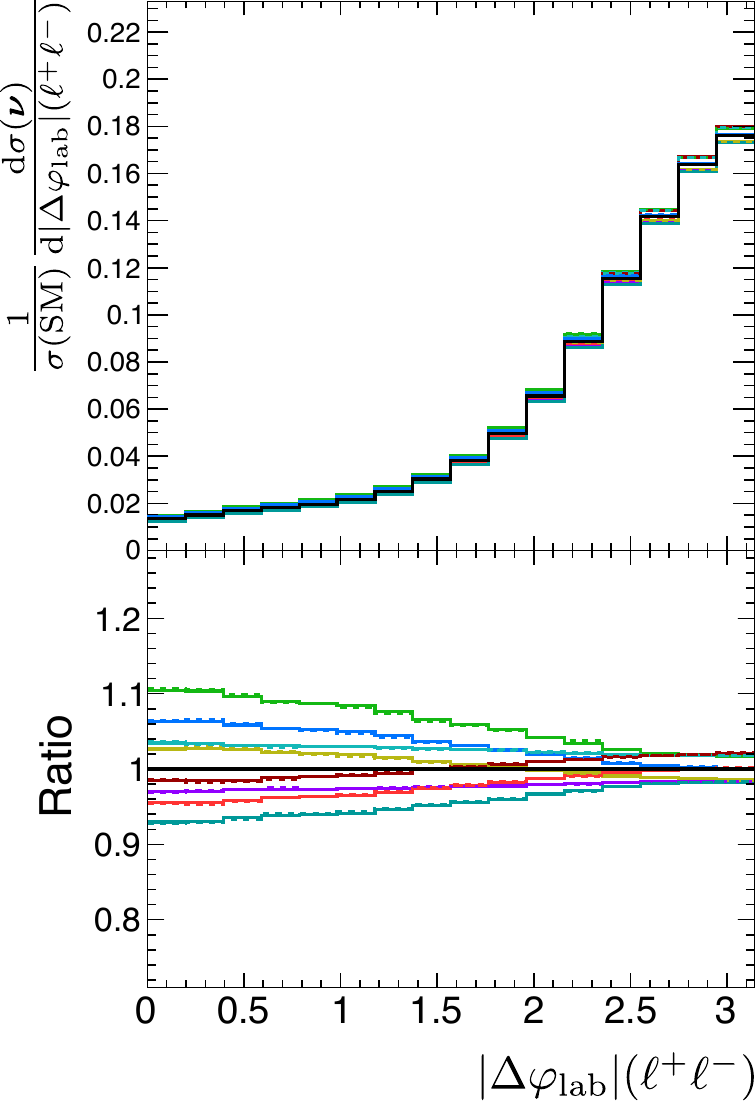}\\\vspace{.3cm}
\includegraphics[width=.32\textwidth]{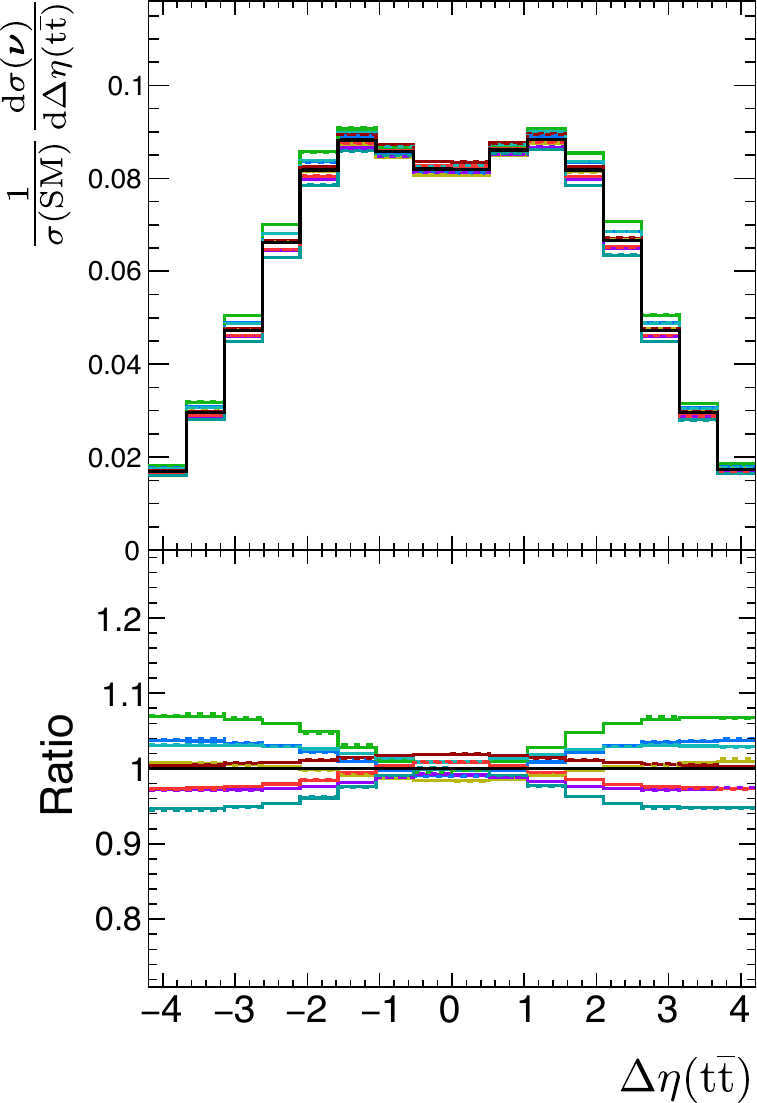}\hfill
\includegraphics[width=.32\textwidth]{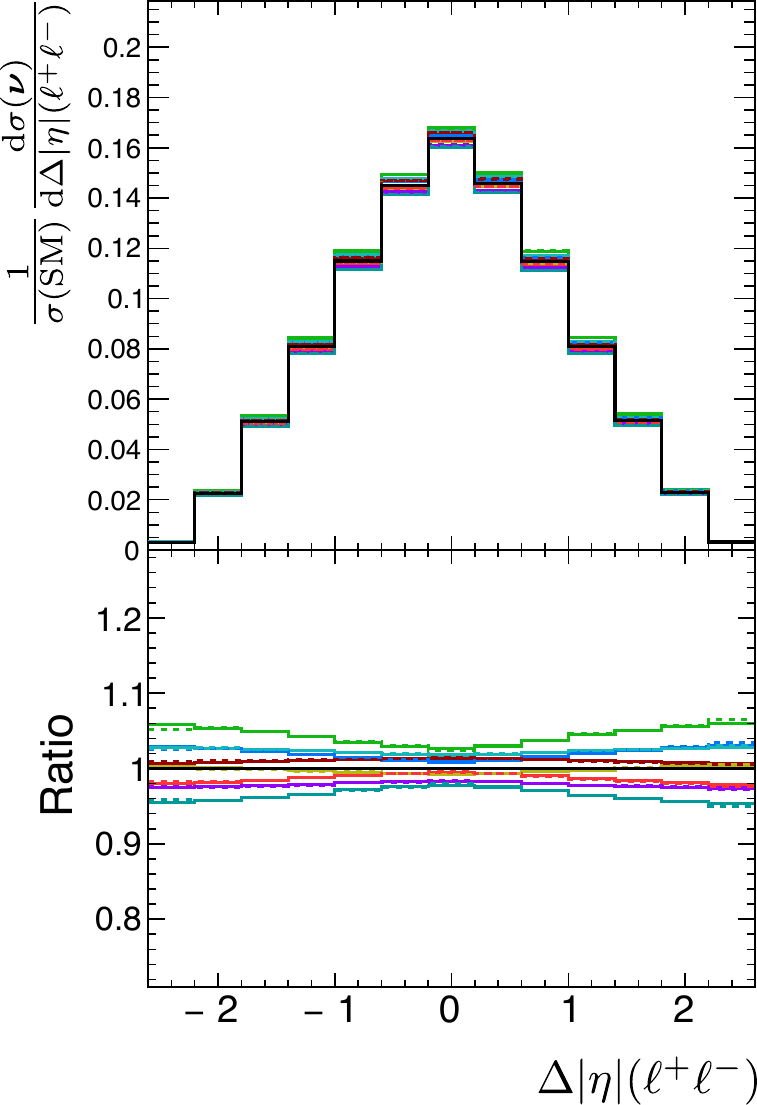}\hfill
\includegraphics[width=.32\textwidth]{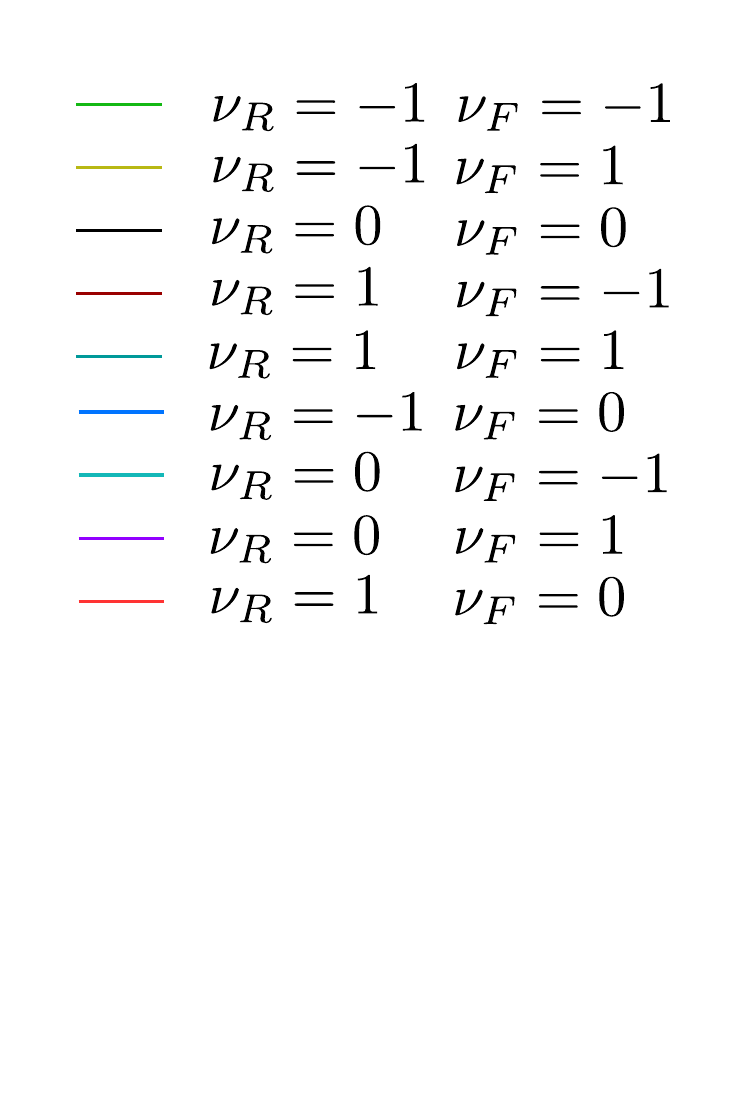}
  \caption{ Variations of the simulation of the $\ttbar(2\ell)$ simulation with the nuisance parameters $\nu_R$ and $\nu_F$, modifying the renormalization and factorization scales, respectively. The dashed lines show the training data, and the solid lines show the prediction from $\hat S_{\textrm{scale}}(\bx|\nu_R,\nu_F)$ as obtained from the BPT. All shapes are normalized to the simulation at the SM parameter point, corresponding to $\nu_R=\nu_F=0$. }
  \label{fig:scale}
\end{figure}

Among the largest systematic effects are uncertainties in the factorization and renormalization scales, as detailed in Sec.~\ref{sec:parton-level-modeling}. Following Eq.~\ref{eq:scale-nuisances}, setting $\nu_R=\pm1$ and $\nu_F=\pm1$ varies the scales $\mu_R$ and $\mu_F$ by a factor of 2. From simulation, we obtain event weights for all eight scale combinations, 
\begin{align}
   (\nu_R,\nu_F)\in\mathcal{V}= \{(-1,-1),\,(-1,0),\,(-1,1),\,(0,-1),\,(0,1),(1,-1),\,(1,0),\,(1,1)\}
\end{align}
with the nominal SM simulation at $\nu_R=\nu_F=0$.
We model the scale uncertainties up to quadratic accuracy in the nuisance parameters with the ansatz
\begin{align}
\hat S_{\textrm{scale}}(\bx|\nu_R,\nu_F)=\exp\left(\nu_R\hat\Delta_{R}(\bx)+\nu_F\hat\Delta_{F}(\bx)+\nu_R^2\hat\Delta_{RR}(\bx)(\bx)+\nu_F^2\hat\Delta_{FF}(\bx)+\nu_R\nu_F\hat\Delta_{RF}(\bx)\right).
\end{align}
The five independent terms, labeled by $A={R,F,RR,FF,RF}$, cover the two linear, two quadratic, and one mixed term. The eight variations in $\mathcal{V}$ thus overconstrain these five functions.

We fit the BPT from Sec.~\ref{sec:bpt} in the standard configuration from Sec.~\ref{sec:tt2l-event-simulation}. The result is shown in Fig.~\ref{fig:scale} with one-dimensional projections for the observables most sensitive to variations in $\nu_R$ and $\nu_F$. Correlated scale variations by a factor of 2 for $\mu_R$ and $\mu_F$, corresponding to $\nu_R=\nu_F=\pm1$ reach 8-10\%. The exception is $p_{\textrm{T}}$, where the tail shows variations exceeding 20\%. In this range, the fit has a small deficit of 1-2\% relative to true variations, likely due to residual inflexibility in the quadratic model. Since $\mu_R$ and $\mu_F$ capture uncertainties from limited perturbative control, we ignore this slight mismatch for now. Other kinematic features show less shape dependence.

\begin{figure}[tph]
\centering
\includegraphics[width=.31\textwidth]{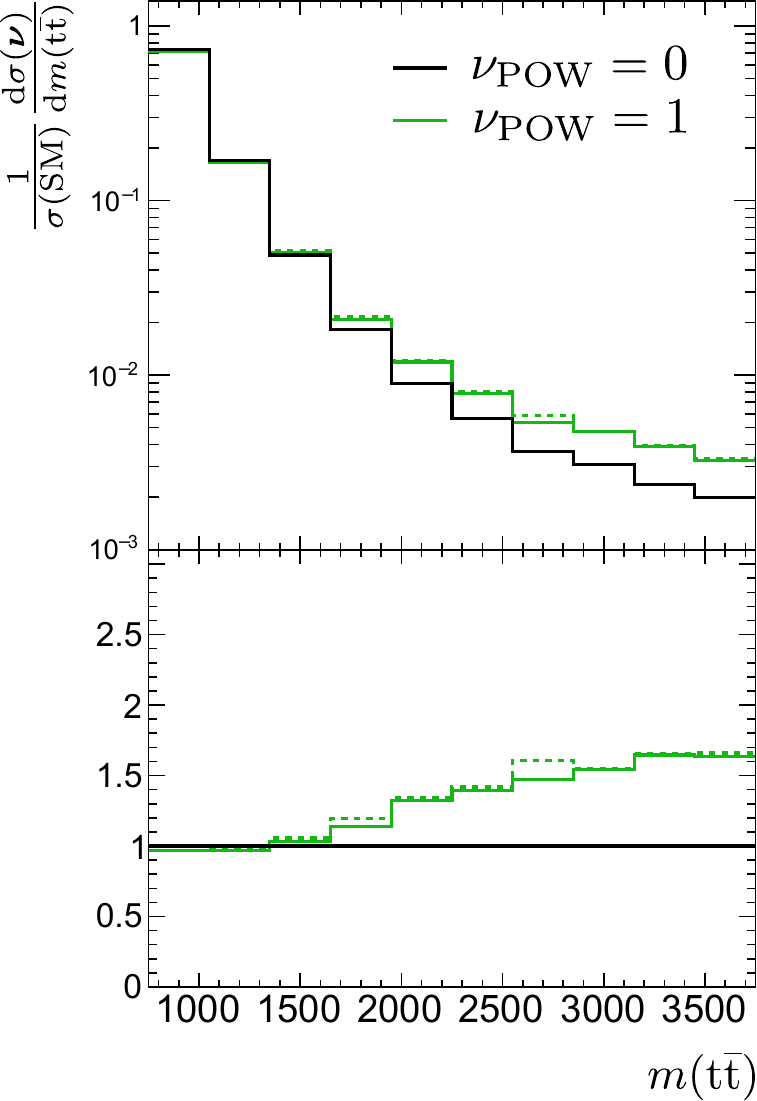}\hfill
\includegraphics[width=.315\textwidth]{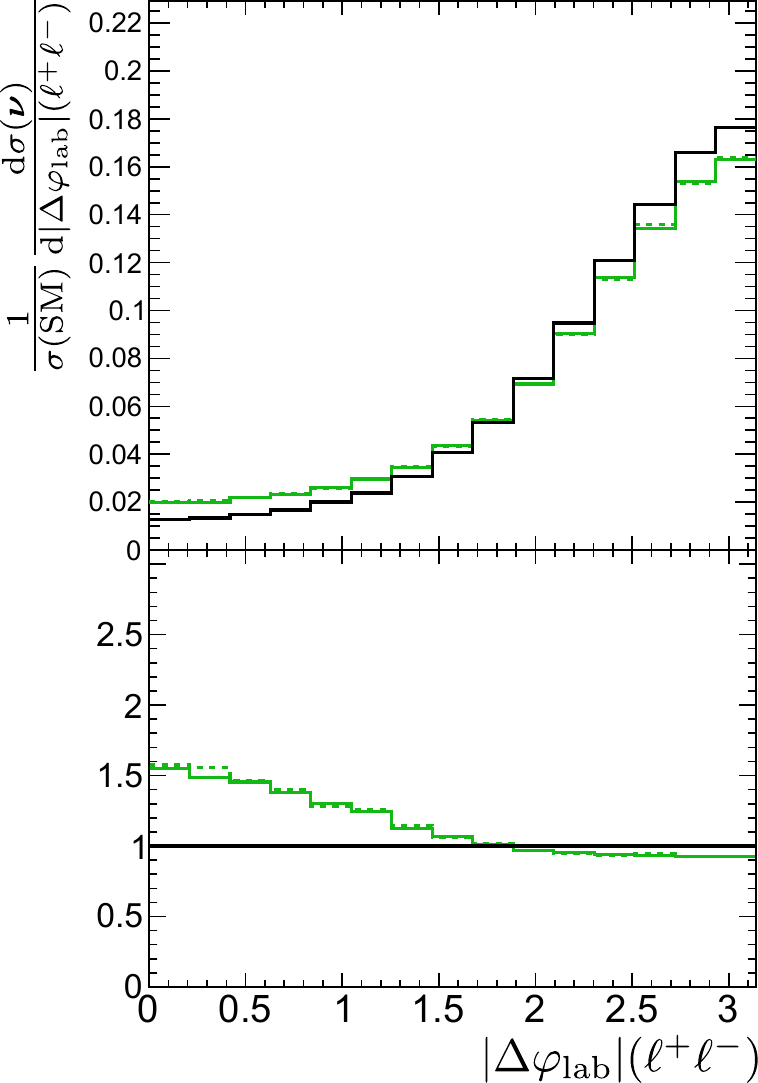}\hfill
\includegraphics[width=.31\textwidth]{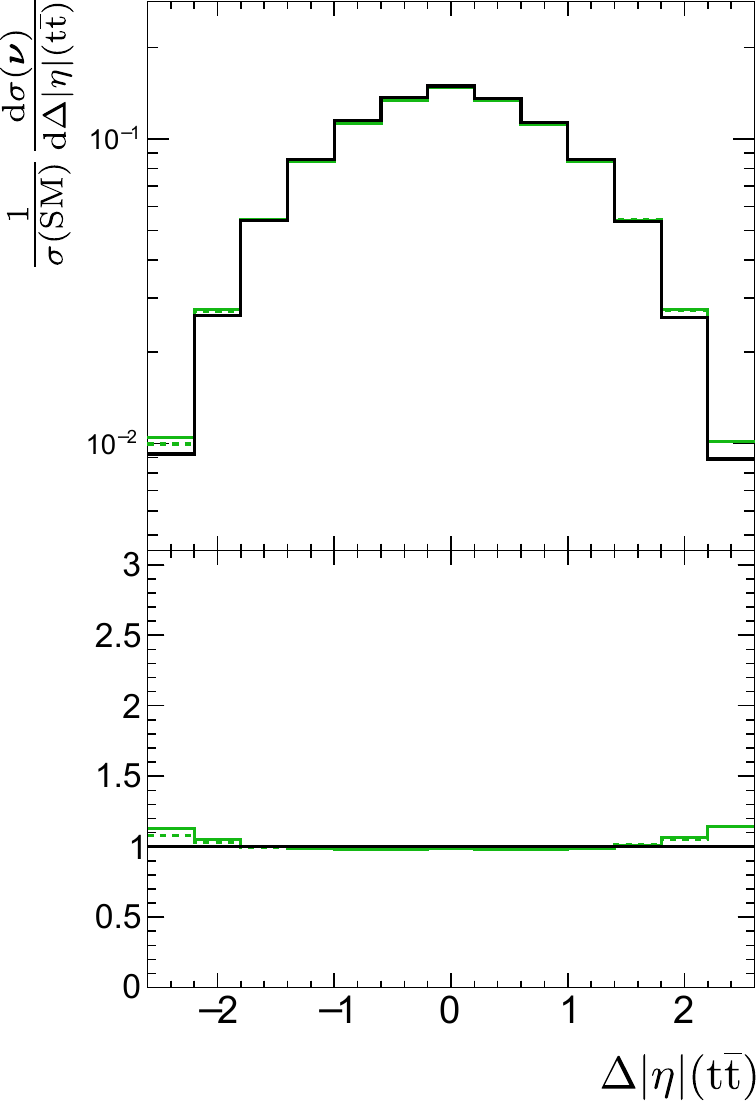}
  \caption{A comparison of \MGvATNLO event simulation~(black) and \POWHEG event simulation~(green, dashed) at the SM parameter point. The solid green line shows the prediction from $\hat S_{\textrm{POW}}(\bx|\bn_{\textrm{POW}})$ for the two-point nuisance parameter $\nu_{\textrm{POW}}$.  }
  \label{fig:pow-vs-MG}
\end{figure}

We also simulate events at the SM parameter point using the alternative \POWHEG generator, producing a similarly sized event sample for the $\ttbar(2\ell)$ process at NLO accuracy in the strong coupling constant. As outlined in Sec.~\ref{sec:2-point-alternatives}, and with the caveats noted there, we assign a nuisance parameter $\nu_{\textrm{POW}}$, with $\nu_{\textrm{POW}}=0$ representing \MGvATNLO and $\nu_{\textrm{POW}}=1$ representing \POWHEG. Here, $\mathcal{V}=\{1\}$ allows us to train a single-parameter linear surrogate for the (log-) DCR, denoted $\hat S_{\textrm{POW}}(\bx|\nu_{\textrm{POW}})$. Figure~\ref{fig:pow-vs-MG} shows one-dimensional projections of the features, revealing shape differences. Minor statistical fluctuations appear in the data tails due to the stochastic independence of the samples, but the BPT averages them out.

Uncertainties in the PDFs, which would require around 100 nuisance parameters for variations along the PDF eigendirections~\cite{Butterworth:2015oua}, are deferred for future treatment. Instead, and to account for uncertainties in the $m(\ttbar)$ selection efficiency, we include a normalization uncertainty of 15\%, setting $\alpha_{\textrm{norm}}=1.15$.

\subsection{Jet energy calibration uncertainties}

\begin{figure}[t]
\centering
\includegraphics[width=.325\textwidth]{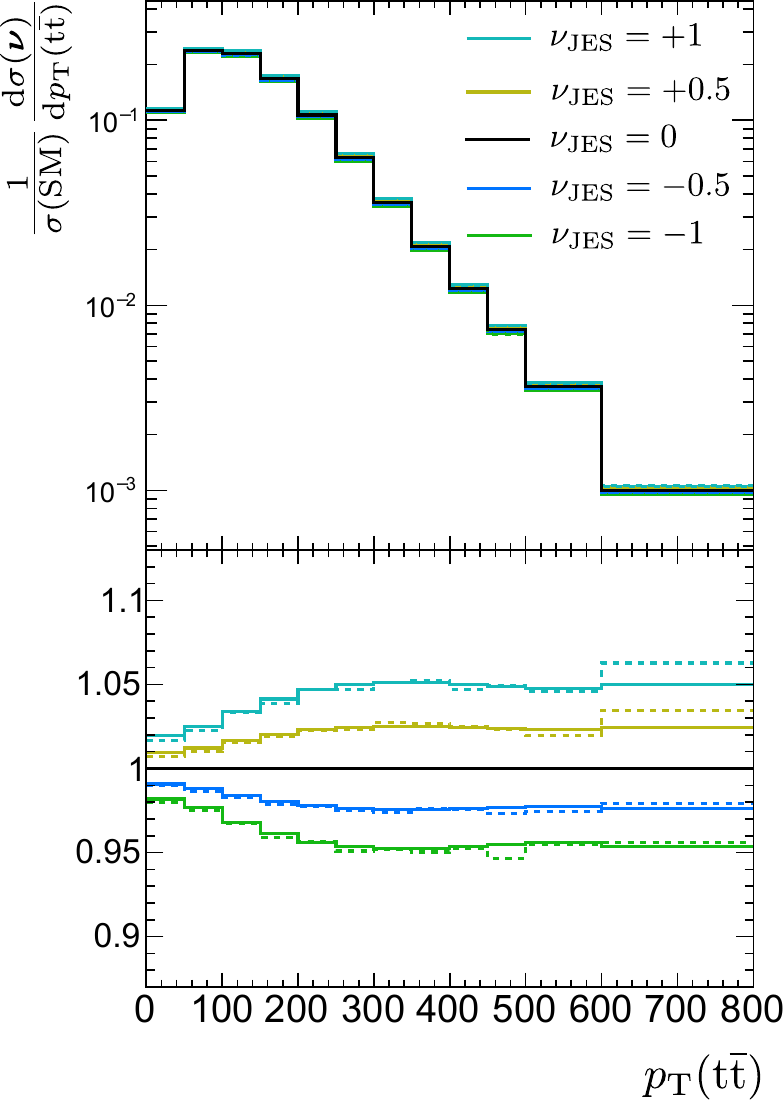}\hfill
\includegraphics[width=.315\textwidth]{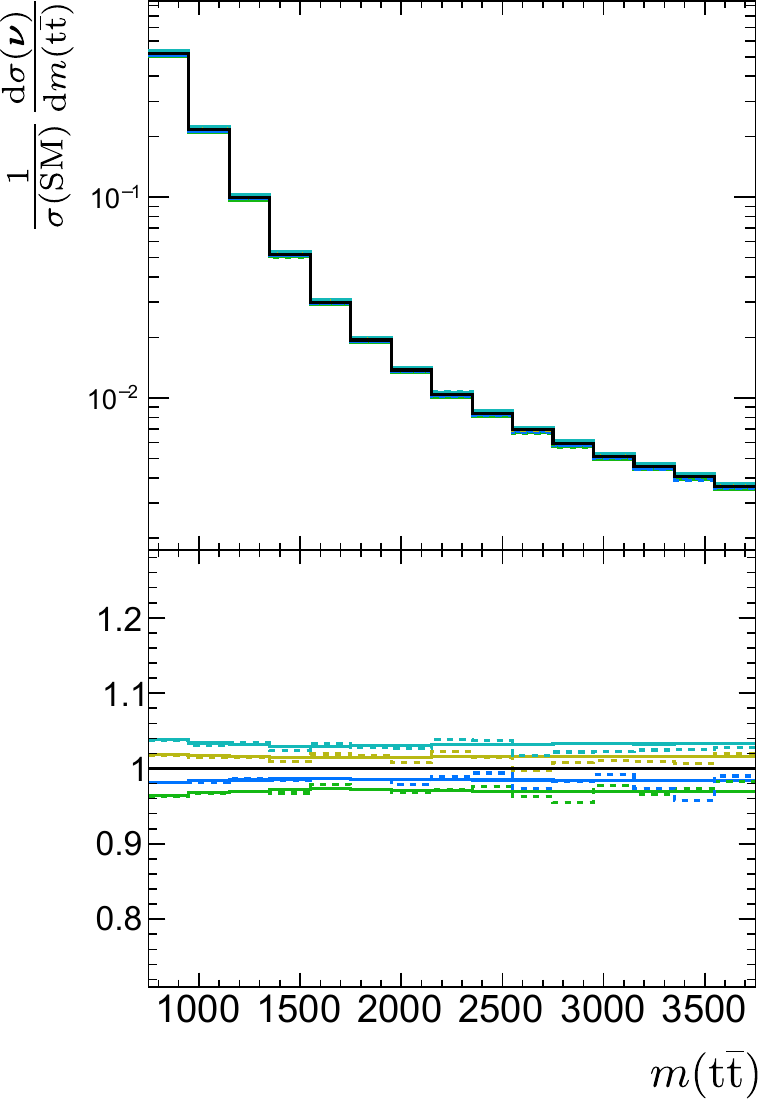}\hfill
\includegraphics[width=.315\textwidth]{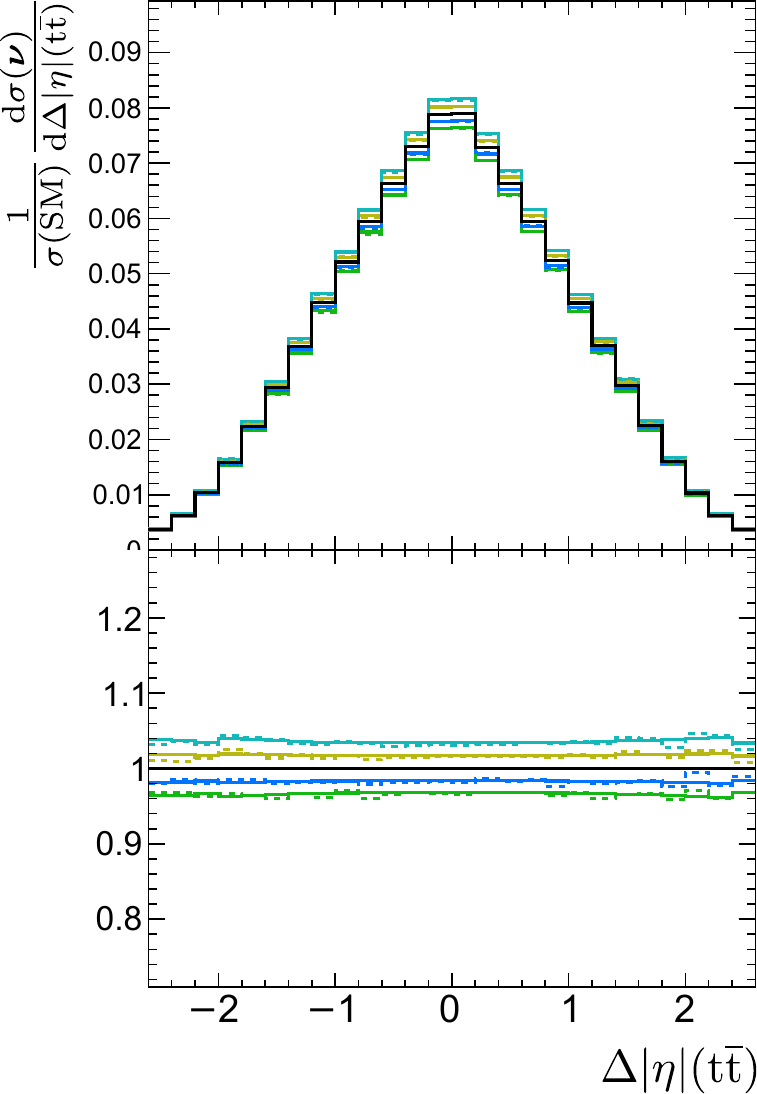}
  \caption{Variations of the nominal simulation with the nuisance parameter for JES uncertainties $\bn_{\textrm{JES}}$. The dashed lines show the training data and the solid lines show the result from the surrogate $\hat S_{\textrm{JES}}(\bx|\nu_{\textrm{JES}})$.   }
  \label{fig:JES}
\end{figure}

To evaluate the impact of uncertainties in the reconstructed transverse jet momenta, we vary the \Delphes-predicted values according to the ``total'' CMS JES uncertainty provided in Ref.~\cite{CMS-Open-Data}. These variations affect the event selection, missing energy, top quark kinematic reconstruction, and other event features in~\bx. Since the per-jet variations depend on the nominal $p_{\textrm{T}}$ and pseudo-rapidity, which are latent (not in \bx), the resulting function $J_{\nu_{\textrm{JES}}}(\bx,\bz)$ is also dependent on the latent event configuration.

Using the method in Sec.~\ref{sec:uncertainty-calibration}, we define synthetic data sets and set $\mathcal{V}=\{-1,-0.5,0.5,1\}$ for the JES nuisance parameter $\nu_{\textrm{JES}}$, parameterizing the per-jet variation effects on \bx in units of the JES uncertainty standard deviations. The half-integer values for $\nu_{\textrm{JES}}$ provide more granularity than the typical $\pm1\sigma$ variations used in binned LHC analyses. We then fit a log-linear surrogate,
\begin{align}
\hat S_{\textrm{JES}}(\bx|\nu_{\textrm{JES}})=\exp\left(\nu_{\textrm{JES}}\hat\Delta_{\textrm{JES}}(\bx)\right)
\end{align}
to model the JES dependence. Figure~\ref{fig:JES} shows an excellent fit of the surrogate to the variations in the training data. Most observables show a flat variation, except for $p_{\textrm{T}}(\ttbar)$, which rises from 2\%--5\%. In the tails of $m(\ttbar)$, slight asymmetries in the training data variations are not captured by the linear model, as it approximately symmetrizes the total uncertainty. Refinement with a higher-degree surrogate is left for future work.

\subsection{Uncertainties in tagging efficiencies}

\begin{figure}[tph]
\centering
\includegraphics[width=.325\textwidth]{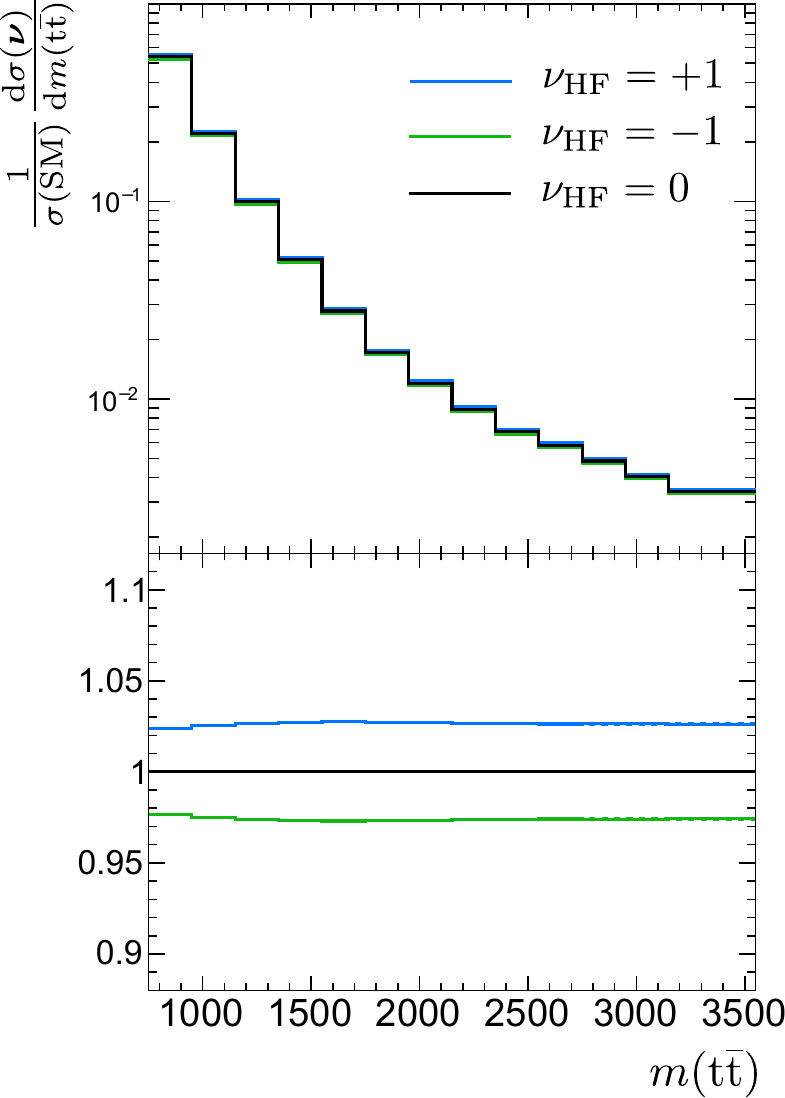}\hfill
\includegraphics[width=.325\textwidth]{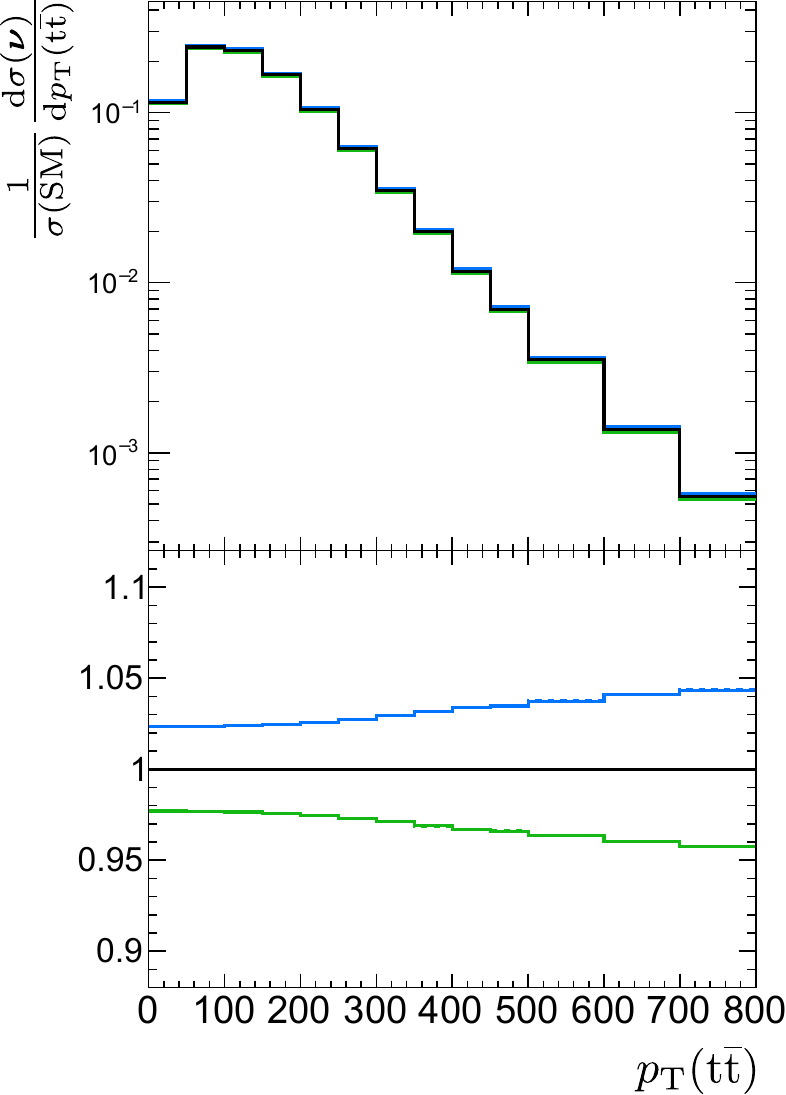}\hfill
\includegraphics[width=.31\textwidth]{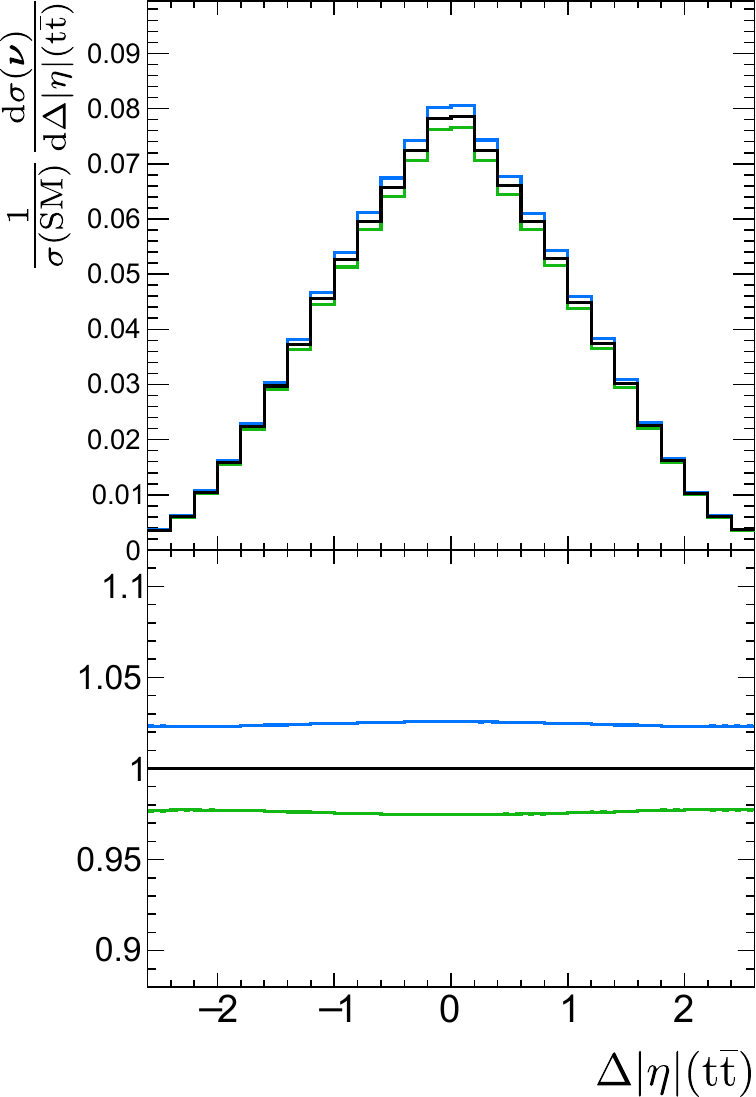}\\\vspace{.3cm}
\includegraphics[width=.325\textwidth]{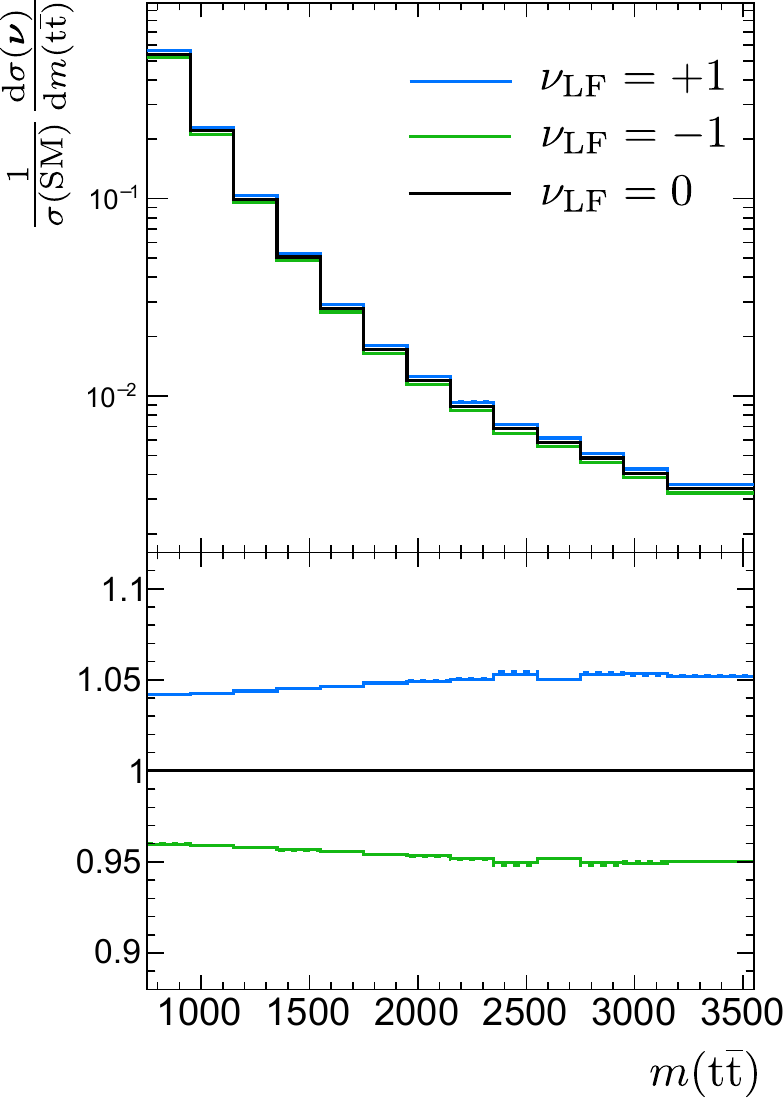}\hfill
\includegraphics[width=.325\textwidth]{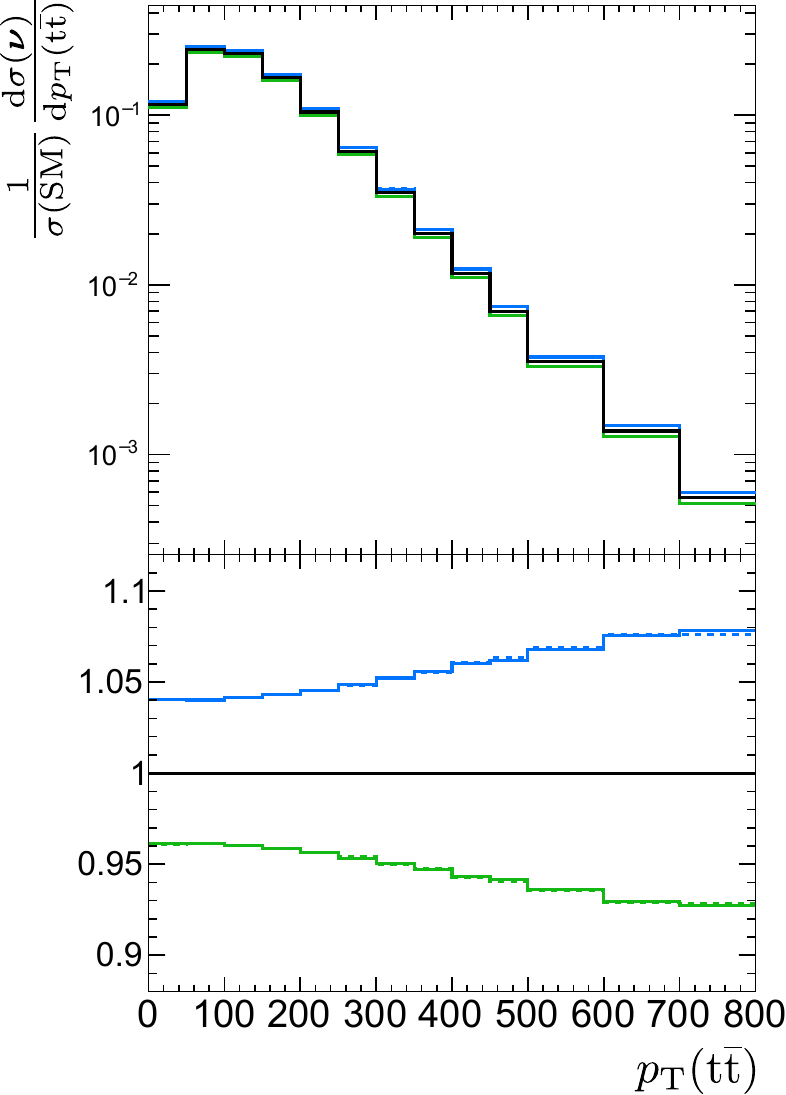}\hfill
\includegraphics[width=.31\textwidth]{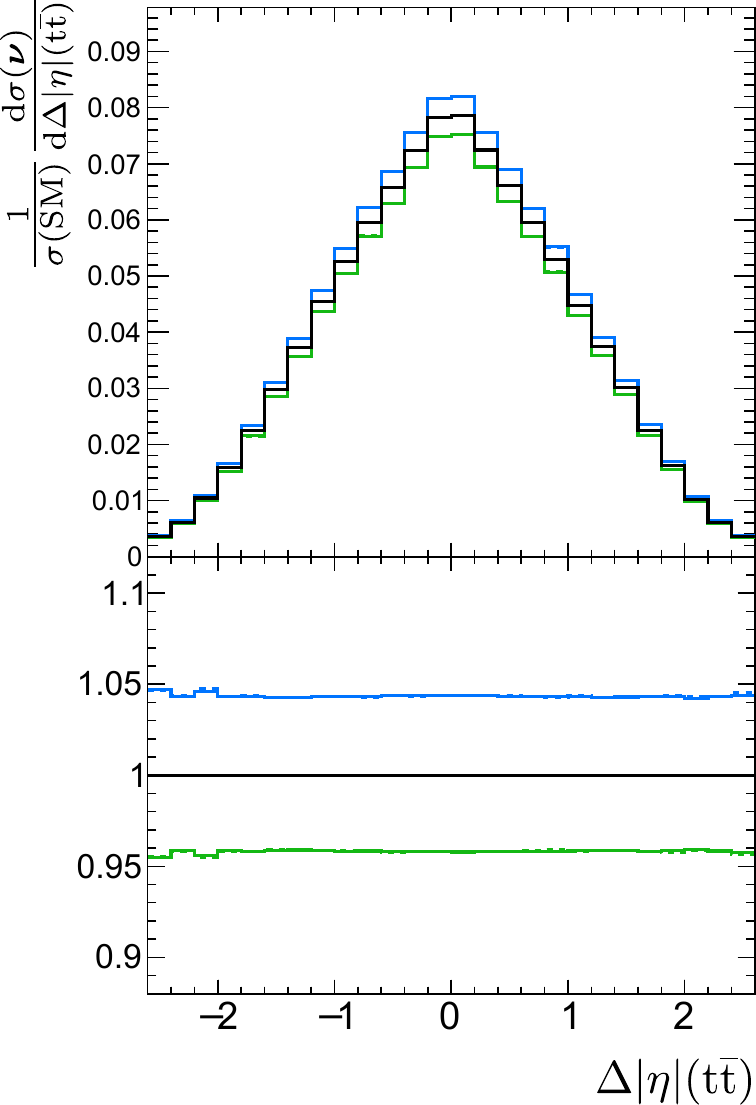}
  \caption{Variations of the nominal simulation with the nuisance parameter for the b-tagging uncertainties $\bn_{\textrm{HF}}$ and $\bn_{\textrm{HF}}$. The dashed lines show the training data and the solid lines show the result from the surrogate $\hat S_{\textrm{HF}}(\bx|\nu_{\textrm{HF}})$ and $\hat S_{\textrm{LF}}(\bx|\nu_{\textrm{LF}})$. }
  \label{fig:b-tag}
\end{figure}

Uncertainties in the b-tagging efficiency for jets, along with their application, are provided in Ref.~\cite{CMS-Open-Data}. This approach relies on $p_{\textrm{T}}$, pseudo-rapidity, and a nominal binary b-tag label from \Delphes. To apply variations, we also need the generator-level jet flavor $f$ within $\{\textrm{udsg}, \Pqc,\Pqb\}$. Using the nominal \Delphes simulation, we parametrize the $p_{\textrm{T}}$ and $\eta$-dependent \Pqb-tagging efficiencies $\varepsilon_f(p_{\textrm{T}},\eta)$ for each flavor.
Two systematic uncertainties are considered with scale factors $\textrm{SF}_f(p_{\textrm{T}},\eta)$ and variations $\Delta\textrm{SF}_f(p_{\textrm{T}},\eta)$. The heavy-flavor~(HF) tagging uncertainty covers the b and c-quark tagging rates, modified in a correlated way for non-zero $\nu_{\textrm{HF}}$. The light-flavor~(LF) mistagging uncertainty addresses tagging rates for light-quark and gluon jets, associated with $\nu_{\textrm{LF}}$.
The reweighting function for synthetic data in Eq.~\ref{eq:synthetic-reweighting} is given by
\begin{align}
r(\bx_i,\bz_i|\nu_k,0)&= \frac{F(\nu_k,\textrm{jets in event }i)}{F(0,\textrm{jets in event }i)}\label{eq-bt-2}
\end{align}
where
\begin{align}
F(\nu_k,\textrm{jets})&=\prod_{\textrm{tagged jets}}\varepsilon_{f}(p_{\textrm{T}},\eta)\left(\textrm{SF}_f(p_{\textrm{T}},\eta)+\nu_k \Delta\textrm{SF}_{f,k}(p_{\textrm{T}},\eta)\right)\nonumber\\
\times&\prod_{\textrm{untagged jets}}\left(1-\varepsilon_{f}(p_{\textrm{T}},\eta)(\textrm{SF}_f(p_{\textrm{T}},\eta)+\nu_k \Delta\textrm{SF}_{f,k}(p_{\textrm{T}},\eta))\right).\label{eq-bt-1}
\end{align}
For $k=\textrm{HF}$, we vary b and c-jet efficiencies, and for $k=\textrm{LF}$, we vary the efficiencies for light-quark and gluon jets. Using $\nu_k=\pm 1$, we construct synthetic data sets and fit linear surrogates
\begin{align}
\hat S_{\textrm{HF}}(\bx|\nu_{\textrm{HF}})=\exp\left(\nu_{\textrm{HF}}\hat\Delta_{\textrm{HF}}(\bx)\right)\quad\textrm{and}\quad \hat S_{\textrm{LF}}(\bx|\nu_{\textrm{LF}})=\exp\left(\nu_{\textrm{LF}}\hat\Delta_{\textrm{LF}}(\bx)\right).
\end{align}

The resulting parametrization is shown in Fig.~\ref{fig:b-tag}. The HF and LF uncertainties exhibit similar shapes. HF variations range from 2\% to 5\%, while LF variations show a slightly larger impact, ranging from 4\% to 8\%. This greater effect of the LF variations is due to the higher light-jet multiplicity following the $m(\ttbar) \geq 750~\GeV$ selection. 

\subsection{Uncertainties in lepton efficiencies}

\begin{figure}[t]
\centering
\includegraphics[width=.31\textwidth]{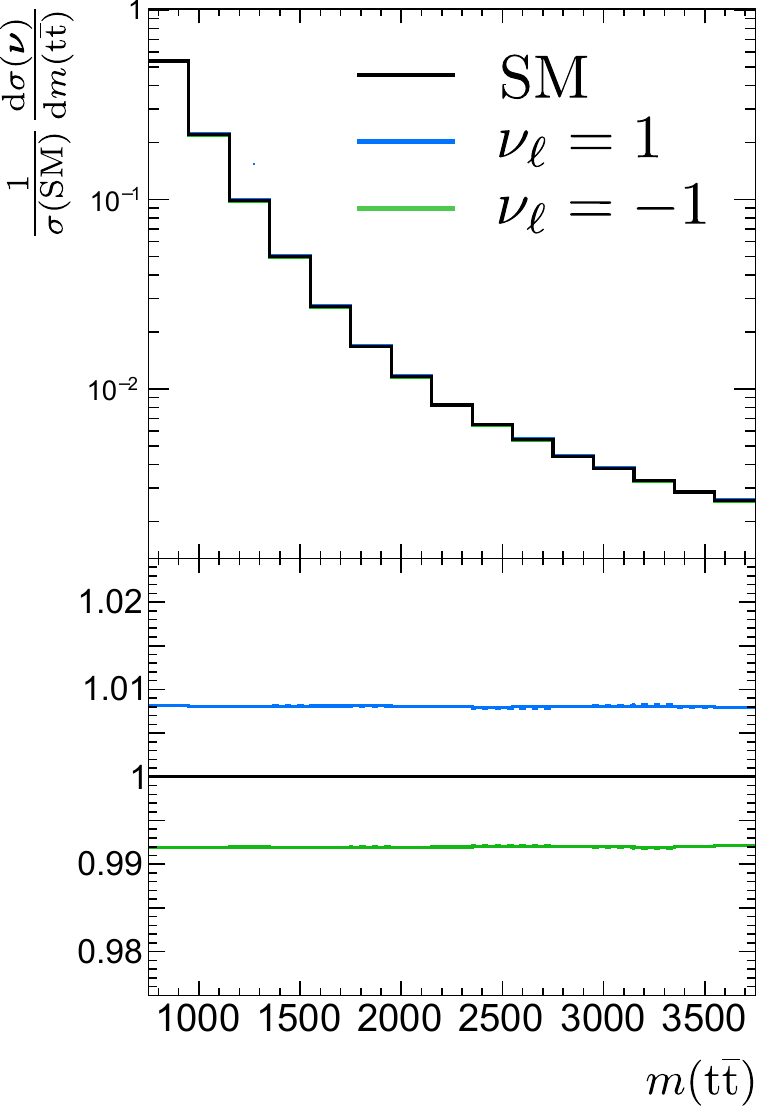}\hfill
\includegraphics[width=.325\textwidth]{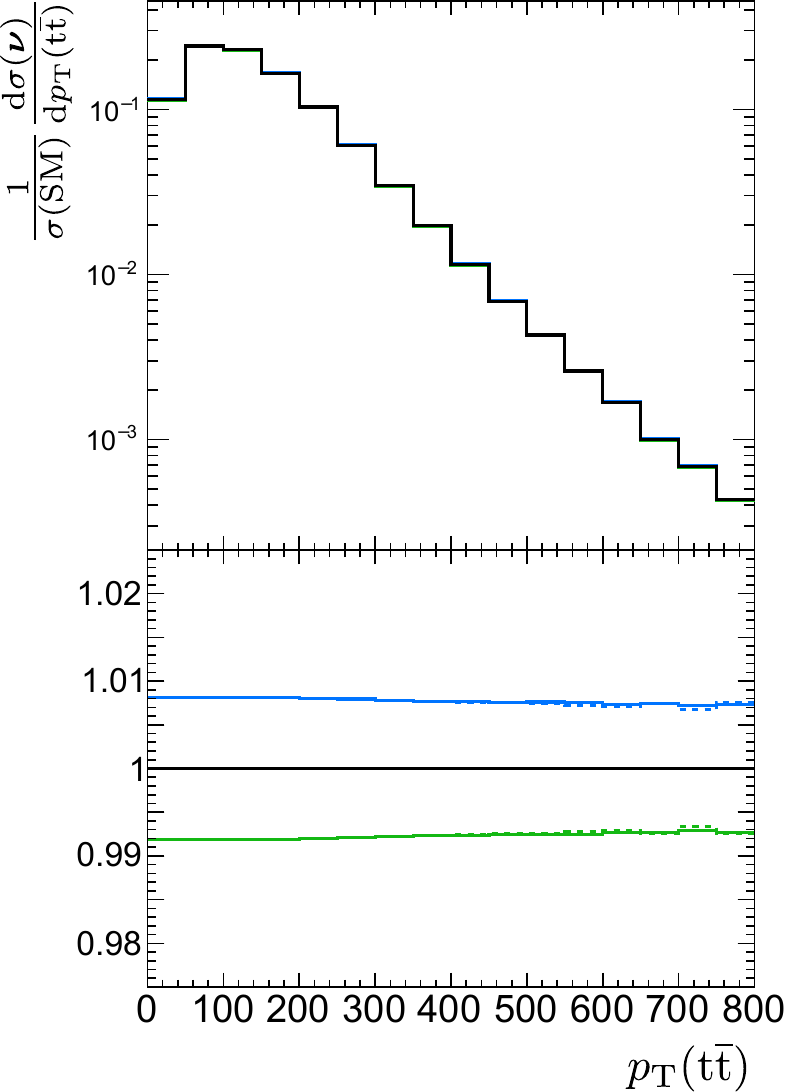}\hfill
\includegraphics[width=.31\textwidth]{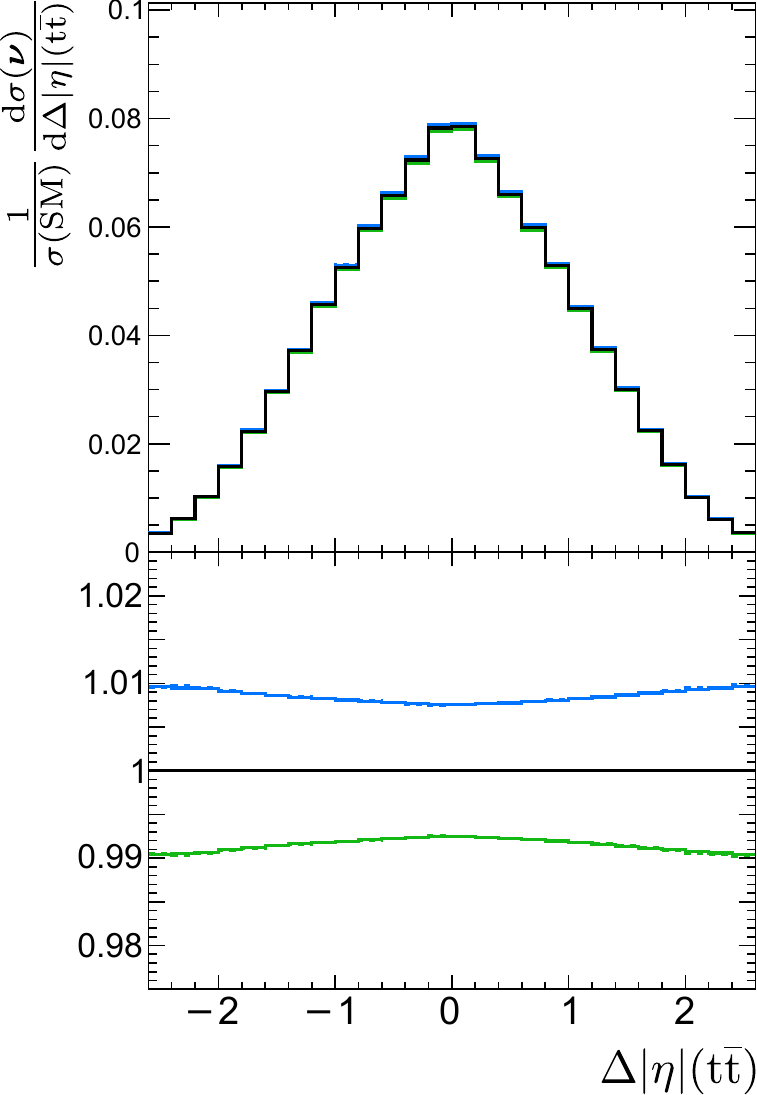}
  \caption{Variations of the nominal simulation with the nuisance parameter for the lepton scale factor uncertainty $\bn_{\ell}$. The dashed lines show the training data, and the solid lines show the result from the surrogate $\hat S_\ell(\bx|\nu_\ell)$.}
  \label{fig:lepton-SF}
\end{figure}

Uncertainties in lepton efficiencies are detailed in Refs.~\cite{open-data-ws-1,open-data-ws-2,open-data-ws-3} and are handled using the weighting function in Eq.~\ref{eq:lepton-sf}. Since the efficiency scale factors and uncertainties depend on the candidate’s pseudo-rapidity, which is not included in $\bx$, we retain the \bz dependence in
\begin{align}    r(\bx_i,\bz_i|\nu_\ell)=\prod_{\ell=1}^{2}\left(1+\frac{\Delta_\ell\textrm{SF}(\ell)}{\textrm{SF}(\ell)}\right)^{\nu_\ell},\label{eq:leptons-2}
\end{align}
used to define two surrogate data sets corresponding to $\pm1\sigma$ variations. Here, $\mathcal{V}=\{-1,1\}$, and we learn a surrogate
\begin{align}
\hat S_\ell(\bx|\nu_\ell)=\exp\left(\nu_\ell\hat\Delta_\ell(\bx)\right).
\end{align}
Variations are under 1\% in all cases, with minimal \bx-dependence, as shown in Fig.~\ref{fig:lepton-SF}.

\subsection{Testing the tree-based estimates with neural networks}\label{sec:C2ST}

The comparisons in previous sections are one-dimensional projections. For a more general check of whether the BPT is fully expressive in the high-dimensional \bx space, we can apply a ``Classifier two-sample test''~(C2ST)~\cite{osti_826696,lopezpaz2018revisiting}.
\begin{figure}[t]
    \centering
    \includegraphics[width=.45\textwidth]{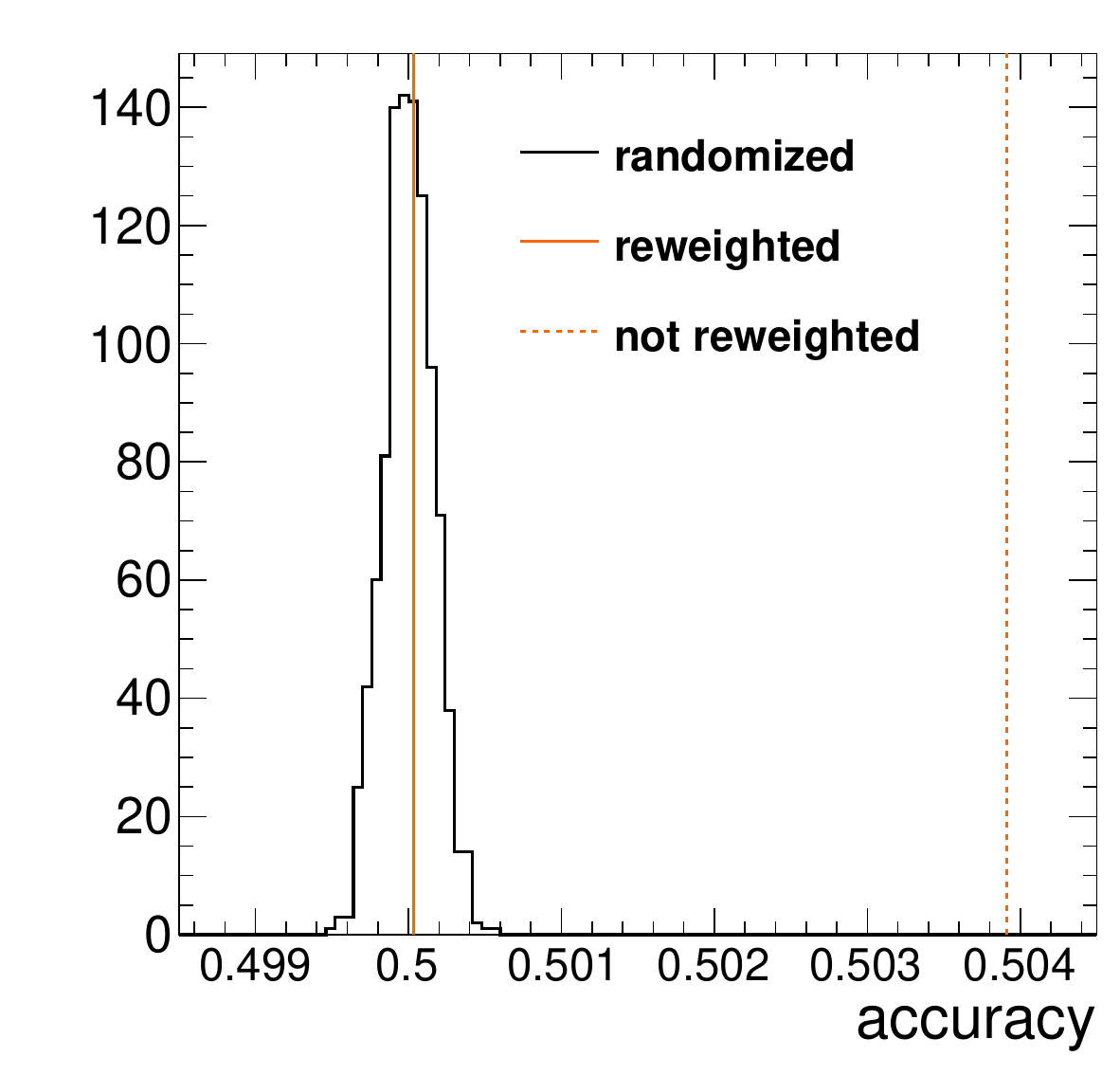}
  \caption{A neural-network classifier tests the surrogate $\hat S_{\textrm{HF}}(\bx|\nu_{\textrm{HF}})$ for $\nu_{\textrm{HF}}=1$. The black contour shows the distribution of the classifier when the two samples are from the same distribution at $\bn_{\textrm{HF}}=0$. The solid orange line indicates the accuracy of a classifier trained with an SM event sample and a sample with $\bn_{\textrm{HF}}=1$ that is reweighted back to the SM using the surrogate under test, suggesting no flaw in the performance. The dashed orange line indicates the accuracy of the test trained with two unequal samples at  $\bn_{\textrm{HF}}=0$ and $\bn_{\textrm{HF}}=1$.}
  \label{fig:C2ST}
\end{figure}

The C2ST is a non-parametric method for assessing if two samples originate from the same distribution. It trains a binary classifier on a combined dataset of the two samples, using labels to indicate sample origin. The classifier's accuracy reveals distribution similarity; accuracy above chance suggests different distributions.

Given a specific nuisance parameter $\bn$ and a pair of synthetic data sets, $\mathcal{D}_{\textrm{SM}}$ and $\mathcal{D}_{\nu}$ with $\nu\neq0$, if a candidate estimate $\hat S(\bx|\bn)$ is accurate and fully expressive, then
\begin{align}
\hat S(\bx|\nu)=\frac{\dd\sigma(\bx|\bn)\;\;\,\,}{\dd\sigma(\bx|\textrm{SM})}
\end{align}
for all $\bx$ and $\bn$.
Thus, reweighting $\mathcal{D}_\bn$ to form 
\begin{align}
    \mathcal{D}_{\textrm{reweighted}}=\left\{w'_i=\hat S(\bx|\bn)^{-1}w_i,\bx_i \;\;\textrm{for all}\;\;w_i,\bx_i\in\mathcal{D}_\bn\right\},
\end{align}
should make $\mathcal{D}_{\textrm{reweighted}}$ indistinguishable from $\mathcal{D}_{\textrm{SM}}$. To test this, a classifier's accuracy in distinguishing $\mathcal{D}_{\textrm{reweighted}}$ from $\mathcal{D}_{\textrm{SM}}$ is used, with a p-value based on the null distribution of the accuracy.
We train a classifier using HF b-tagging with $\nu=\nu_{\textrm{HF}}=1$ to test $\hat S_{\textrm{HF}}(\bx|\nu_{\textrm{HF}}=1)$. The classifier, a neural network in \texttt{pytorch} with sigmoid activation and three hidden layers (512, 512, 256 units), is optimized with \texttt{Adam} on half of the data. Its accuracy is 0.5001, suggesting near-perfect agreement. To evaluate this result, we merge $\mathcal{D}_{\textrm{reweighted}}$ and $\mathcal{D}_{\textrm{SM}}$, randomize labels, and train 1000 classifiers on pairs of identical subsets. The null distribution peaks at 0.5, as shown in Fig.~\ref{fig:C2ST}. For comparison, distinguishing $\mathcal{D}_{\nu_{\textrm{HF}}=1}$ from $\mathcal{D}_{\textrm{SM}}$ yields 0.504, a significant deviation (Fig.~\ref{fig:C2ST}).
This deviation is small due to the mild $\bx$-dependence of the DCR, yet the neural network accurately detects the difference. In summary, removing $\nu_{\textrm{HF}}$-dependence with our surrogate makes it impossible for a high-sensitivity neural network to distinguish from the SM, indicating strong performance across the feature space.

\subsection{Expected Limits from unbinned Asimov data}

The Asimov dataset~\cite{Cowan:2010js} is commonly used to derive expected exclusion limits from binned Poisson likelihoods~\cite{CMS:2024onh}. Ref.~\cite{GomezAmbrosio:2022mpm} extends this to the unbinned case, enabling sampling-free exclusions within continuous parametric models. Here, we consider composite hypotheses involving two Wilson coefficients, which we denote by \bt. Under the exclusion scenario, \bt represents the null hypothesis with $N_{\bt}=2$, while other Wilson coefficients are profiled as nuisance parameters. The alternative hypothesis assumes $\bt=\bzero$.

Wilks' theorem states that if the data are distributed under the null hypothesis $\boldsymbol{\theta}$, the test statistic \( p(q_\bt|\bt,\bn) \) asymptotically follows a central \(\chi^2\) distribution with \( N_{\bt} \) degrees of freedom. This distribution is independent of the true values of the nuisance parameters. Given that our POIs primarily influence the predictions linearly~\cite{Elmer:2023wtr}, we assume any minor quadratic terms do not invalidate Wilks’ theorem~\cite{Bernlochner:2022oiw}. However, in practical applications, this assumption should be verified, as shown in Ref.~\cite{GomezAmbrosio:2022mpm}, where good agreement was observed.
Since \( q_\bt \) is monotonic with the p-value, it can define acceptance regions for \bt at confidence levels of 68\% (\(\alpha=32\%\)) or 95\% (\(\alpha=5\%\)). We anticipate excluding a hypothesis \bt at a given CL if there’s a 50\% or greater probability for \( q_\bt \) to fall outside the corresponding acceptance region when the alternate hypothesis \(\bt=\bzero\) is true.
Therefore, we must solve 
\begin{align}
\int_{q_{\bt,\textrm{med}}}^{\infty}p(q_\bt|\bt)\dd q_{\bt}=\alpha\quad\textrm{and}\quad q_{\bt,\textrm{med}}=\textrm{Med}(q_\bt|\bt=\bzero).
\end{align}
The final ingredient is Wald's theorem~\cite{Wald}, which implies that the distribution $\textrm{p}(q_\bt|\bzero)$ asymptotically follows a non-central $\chi^2$ distribution with $N_\bt$ degrees of freedom and non-centrality parameter $\Lambda$. This parameter can be computed (see Ref.~\cite{GomezAmbrosio:2022mpm} for details) for the unbinned likelihood ratio. As in the binned case, it corresponds to the Asimov expectation of Eq.~\ref{eq:likelihoodratio-normalized} for the alternate hypothesis, multiplied by $-2$. In our notation, the result is
\begin{align}
-\frac{1}{2}\Lambda&=-\mathcal{L}(\bn)\,\sigma(\bt,\bn)+\mathcal{L}_0\,\sigma(\textrm{SM})+\mathcal{L}_0\left\langle\log\left(\mathcal{L}(\bn)\mathcal{R}(\bx_i|\bt,\bn)/\mathcal{L}_0\right)\right\rangle_{\textrm{SM}}-\frac{1}{2}\sum_{k=1}^K\nu_k^2\nonumber\\
&=\sum_{\bx_i,w_i\in\mathcal{D}_0\cap\mathcal{X}}w_i\left(-\mathcal{L}(\bn)  \mathcal{R}(\bx_i|\bt,\bn)+\mathcal{L}_0 +\mathcal{L}_0 \log\left(\mathcal{L}(\bn)\mathcal{R}(\bx_i|\bt,\bn)/\mathcal{L}_0\right)\right)-\frac{1}{2}\sum_{k=1}^K\nu_k^2
\end{align}
with the integrated luminosity from Eq.~\ref{eq:lumi-example} and the model's DCR $\mathcal{R}(\bx|\bt,\bn)$ from Eq.~\ref{eq:tt2l-diff-xsec-model}. The sum is over all events in the nominal $\ttbar(2\ell)$ sample passing the event selection. The ratio $\mathcal{R}(\bx|\bt,\bn)$ appears in both the logarithm and the ``extended'' term for the total fiducial cross-section. This expression provides the test statistic under the alternate hypothesis, allowing us to obtain the expected exclusion contour from the profiled likelihood test statistic. The minimization is performed with the \textsc{iminuit} package~\cite{iminuit}.

\begin{figure}[tph]
    \centering
    \includegraphics[width=.99\textwidth]{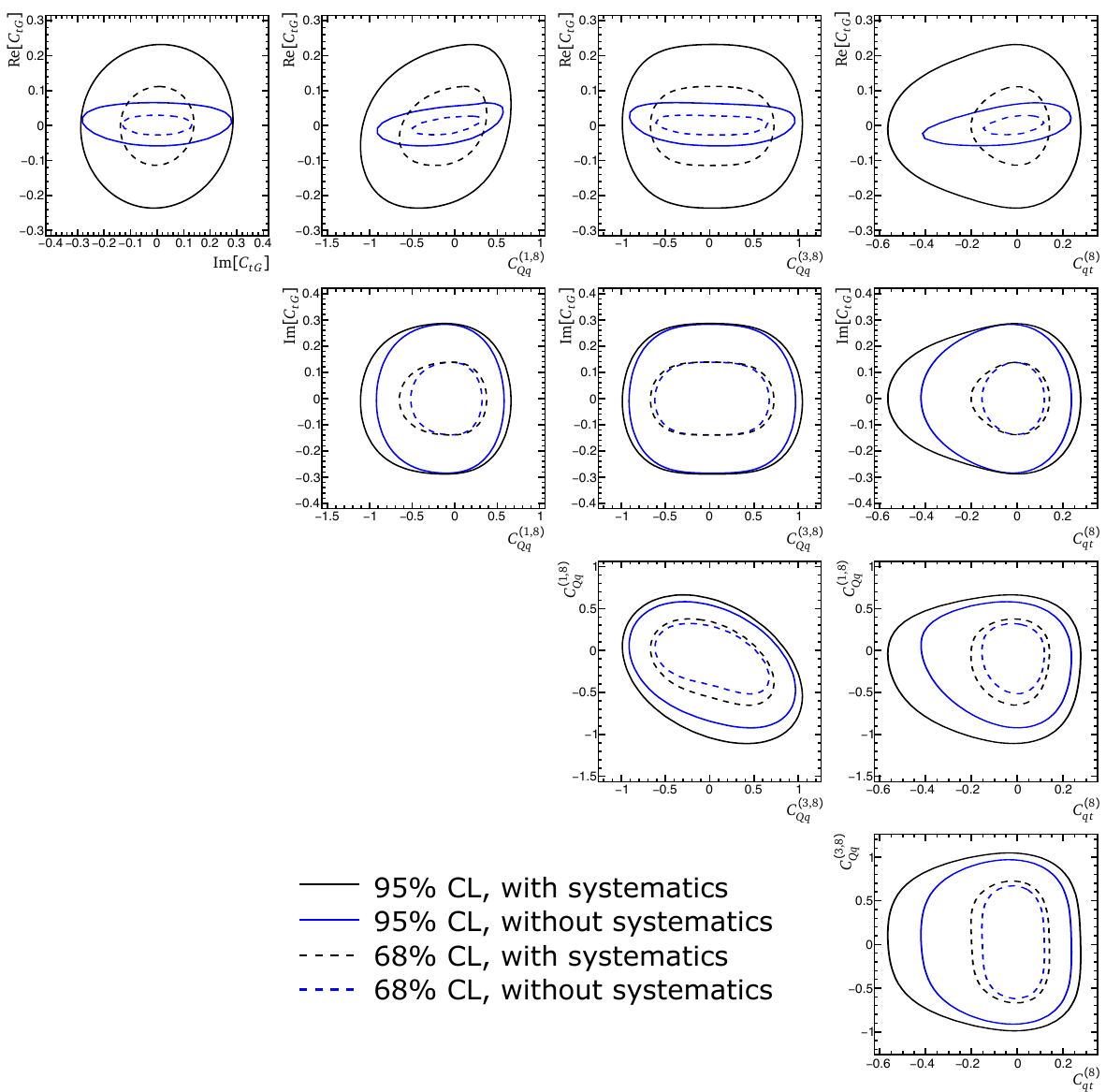}
  \caption{Asimov expected exclusion contours at 68\% CL~(dashed) and at 95\% CL~(solid). The three other Wilson coefficients are profiled for the blue contours, and all nuisance parameters are frozen at zero. For the black contours, the three other Wilson coefficients and all the nuisance parameters are profiled.}
  \label{fig:results}
\end{figure}

\subsection{Results}
Figure~\ref{fig:results} shows the Asimov expected exclusion contours at 68\% CL~(dashed) and 95\% CL~(solid). The SMEFT effects are simulated up to quadratic order in the POIs. For the blue contours, the other three Wilson coefficients are profiled, while nuisance parameters are set to zero. For the black contours, nuisance parameters are also profiled. Systematic uncertainties significantly impact $\textrm{Re}[C_{tG}]$, which strongly affects the total yield and is sensitive to integrated luminosity, renormalization and factorization scales, and normalization uncertainties.
For non-zero $\textrm{Im}[C_{tG}]$ and $\textrm{Re}[C_{tG}]\approx 0$, effects from the three remaining four-fermion operators outweigh those of systematic uncertainties, explaining why the contours degrade only slightly when including systematics.
A comprehensive $\ttbar(2\ell)$ sensitivity analysis would require analyzing all uncertainties, some of which are not publicly available. This study, however, shows how systematic effects can be captured in machine-learned surrogates and applied in limit setting.

\section{Conclusion}\label{sec:conclusion}

This paper presents a comprehensive, scalable framework for modeling the effects of systematic uncertainties in unbinned analyses of collider data. By factorizing systematic effects across parton, particle, and detector levels, we make them accessible for machine learning. With a highly granular factorization of the various dependencies, we leverage the extensive knowledge gained from binned LHC data analyses and fully capitalize on high-quality Monte Carlo simulation.
A flexible approach facilitates the progressive refinement of unbinned models, including but not restricted to applications in SMEFT. It accommodates new systematic effects or background contributions without invalidating previously trained surrogates.

A significant technical innovation introduced is the Boosted Parametric Tree~(BPT), an extension of tree-boosting algorithms designed to learn accurate parametrizations of systematic dependencies. BPTs offer a robust and efficient alternative to neural networks for modeling systematic effects, providing reliable surrogate models for complex, high-dimensional parameter spaces in unbinned hypothesis testing.

Our work thus bridges a critical gap in the methodological toolbox for SMEFT analyses, searches for other non-resonant effects beyond the standard model, and similar inference problems. We demonstrate the practical application through a semi-realistic case study of top quark pair production in the dilepton channel, which underscores the effectiveness of our approach in learning and incorporating systematic effects.
Overall, the new techniques pave the way for more refined and adaptable unbinned hypothesis tests, enhancing the accuracy and reliability of SMEFT analyses. We anticipate these advancements will be instrumental in exploiting the data from future collider experiments.
Finally, we believe that publicly available refined models would be useful for future SMEFT combinations and for providing legacy LHC results. 

\section*{Acknowledgements}

The computational results were obtained using the Vienna Bio Center and the CLIP computing cluster of the Austrian Academy of Sciences  at~\url{https://www.clip.science/}.
I am indebted to Suman Chatterjee, Claudius Krause, Tilman Plehn, Dennis Schwarz, Nick Smith, and Nicholas Wardle for many useful discussions. 

\newpage
\section*{Appendices}
\begin{appendix} \numberwithin{equation}{section}

\section{Per-event SMEFT weights}\label{sec:EFT-weights}
We show how to efficiently obtain a polynomial per-event SMEFT parametrization from generator weights obtained at a sufficient number of different values $\bt$ with dimension $N_\bt$. The procedure can be extended to arbitrary fixed polynomial order, but for simplicity, we truncate after the quadratic term,
\begin{align}
w_i(\bt)=\omega_{i,0}+\omega_{i,m}\theta_m+\omega_{i,mn}\theta_m\theta_n.\label{eq:SMEFT-poly-param1}
\end{align}
We can take the quadratic coefficients for each event as an upper triangular matrix,  $\omega_{i,mn}=0$ for $n<m$ for all events $i$. From the generator, we can obtain the r.h.s. of Eq.~\ref{eq:SMEFT-poly-param1} as 
\begin{align}
    w_i(\bt)\propto|\mathcal{M}_{\textrm{SMEFT}}(\bz_{p,i}|\bt)|^2,
\end{align}
which we evaluate for $M=1,\ldots,|M|$ base points. We denote those parameter values by $\bt^M$ and the resulting base point weights by $w_{i}{}^M=w_i(\bt^M)$.
We chose $|M|$ to correspond to the maximum number of independent per-event coefficients so that we have exactly enough base point weights to specify the general polynomial dependence. Therefore, 
\begin{align}
    |M|=1+N_\bt+\frac{1}{2}N_\bt(N_\bt+1),
\end{align}
where the three terms in the sum correspond to the number of independent coefficients corresponding to the constant, the linear, and the quadratic per-event SMEFT dependence. To be explicit,  $\theta_{M,m}$ is the value of the m-th Wilson coefficient at the M-th base point. For each event $i$, this gives us the $|M|$ equations 
\begin{align}
w_{i}{}^M=\omega_{i,0}+\omega_{i,m}\theta^M_{m}+\omega_{i,mn}\theta^M_{m}\theta^M_{n}\label{eq:SMEFT-poly-param2}
\end{align}
where we use Einstein summation for $m$ and $n$.
This is an $|M|\times|M|$ linear equation in $\omega_{i,0}$,  $\omega_{i,m}$,  and $\omega_{i,mn}$ with coefficients 1,  $\theta^M_{m}$,  and $\theta^M_{m}\theta^M_{n}$. Equation~\ref{eq:SMEFT-poly-param2} suggests to relabel the indices $\{(1),(m),(mn)\}$ by a multi-index K=$1,\ldots,|M|$ where the 1 represents the constant piece, $m$ the $N_\bt$ linear terms, and the ordered pair $(mn)$ the $1/2N_\bt(N_\bt+1)$ different quadratic terms. Any value of $\bt$ can then also be represented as an $|M|$-component vector $\theta_K=\{1,\theta_m,\theta_{m}\theta_n\}_K$ and the $|M|$ base points $\bt^M$ provide the $|M|\times|M|$ matrix 
\begin{align}
    C^M{}_K=\left\{1,\theta^M_m,\theta^M_{m}\theta^M_n\right\}_K.
\end{align}
Concretely, when $N_\bt=15$, we have 136 values that the indices $M$ and $K$ can take. 
Equation~\ref{eq:SMEFT-poly-param2} then reads
\begin{align}
w_i{}^M=C^M{}_K \omega_{i}{}^K
\end{align}
The matrix $C$ and its inverse do not depend on the event as long as the base points are kept when running the generator. The base points must be chosen such that $C^{-1}$ exists.
We can now compute the per-event polynomial weight coefficients as
\begin{align}
\omega_{i}{}^K=C^{-1}{}^K{}_M\,w_i{}^M\label{eq:SMEFT-coeffs}
\end{align}
from the per-event base-point weights $w_i{}^M$. With these coefficients, we can now evaluate Eq.~\ref{eq:SMEFT-poly-param1} for variable $\bt$ as 
\begin{align}
    w_i(\bt)=\theta_K \, C^{-1}{}^K{}_M\,w_i{}^M.
\end{align}
Finally, we can break up the index $K$ again, i.e.,  $K=1$ will give us the constant coefficient, the $N_\bt$ terms $K=m$ will give us the linear event-weight dependence,  and the $1/2N_\bt(N_\bt+1)$ terms $K=(mn)$ provide the quadratic coefficients. For the SMEFT coefficients in Eq.~\ref{eq:r-SMEFT} we find from Eq.~\ref{eq:SMEFT-coeffs}
\begin{align}
r^{(m)}(\bz_{\textrm{p},i})&=\omega_{i}{}^{(m)}/\omega_{i,0},\nonumber\\
r^{(mn)}(\bz_{\textrm{p},i})&=\omega_{i}{}^{(mn)}/\omega_{i,0}.
\end{align}

\section{Alternative loss functions}\label{sec:alternate-losses}
The solution in Eq.~\ref{eq:generic-minimum} can be obtained from other loss functions that differ in behavior away from the minimum. An example is the quadratic loss
\begin{align}
L_{\textrm{Q}}[\hat f]&=\left\langle\hat f(\bx)^2\right\rangle_{\bx,\bz|\bto,\bno}+\left\langle (1-\hat f(\bx))^2\right\rangle_{\bx,\bz|\btn,\bnn},
\end{align}
which can be used with synthetic data sets, either with or without reweighting, by following the same steps as in Sec.~\ref{sec:likelihood-ratio-trick-and-CE}. For the latter case, the result is
\begin{align}
L_{\textrm{Q}}[\hat f]&=\int\dd\bx\dd\bz\;p(\bx,\bz|\btn,\bnn)\left(r(\bx,\bz|\bto,\bno,\btn,\bnn)\hat f(\bx)^2 + (1-\hat f(\bx))^2\right).
\end{align}
The same is true for the mean-squared-error loss function 
\begin{align}
L_{\textrm{MSE}}[\hat f]&=\left\langle \left(\hat f(\bx)-r(\bx,\bz|\bto,\bno,\btn,\bnn)\right)^2\right\rangle_{\bx,\bz|\btn,\bnn}.
\end{align}
Its minimum satisfies
\begin{align}
 \frac{\dd\sigma(\bx|\bto,\bno)}{\dd\sigma(\bx|\btn,\bnn)}=f^{\ast}_{\textrm{MSE}}(\bx).\label{eq:generic-minimum-MSE}
\end{align}
If needed, a version with separate samples is obtained by expanding the square and keeping the $\hat f$-dependent terms. The result is
\begin{align}
L_{\textrm{MSE}}[\hat f]&=\left\langle\hat f(\bx)^2\right\rangle_{\bx,\bz|\bto,\bno}-2\left\langle \hat f(\bx)\right\rangle_{\bx,\bz|\bnn,\bnn}.
\end{align}
More loss functions can be obtained from the general ansatz
\begin{align}
L[\hat f]&=\left\langle L_1[\hat f(\bx)]\right\rangle_{\bx,\bz|\bto,\bno}+\left\langle L_2[\hat f(\bx)]\right\rangle_{\bx,\bz|\btn,\bnn},\label{eq:loss-general}
\end{align}
where $L_1$ and $L_2$, typically, are simple functions of $\hat f$.
Because it is a sum of expectations over the joint space, this general form allows using the joint-likelihood-ratio in the same way as done for Eq.~\ref{eq:CE-1samp}. Moreover, because $\hat f(\bx)$ does not depend on $\bz$, it is minimized by a function of the ratio of two $\bz$-integrals~(Eq.~\ref{eq:fiducial-pdf-int}) and, therefore, is in one-to-one correspondence with the regression target\footnote{I thank Giuliano Panico for pointing this out.}. If we view the two terms $L_1$ and $L_2$ as (standard) functions of $\hat f$ and denote the (standard)  derivative by $L'$, it is straightforward to show that the conditions
\begin{align}
-\frac{L_2'}{L_1'}=\frac{1}{\hat f}-1 \qquad\textrm{and}\qquad
-\frac{L_2'}{L_1'}=\hat f\label{eq:general-loss}
\end{align}
lead to loss functions minimized by Eq.~\ref{eq:generic-minimum} 
and  Eq.~\ref{eq:generic-minimum-MSE}, respectively. The loss functions discussed so far are special cases of Eq.~\ref{eq:general-loss}. To control the loss behavior away from the minimum, one can choose, therefore, an appropriate $L_1[\hat f]$ or $L_2[\hat f]$ and compute the other term from Eq.~\ref{eq:general-loss}.

\section{Construction of the BPT algorithm}\label{sec:parametric-regression-trees-details}

This section provides a step-by-step derivation of the BPT algorithm.
A summary of the resulting procedures is described in Sec.~\ref{sec:bpt}.

\subsection{Tree-boosting of parametric regressors}\label{sec:tree-boosting-details}
It is instructive to discuss boosting for generic non-parametric estimators based on the cross-entropy loss function $L_{\textrm{CE}}[\hat f]$ in Eq.~\ref{eq:CE-2samp}. 
After the replacement in Eq.~\ref{eq:parametric-ansatz}, we have
\begin{align}
L[\hat T]&=\left\langle\textrm{Soft}^+(\hat T(\bx))\right\rangle_{\bx,\bz|\bzero}+\left\langle\textrm{Soft}^+(-\hat T(\bx))\right\rangle_{\bx,\bz|\bn},\label{eq:loss-analytic-single-details}
\end{align}
where we do not yet specify the implementation of $\hat T(\bx)$.
The loss would attain its minimum at
\begin{align}
     T^\ast(\bx)=\log\frac{\dd\sigma(\bx|\bn)}{\dd\sigma(\bx|\bzero)\,},
\end{align}
but instead of obtaining this result in a single fit, we chose a number $B$ of boosting iterations and corresponding learning rates $0<\eta^{(b)}< 1$ for $b=1,\ldots,B$. We use an additive expansion of $\hat T(\boldsymbol{x})$ in terms of the weak learners $\hat t^{(b)}(\bx)$. To this end, we iterate the boosting relations
\begin{align}
 t^{(b)}{}^\ast(\bx)&=\argmin_{ \hat t^{(b)}} L\left[\hat t^{(b)}{} (\bx)+ \hat T^{(b-1)}(\bx) \right],\label{eq:boosting-1-details}\\
\hat T^{(b)}(\bx)&=\hat T^{(b-1)}(\bx)+\eta^{(b)}\hat t^{(b)}{}^\ast(\bx)\label{eq:boosting-2-details}
\end{align}
a number of $B$ times, starting with the initial choice $\hat T^{(0)}(\bx)=0$. 
Equation~\ref{eq:boosting-1-details} obtains the weak learner $\hat t^{(b)}(\bx)$ when the result of the preceding iteration $\hat T^{(b-1)}(\bx)$ is known. Equation~\ref{eq:boosting-2-details} updates the additive model with a fraction $\eta^{(b)}$ of this weak learner's prediction.
After $B$ iterations, the boosted prediction $\hat T^{(B)}$ can be expressed as
\begin{align}
    \hat T^{(B)}(\bx)&=\sum_{b=1}^B \eta^{(b)} t^{(b)}{}^\ast(\bx).\label{eq:boosting-solution-details}
\end{align}
An important practicality for boosting learners that are fit to synthetic data sets follows from the minimum condition in Eq.~\ref{eq:boosting-1-details}. It implies that the minimum at iteration $b$ satisfies
\begin{align}
   t^{(b)}{}^{\ast}(\bx)+\hat T^{(b-1)}(\bx) \simeq \log \frac{\dd\sigma(\bx|\bn)}{\dd\sigma(\bx|\bzero)}=\log\frac{\sigma(\bn)}{\sigma(\bzero)}\frac{\int\dd\bz\,p(\bx,\bz|\bn)}{\int\dd\bz\,p(\bx,\bz|\bzero)}
\end{align} 
which we rearrange to
\begin{align}
   t^{(b)}{}^\ast(\bx) \simeq \log \frac{\sigma(\bn)}{\sigma(\bzero)}\frac{\int\dd\bz\,p(\bx,\bz|\bn)\times\,\exp\left(-\hat T^{(b-1)}(\bx)\right)}{\int\dd\bz\,p(\bx,\bz|\bzero)\phantom{\times\,\exp\left(-\hat S^{(b-1)}(\bx)\right)}\;}.\label{eq:reweighting-boosting-joint-details}
\end{align}
By reading this equation as an \bx-dependent scaling of the joint-space integration measure $\dd \sigma(\bx,\bz|\bn)=\sigma(\bn)\,p(\bx,\bz|\bn)\dd\bx\dd\bz$ by the reciprocal of the estimate of the preceding boosting iteration, we find that $t^{(b)}{}^{\ast}$ can also be obtained if, instead of using the additive expansion, we replace $\sigma(\bn)\,p(\bx,\bz|\bn)\rightarrow \exp\left(-\hat T^{(b-1)}(\bx)\right)\,\sigma(\bn)\,p(\bx,\bz|\bn)$.
Because Eq.~\ref{eq:boosting-2-details} provides the exponent $\hat T$ iteratively, we only have to multiply the cross-section by $\exp(-\eta^{(b-1)}t^{(b-1)}{}^\ast(\bx))$ when moving from iteration $b-1$ to iteration $b$. This way, the boosting equations read
\begin{align}
 \sigma(\bn)p^{(b)}(\bx,\bz|\bn)&=\exp\left(-\eta^{(b-1)}t^{(b-1)}{}^\ast(\bx)\right)\,\sigma(\bn) p^{(b-1)}(\bx,\bz|\bn)\label{eq:boosting-2-1-details}\\
 t^{(b)}{}^\ast(\bx)&=\argmin_{ \hat t^{(b)}} L\left[\hat  t^{(b)}{} (\bx)\right],\label{eq:boosting-2-2-details}\\
\hat T^{(b)}(\bx)&=\hat T^{(b-1)}(\bx)+\eta^{(b)}\hat t^{(b)}{}^\ast(\bx)\label{eq:boosting-2-3-details},
\end{align}
initialized by $\hat T^{(0)}(\bx)=t^{(0)}{}^\ast(\bx)=0$.
The advantage of this formulation is that Eq.~\ref{eq:boosting-2-2-details} is a standard loss function minimization without the additive model appearing in the argument as in Eq.~\ref{eq:boosting-1-details}. The update of the synthetic data set $\mathcal{D}_{\bn}^{(b)}=\{w_i^{(b)},\bx_{\bn,i},\bz_i\}$, now also defined for each iteration $b$, follows from Eq.~\ref{eq:boosting-2-1-details} as  
\begin{align}
w_i^{(b)}=\exp\left(-\eta^{(b-1)}t^{(b-1)}{}^\ast(\bx_i)\right)w^{(b-1)}_{i}.\label{eq:reweighting-empirical-details}
\end{align}
This prescription can be interpreted as a recursive weighting of the differential cross-section of $\mathcal{D}_{\bn}$ in the second term in Eq.~\ref{eq:loss-analytic-single-details} towards $\mathcal{D}_{\bzero}$ in the first term.
The boosting algorithm removes the learned approximation from the training data as the regressor learns to approximate the DCR more accurately. 
It is customary to chose $\eta^{(b)}$ independently of $b$, and values between $10^{-3}$ and $3\cdot10^{-1}$ for this universal learning rate have proven efficient. 

The sample $\mathcal{D}_{\bzero}$ stays unchanged in the boosting procedure because we decided to write the \bx-dependent scaling in Eq.~\ref{eq:boosting-2-1-details} in the numerator. The choice of only reweighting the sample $\mathcal{D}_\bn$ is a critical detail. It holds the key to a boosting algorithm that works for the fully parametric regressor, including the \bn dependence. 
We can construct the loss for a parametric tree-based algorithm from the general parametric loss function in Eq.~\ref{eq:sum-loss-generic}, which is a sum of equally structured terms
\begin{align}
L&=\sum_{\bn\in\mathcal{V}}L_{\textrm{CE}}[\hat T(\bx|\bn)]=\sum_{\bn\in\mathcal{V}}\left(\left\langle\textrm{Soft}^+(\hat T(\bx|\bn))\right\rangle_{\bx,\bz|\bzero}+\left\langle\textrm{Soft}^+(-\hat T(\bx|\bn))\right\rangle_{\bx,\bz|\bn}\right)\label{eq:loss-analytic-single-2-details}.
\end{align}
The synthetic data set in the first expectation value in each sum term is always $\mathcal{D}_\bzero$, irrespective of the value of \bn.  The synthetic data set in the second expectation is $\mathcal{D}_{\bn}$ and is different for each $\mathcal{\bn}\in\mathcal{V}$. 
It is this term whose synthetic data set changes during the boosting algorithm, and because there is one such set for each $\bn\in\mathcal{V}$, the reweighting can be done simultaneously for each $\bn$ in the sum over $\mathcal{V}$ in Eq.~\ref{eq:loss-analytic-single-2-details}.
Repeating the steps starting at Eq.~\ref{eq:loss-analytic-single-details} with a sufficiently expressive \bn-dependent function $\hat T(\bx|\bn)$, it is straightforward to show that aside from the extra $\bn$-dependence in the notation nothing else changes. Concretely, we only need to modify Eq.~\ref{eq:boosting-solution-details} to notate the \bn-dependence in the weak learner $\hat t(\bx|\bn)$. The other steps follow analogously, and Eq.~\ref{eq:reweighting-empirical-details} generalizes to
\begin{align}
    \mathcal{D}_{\bn}^{(b)}= \left\{\exp(-\eta^{(b-1)} t^{(b-1)\ast}(\bx_i|\bn))\,w^{(b-1)}_i, \bx_i,\bz_i\right\} \;\; \textrm{for all} \;\; \{w^{(b-1)},\bx_i,\bz_i\}\in\mathcal{D}_{\bn}^{(b-1)}\;\;\textrm{for all}\;\;\bn\in\mathcal{V},
\end{align}
which is the same as Eq.~\ref{eq:reweighting-empirical-details} except for that it is carried out simultaneously for each $\bn\in\mathcal{V}$. This completes the boosting algorithm for generic weak learners, and we can proceed with constructing the tree-based implementation.

\subsection{Learning the phase-space partitioning}\label{sec:phase-space-partitioning-details}
We construct the parametric weak learner $\hat t^{(b)}(\bx|\bn)$ in two steps. Because the procedure is identical at each boosting iteration, we drop the superscript $(b)$ in this section in favor of readability and write $\hat t(\bx|\bn)$ in place of $\hat t^{(b)}(\bx|\bn)$.
We first specify the non-linearity in \bx while keeping a parametric \bn-dependence fully general. 

We decompose the phase space $\mathcal{X}$ into non-overlapping regions $\Delta\bx_j$, collectively denoted by $\mathcal{J}$. Such a phase-space partitioning satisfies
\begin{align}
\mathcal{X}=\bigcup_{j\in\mathcal{J}}\Delta\bx_j\quad\textrm{and} \quad \Delta\bx_j\cap\Delta\bx_{j'}=\emptyset\;\;\longleftrightarrow \;\;j\neq j'.\label{eq:def-partitioning-details}
\end{align}
The nonlinearity in a tree ansatz can always be expressed via the index function 
\begin{align}
  \mathbbm{1}_j(\bx)=\begin{cases} 
      1 & \textrm{if}\; \bx\in\Delta\bx_j \\
      0&\textrm{otherwise}
   \end{cases}
\end{align}
in terms of, for now, arbitrary functions $\hat t_{j}(\bn)$ that have no $\bx$-dependence,
\begin{align}
    \hat t(\boldsymbol{x}|\bn)
=\sum_{j\in\mathcal{J}}\mathbbm{1}_j(\bx)\,\hat t_{j}(\bn).\label{eq:x-nonlinearity-G-details}
\end{align}
The function $\hat t_{j}(\bn)$ should describe the DCR in bin $j$.

Because the $\bx$-dependence is only in the index function, we can insert Eq.~\ref{eq:x-nonlinearity-G-details} into Eq.~\ref{eq:loss-analytic-single-2-details} and use Eq.~\ref{eq:CE-2samp-empirical} to carry out the event sums over the synthetic data sets. The result is
\begin{align}
L[\mathcal{J},\hat t_j]=\sum_{j\in\mathcal{J}}L_j[\hat t_j]=\sum_{j\in\mathcal{J}}\sum_{\bn\in\mathcal{V}}\left[\sigma_{j,0}\,\textrm{Soft}^+(\hat t_j(\bn))+\sigma_{j,\bn}\,\textrm{Soft}^+(-\hat t_{j}(\bn))\right].\label{eq:softPlus-loss-details}
\end{align}

The $\sigma_{j,0}$ and $\sigma_{j,\bn}$ are given in terms of the training data as
\begin{align}
\sigma_{j,0}=\sum_{(\bx_i,w_i)\,\in\,\mathcal{D}_\bzero\cap\Delta\bx_j}w_i\qquad\textrm{and}\qquad\sigma_{j,\bn}=\sum_{(\bx_i,w_i)\,\in\,\mathcal{D}_\bn\cap\Delta\bx_j}w_i\label{eq:sigma-sums-details}
\end{align}
and can be understood as the synthetic predictions for the cross-section in bin $j\in\mathcal{J}$ for nuisance parameters $\bzero$ and $\bn$, respectively.
We have now decomposed our problem into two related problems that each pertain to different trainable parameters: the phase space partitioning $\mathcal{J}$ and, independently in each region of the partitioning, a function $\hat t_{j}(\bn)$ whose $\bn$-dependence we still have to specify. 

Before we tackle these problems, it is instructive to develop an intuition for the loss function in Eq.~\ref{eq:softPlus-loss-details}.
We assume an infinitely expressive $\hat t_j(\bn)$ and functionally differentiate 
Eq.~\ref{eq:softPlus-loss-details} to arrive at
\begin{align}
    0=\frac{\delta L_j}{\delta\hat t_{j}}=\sum_{\bn\in\mathcal{V}}\left[\frac{\sigma_{j,\bzero}}{1+\exp\left(-\hat t_{j}(\bn)\right)}-\frac{\sigma_{j,\bn}}{1+\exp\left(\hat t_{j}(\bn)\right)}\right].
\end{align}
This equation is satisfied exactly if
\begin{align}
\hat t_{j}(\bn)=\log\frac{\sigma_{j,\bn}}{\sigma_{j,\bzero}}\quad\textrm{for all}\quad \bn\in\mathcal{V}, \label{eq:predictor-exact-details}
\end{align}
which we can always fulfill for sufficiently expressive $\hat t_{j}(\bn)$; a perfect representation of the $\bn$-dependence in each tree node $j\in\mathcal{J}$ reproduces the logarithm of the DCR for those values of $\bn$ whose synthetic data sets we included in the loss function.
The predictive function of a tree is finitely expressive. Hence, we will seek an approximation for Eq.~\ref{eq:predictor-exact-details} in the next section. But we can meanwhile use the result to shed light on the loss function Eq.~\ref{eq:softPlus-loss-details}. Formally eliminating $\hat t_{j}(\bn)$ in favor of its predictions at the points $\bn\in\mathcal{V}$, we express the loss solely in terms of the per-bin cross-sections  $\sigma_{j,\bzero}$ and $\sigma_{j,\bn}$,
\begin{align}
L[\mathcal{J}]&=\sum_{j\in\mathcal{J}}\sum_{\bn\in\mathcal{V}}\left[\sigma_{j,\bzero}\log\left(1+\frac{\sigma_{j,\bn}}{\sigma_{j,\bzero}}\right)+\sigma_{j,\bn}\log\left(1+\frac{\sigma_{j,\bzero}}{\sigma_{j,\bn}}\right)\right].\label{eq:boostingloss-details}
\end{align}
This equation would provide a loss function for finding the optimal phase space partitioning if we did not need to use finitely expressive $\hat t_j(\bn)$.
If we assume small $\bn$ such that a Taylor expansion of $\sigma_{j,\bn}$ around $\bn=\bzero$ is a good approximation, we get
\begin{align}
 L[\mathcal{J}]&=-\frac{1}{4}\sum_{j\in\mathcal{J}}\sum_{\bn\in\mathcal{V}}\nu_a\nu_b I_{(ab),j}+\ldots
\end{align}
where the ellipsis comprise $\mathcal{O}(\nu^3)$ terms and $\mathcal{J}$-independent contributions. The quantity 
\begin{align}
I_{(ab),j}=\frac{1}{\sigma_{j,\bzero}}\frac{\partial \sigma_{j,\bn}}{\partial \nu_a}\frac{\partial \sigma_{j,\bn}}{\partial \nu_a}\Bigg|_{\bn=\bzero}\label{eq:Fisher-expansion-details}
\end{align}
is the leading contribution in $L[\mathcal{J}]$ and represents the Fisher information matrix of a Poisson measurement in bin $j$ regarding the model parameters.
We thus show that our loss function will guide the algorithm towards finding a partitioning $\mathcal{J}$ that maximizes the sum of the Fisher information over all terminal tree nodes. 

\subsection{Terminal node predictions}\label{sec:terminal-node-details}

The second and final constructive step is to curtail the $\bn$-dependence of $\hat t_j$ for each node in the weak learner. We can choose it in analogy to the binned case as it will turn out.
We use the ansatz
\begin{align}
\hat t_j(\bn)&=\nu_a\hat\Delta_{a,j}+\nu_a\nu_b\hat\Delta_{ab,j}+\nu_a\nu_b\nu_c\hat\Delta_{abc,j}+\cdots=\nu_A\hat\Delta_{A,j}\label{eq:poly-expansion-details}
\end{align}
with the multi-index notation as explained in Sec.~\ref{sec:ml-parametrizations}. The polynomial order and the coefficients at each polynomial order are truncated to the application's required accuracy. We also 
allow for its fine-tuning by excluding some of the terms in the polynomial for application-specific reasons.
If the node $j$ is small enough that the DCR does not significantly vary with \bx, we get for the first term
\begin{align}
\hat \Delta_{j,a} \approx \frac{\partial}{\partial\nu_a}t_j(\bn)\Bigg|_{\bn=\bzero}=\frac{\partial}{\partial\nu_a}\log \frac{\dd\sigma(\bx|\bn)}{\dd\sigma(\bx|\bzero)}\Bigg|_{\bn=\bzero}= s_a+\frac{\partial}{\partial\nu_a}\log\sigma(\bn)\Bigg|_{\bn=\bzero}\quad\textrm{for}\quad\bx\in\Delta\bx_j.\label{eq:param-expansion-details}
\end{align} 
The last expression relates $\hat \Delta_{j,a}$ to the well-known score vector $s_a$, a sufficient statistic for small $\bn$ and, therefore, an optimal observable. The log-derivative of the inclusive cross-section in the last term does not depend on the phase-space partitioning and, thus, is irrelevant to the optimization. The algorithm will, therefore, aim to reduce the expectation of the variance of the score in the training sample. This is, by definition, the negative value of the Fisher information matrix, consistent with the interpretation in the preceding section. Depending on the concrete problem and the desired accuracy, the higher-order terms in Eq.~\ref{eq:param-expansion-details} can improve the parametrization for larger values of~$\bn$.

We now determine the terminal node predictions of a weak learner up to a fixed arbitrary polynomial ordering $\bn$ by computing $\hat\Delta_{A,j}$.
The parametric tree ansatz is
\begin{align}
    \hat t(\boldsymbol{x}|\bn)= \sum_{j\in\mathcal{J}}\mathbbm{1}_j(\bx)\left(\nu_A\hat\Delta_{A,j}\right).\label{eq:tree-ansatz-poly-details}
\end{align}
An exact solution cannot be obtained for the optimal values of $\hat\Delta$ in the general case because the resulting equations 
\begin{align}
\nu_A\hat \Delta_{A,j}&=\log\frac{\sigma_{j,\bn}}{\sigma_{j,\bzero}}\quad\textrm{for all}\quad \bn\in\mathcal{V}\label{eq:predictor-details}
\end{align}
are overdetermined if $|\mathcal{V}|>N_\Delta$. We note that Eq.~\ref{eq:predictor-details} has the same form as Eq.~\ref{eq:predictor-exact-details} except for the finite expressivity on the l.h.s. We are content with an approximate solution of the per-node parametrization because boosting the weak learner will iteratively reduce the shortcomings either way. Any deficiency of a concrete weak learner will be reduced in the subsequent boosting iteration. 
For $|\mathcal{V}|<N_\Delta$, the training data cannot provide a unique estimate, and more data sets must be obtained.
 
The simplest approach for approximately solving Eq.~\ref{eq:predictor-details} is by minimizing the mean-squared error separately for each node $j\in\mathcal{J}$,
\begin{align}
L_{j,\textrm{MSE}}[\hat\Delta]&=\sum_{\bn\in\mathcal{V}}\left(\nu_A\hat\Delta_{A,j}-\log\frac{\sigma_{j,\bn}}{\sigma_{j,\bzero}}\right)^2.
\end{align}
It is solved by
\begin{align}
\hat\Delta_{A,j}=\left[\;\sum_{\nu\in\mathcal{V}}\bn\bn^{\textrm{T}}\right]_{AB}^{-1} \left[\sum_{\;\bn\in\mathcal{V}}\bn\log\frac{\sigma_{j,\bn}}{\sigma_{j,\bzero}}\right]_B.\label{eq:weak-learner-prediction-details}
\end{align}
The matrix 
\begin{align}
V_{AB}=\left[\sum_{\nu\in\mathcal{V}}\bn\bn^{\textrm{T}}\right]_{AB},\label{eq:V-details}
\end{align}
appearing in the approximate solution, is invertible if the base point coordinate matrix has full rank, as we have assumed in Sec.~\ref{sec:ml-parametrizations}.

It is instructive to check that the weak learner appropriately responds to training data that is perfectly, not just approximately, consistent with the polynomial ansatz. If we take constants $\delta_{A}$ and consider a model that predicts $\sigma_{j,\bn}=\exp(\nu_A\delta_{A})\sigma_{j,\bzero}$ in a given region, we can insert into Eq.~\ref{eq:weak-learner-prediction-details} and find $\hat\Delta_{A}=\delta_{A}$, confirming that the algorithm learns the exact solution if it has the chance.

To complete the construction of the weak learner, we insert the ansatz Eq.~\ref{eq:tree-ansatz-poly-details} into Eq.~\ref{eq:softPlus-loss-details} and get 
\begin{align}
L[\mathcal{J}]&=\sum_{j\in\mathcal{J}}\sum_{\bn\in\mathcal{V}}\left[\sigma_{j,\bzero}\,\textrm{Soft}^+\left(\nu_A\hat\Delta_{j,A}\right)+\sigma_{j,\bn}\,\textrm{Soft}^+\left(-\nu_A\hat\Delta_{j,A}\right)\right],\label{eq:weak-learner-loss-details}
\end{align}
where $\hat\Delta_{A,j}$ are obtained from Eq.~\ref{eq:weak-learner-prediction-details} and $\sigma_{j,0}$ and $\sigma_{j,\bn}$ from Eq.~\ref{eq:sigma-sums-details}.
The data samples $\mathcal{D}_\bzero$, used for the prediction of $\sigma_{j,\bzero}$ in the first term, can either taken to be the same or statistically independent samples. 
This loss function is amenable to standard tree algorithms, for example, the CART  algorithm or the ``Tree Alternate Optimization''~(TAO)~\cite{TAO-1,TAO-2,TAO-3,TAO-4} algorithm, both providing tree structures with a hierarchical selection using the features $\bx$ and that satisfy the requirements in Eq.~\ref{eq:def-partitioning-details}.  
These algorithms proceed by recursively splitting the training data along either axis-aligned~(for CART) or linear combinations of the input features~(for TAO), reducing the loss at each iteration. The maximum iteration depth and the minimum number of events in each node are hyperparameters that regularize the fit. If no more splits can be performed, the terminal selections~(nodes) represent a phase space partitioning $\mathcal{J}$ of the form in Eq.~\ref{eq:def-partitioning-details} and the quantities $\hat\Delta_j$ can be computed from Eq.~\ref{eq:weak-learner-prediction-details}. The tree then estimates the (log-)DCR as in Eq.~\ref{eq:tree-ansatz-poly-details}.
As a function of $\bx$, the prediction of a single tree changes discontinuously if $\bx$ traverses a boundary between nodes, and the possibly poor approximation close to the boundaries weakens the learner. Utilizing boosting, i.e., using a sequence of trees in Eq.~\ref{eq:boosting-1-details} and Eq.~\ref{eq:boosting-2-details}, we recover the smooth behavior in \bx of an arbitrarily expressive regressor. Because each tree in the boosted result in Eq.~\ref{eq:boosting-solution-details} is parametric in $\bn$, so is the final parametric regression tree.

\subsection{Algorithm summary}\label{sec:bpt-summary}

We can combine the steps in Secs.~\ref{sec:tree-boosting-details}--\ref{sec:terminal-node-details} to summarize the BPT algorithm. It is an iterative fit of a tree-based weak learner with the loss in Eq.~\ref{eq:weak-learner-loss-details} to the residuals of the preceding boosting iteration whose predictions are obtained from Eqs.~\ref{eq:boosting-2-1-details}--\ref{eq:reweighting-empirical-details}.
Concretely, we start with training data $\mathcal{D}_0$ at a reference point and several synthetic data sets associated with model parameters $\bn\in\mathcal{V}$. We must have enough data so $V_{AB}$ has full rank. We fit a weak learner using the CART algorithm. At each iteration, the CART algorithm recursively divides the feature space by greedily selecting the dimension and cut value combination that minimizes the loss function. 
Overfitting is mitigated by enforcing a maximum tree depth and a minimum number of events in each terminal node. 
We construct new synthetic data from the weak learner's prediction using Eq.~\ref{eq:reweighting-empirical-details} to replace $\mathcal{D}_\bn$. This reweighting procedure brings the samples $\mathcal{D}_\bn$ closer to $\mathcal{D}_0$ by an amount controlled by the learning rate $\eta$. We iterate the whole procedure $B$ times and obtain the final result from Eq.~\ref{eq:boosting-solution-details} as
\begin{align}
\hat T(\bx|\bn)=\log\hat S(\bx|\bn)=\sum_{b=1}^B\eta^{(b)}\sum_{j\in\mathcal{J}^{(b)}}\mathbbm{1}_j(\bx)\nu_A\hat\Delta^{(b)}_{A,j},
\end{align}
where \bn is the model parameter we like to predict for and $\mathcal{J}^{(b)}$ is the phase-space partitioning obtained from the CART algorithm. The $\hat \Delta^{(b)}_{A,j}$ are the polynomial coefficients of the DCR parametrization in the terminal node $j$ at boosting iteration $b$.
Algorithm~\ref{alg:boosting} is a pseudo-code summary of these steps and defines the parametric regression tree algorithm. 
It is efficiently implemented using the {\texttt{Numpy}} package~\cite{harris2020array} and available at~\cite{code}.

\subsection{An analytic toy example}\label{sec:analytic-toy}
To illustrate the BPT, we consider an arbitrarily chosen one-dimensional two-parameter model 
\begin{align}
\dd\sigma(\bx|\nu_1,\nu_2)&=N \exp\left(0.25\left(\nu_1\sin(x)+\nu_2\cos(0.5x)\right)^2\right)\dd x
\end{align}
with support $x\in[-\pi,\pi]$. The logarithm of the cross-section is a quadratic polynomial in $\bn$ for all \bx, suggesting a perfect fit with a two-parameter parametric tree at quadratic accuracy. For the training, we chose five base points $\mathcal{V}=\{(0.5, 0),\, (0, 0.5),\, (1, 0),\, (0, 1),\, (0.5, 0.5)\}$ that lead to a full-rank matrix $V$ in Eq.~\ref{eq:V-details}. With a nominal data set at $(\nu_1,\nu_2)=(0,0)$, we have six synthetic data sets, each with $5\cdot10^5$ events, sufficient to train the algorithm. We fit $B=100$ boosting iterations and require a maximum tree depth of $4$ and a minimum requirement for the number of events in each terminal node, which is 50 events. The learning rate is set to 0.2 for all boosting iterations.

\begin{figure}[t]
    \centering
    \includegraphics[width=.95\textwidth]{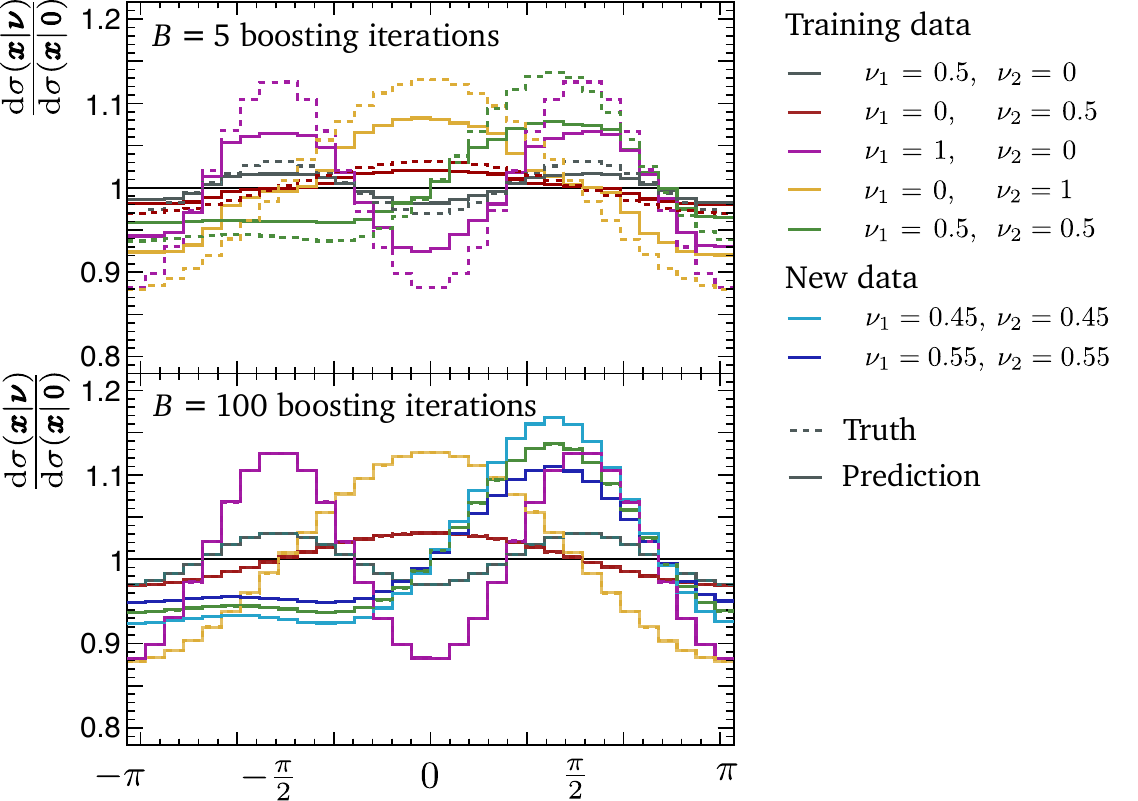}
  \caption{The DCR for various model parameters for the toy study described in the text. Dashed lines show the true DCR from the training data; solid lines show the prediction obtained from the BPT algorithm. The top~(bottom) panel compares after $B=5$~($B=100$) boosting iterations. The bottom panel also includes new model parameter points not present during training. In this panel, the prediction is drawn after the truth, and the dashed histograms are not separately visible because of the good quality of the fit with $B=100$.  }
  \label{fig:analytic-toy}
\end{figure}

In Fig.~\ref{fig:analytic-toy}, we compare the true and predicted values for the DCR for various model parameters. The model parameter configurations include the training and new synthetic data, which are absent during training. After only five iterations, the prediction begins to resemble the true DCRs. After 100 iterations, the fit is nearly perfect; dashed lines show the true DCR from the training data and are not separately visible because of the fit's quality, including the parameter configurations not used during training. In realistic applications, the logarithms of the true DCRs will not be exactly polynomial, mandating some degree of validation of the fit quality on unseen data.

\section{Additional angular observables in the \texorpdfstring{$\ttbar(2\ell)$}{tt2l} final state}\label{sec:ttbar-xi}

\begin{figure}[p!]
\centering
\includegraphics[width=.32\textwidth]{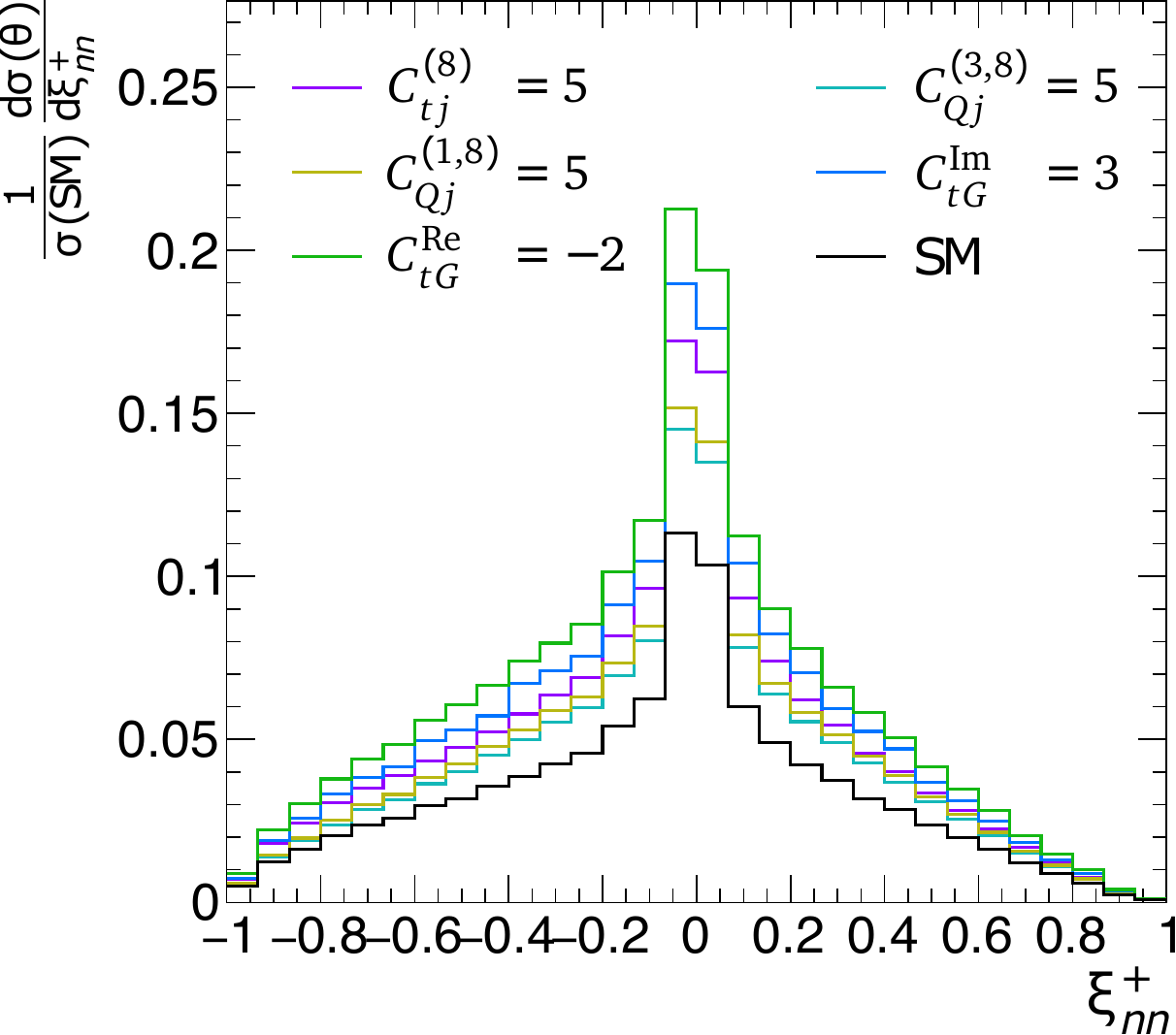}\hfill
\includegraphics[width=.32\textwidth]{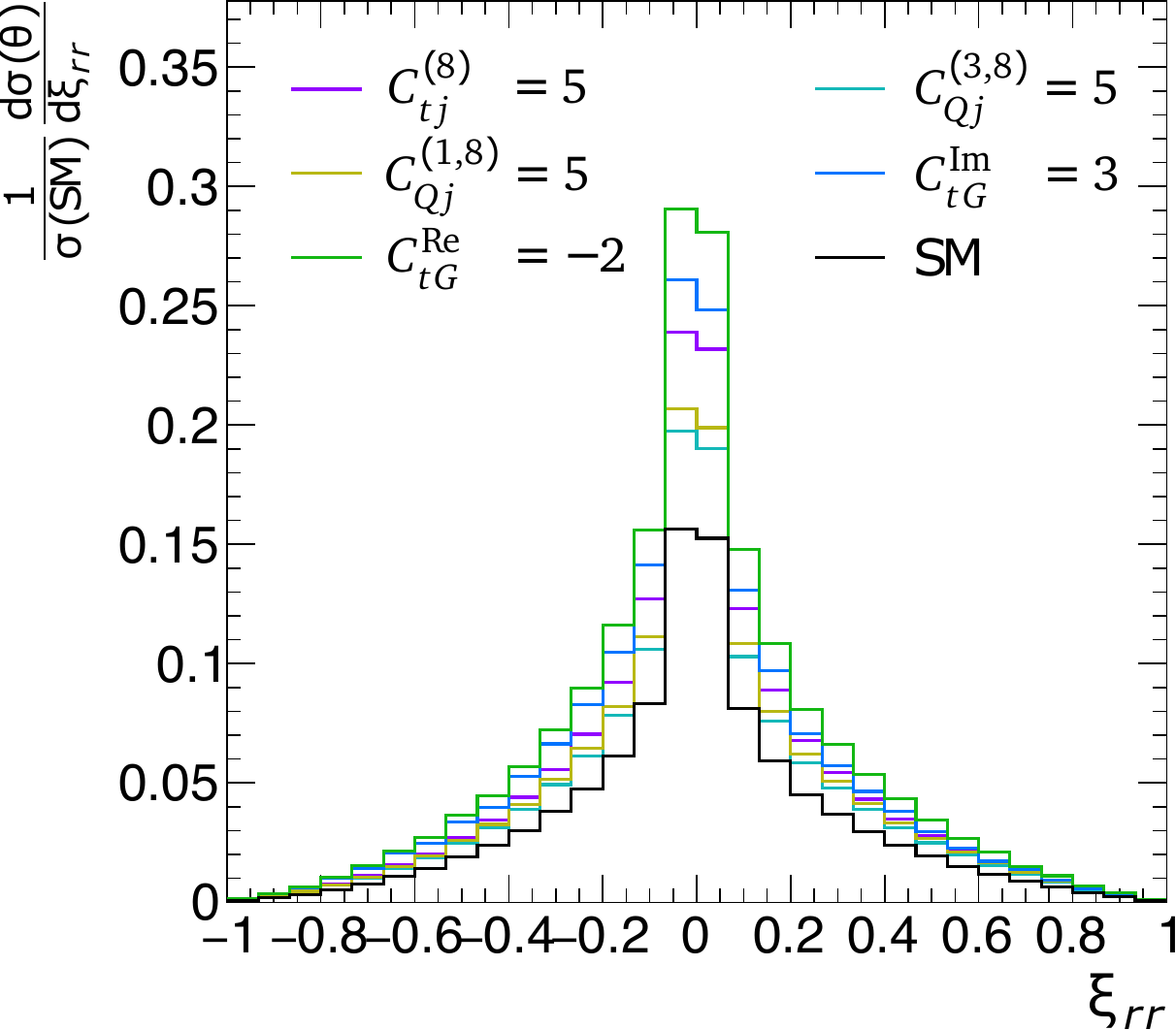}\hfill
\includegraphics[width=.32\textwidth]{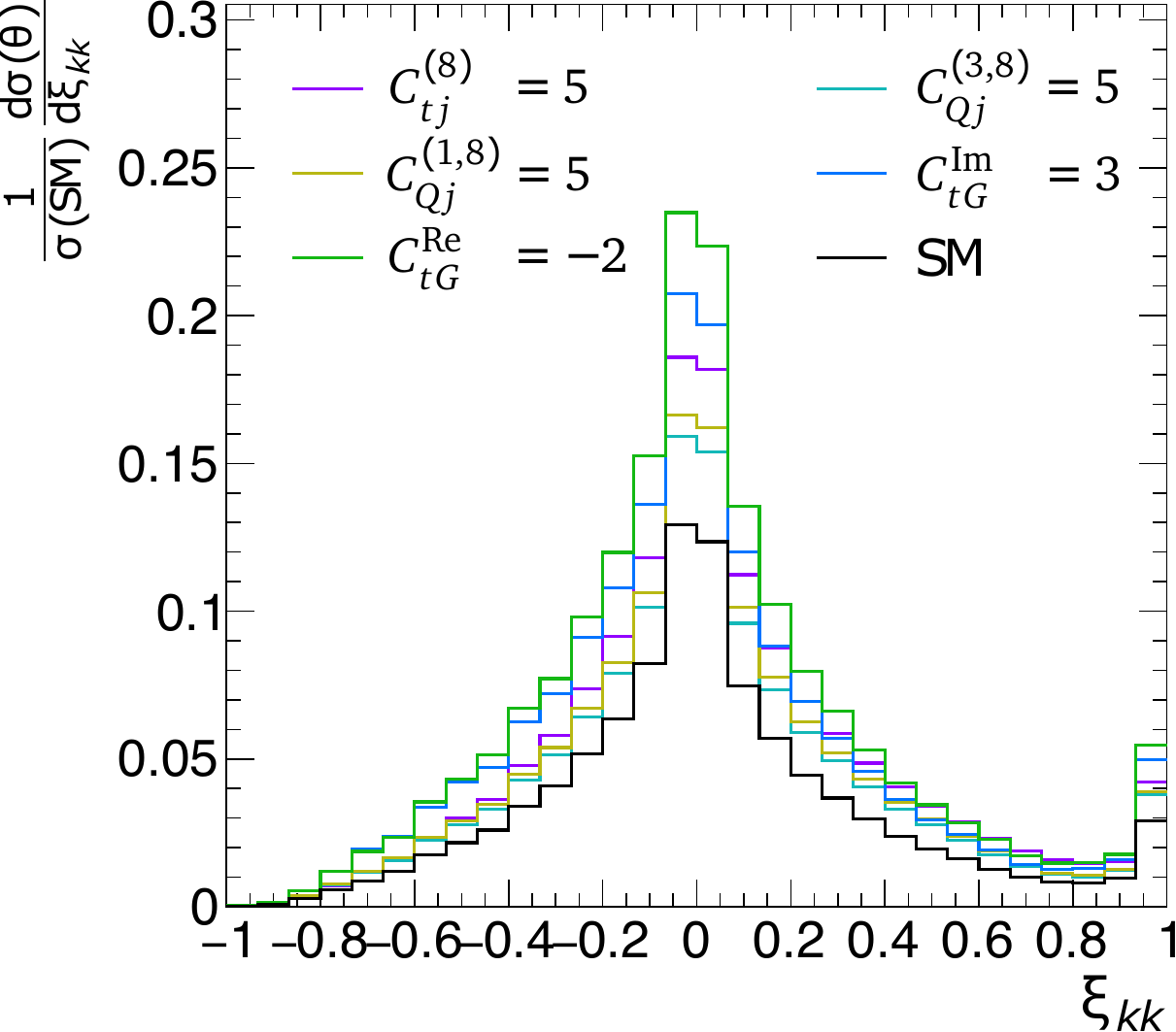}\\\vspace{.2cm}
\includegraphics[width=.32\textwidth]{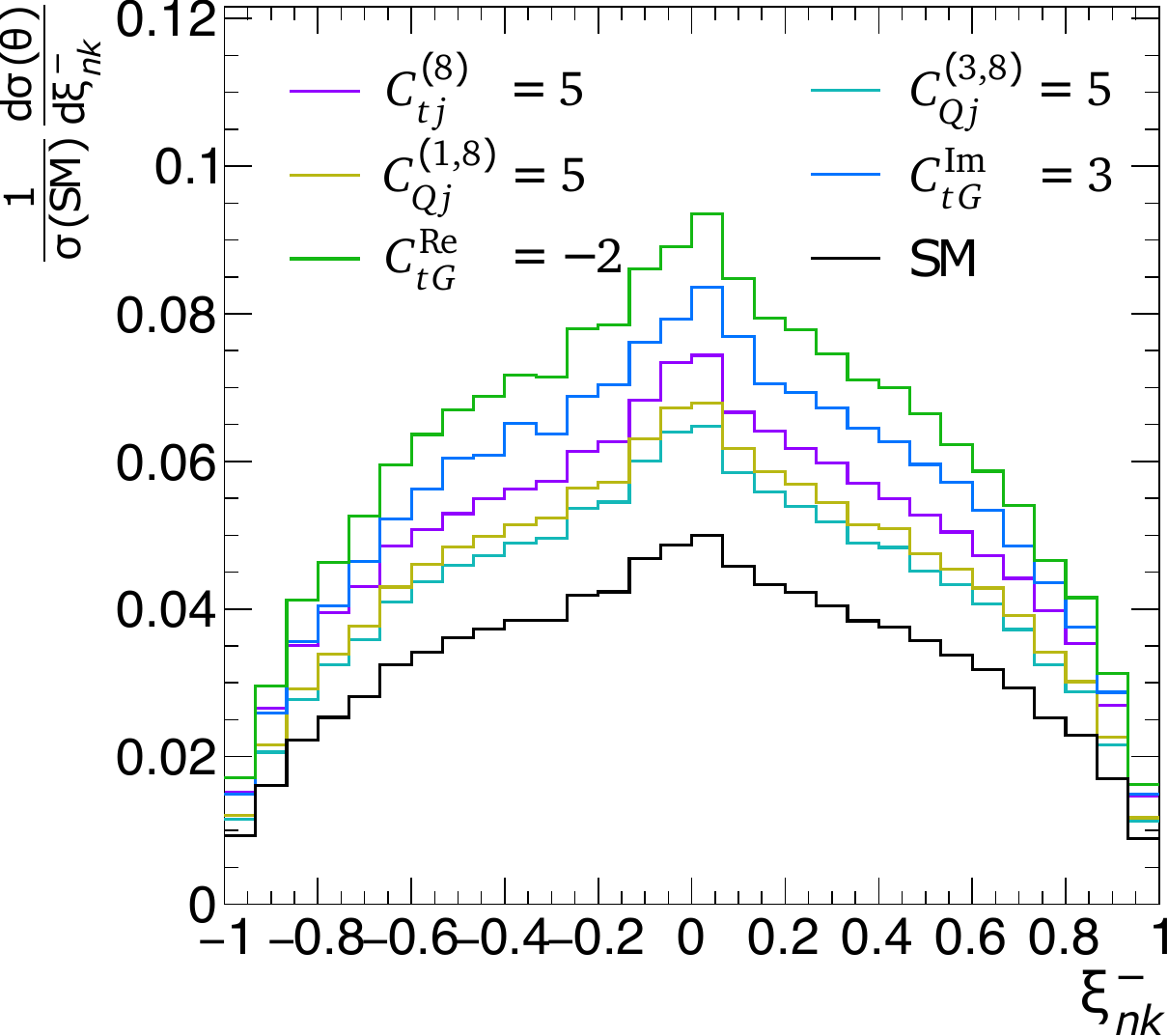}\hfill
\includegraphics[width=.32\textwidth]{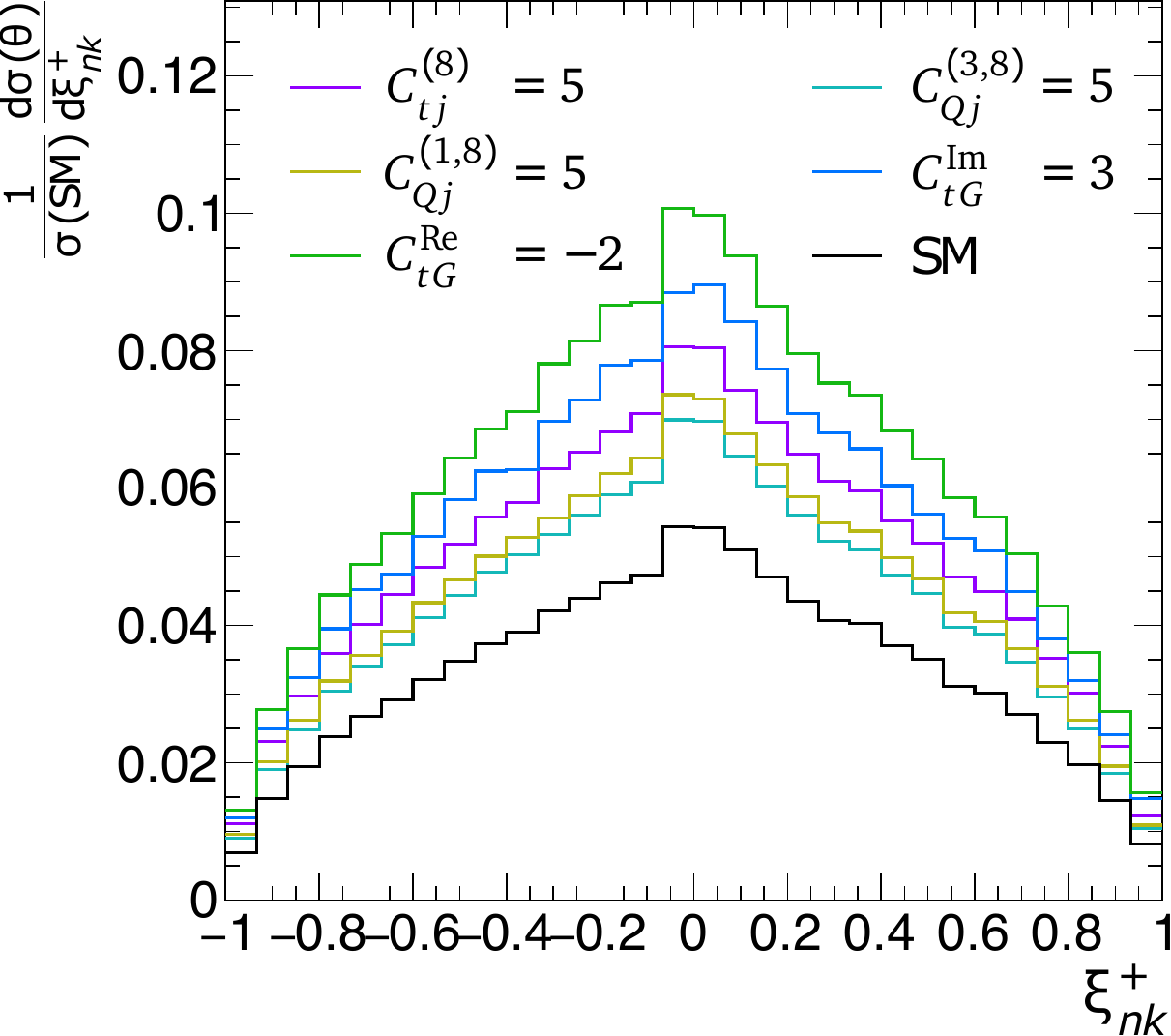}\hfill
\includegraphics[width=.32\textwidth]{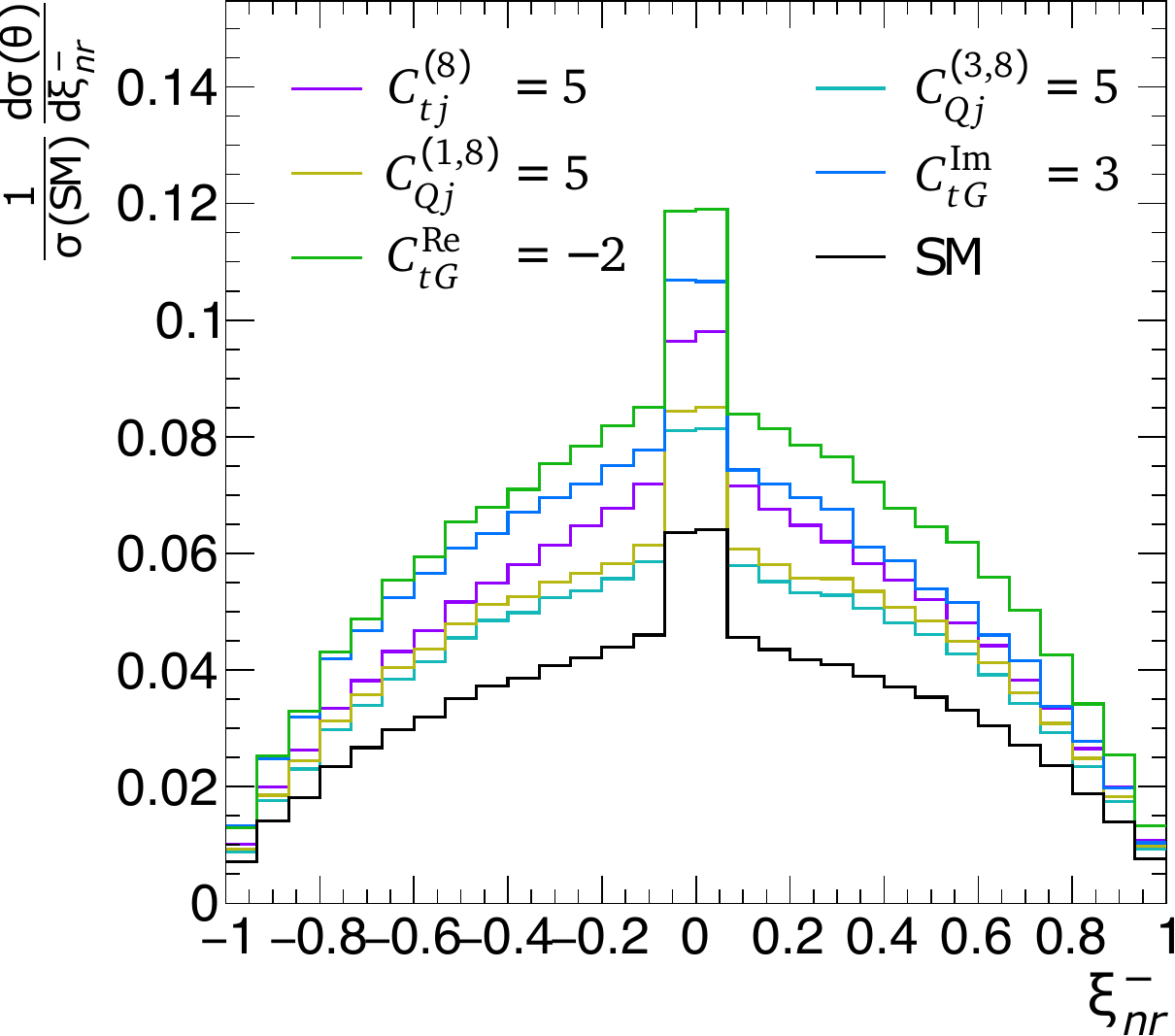}\\\vspace{.2cm}
\includegraphics[width=.32\textwidth]{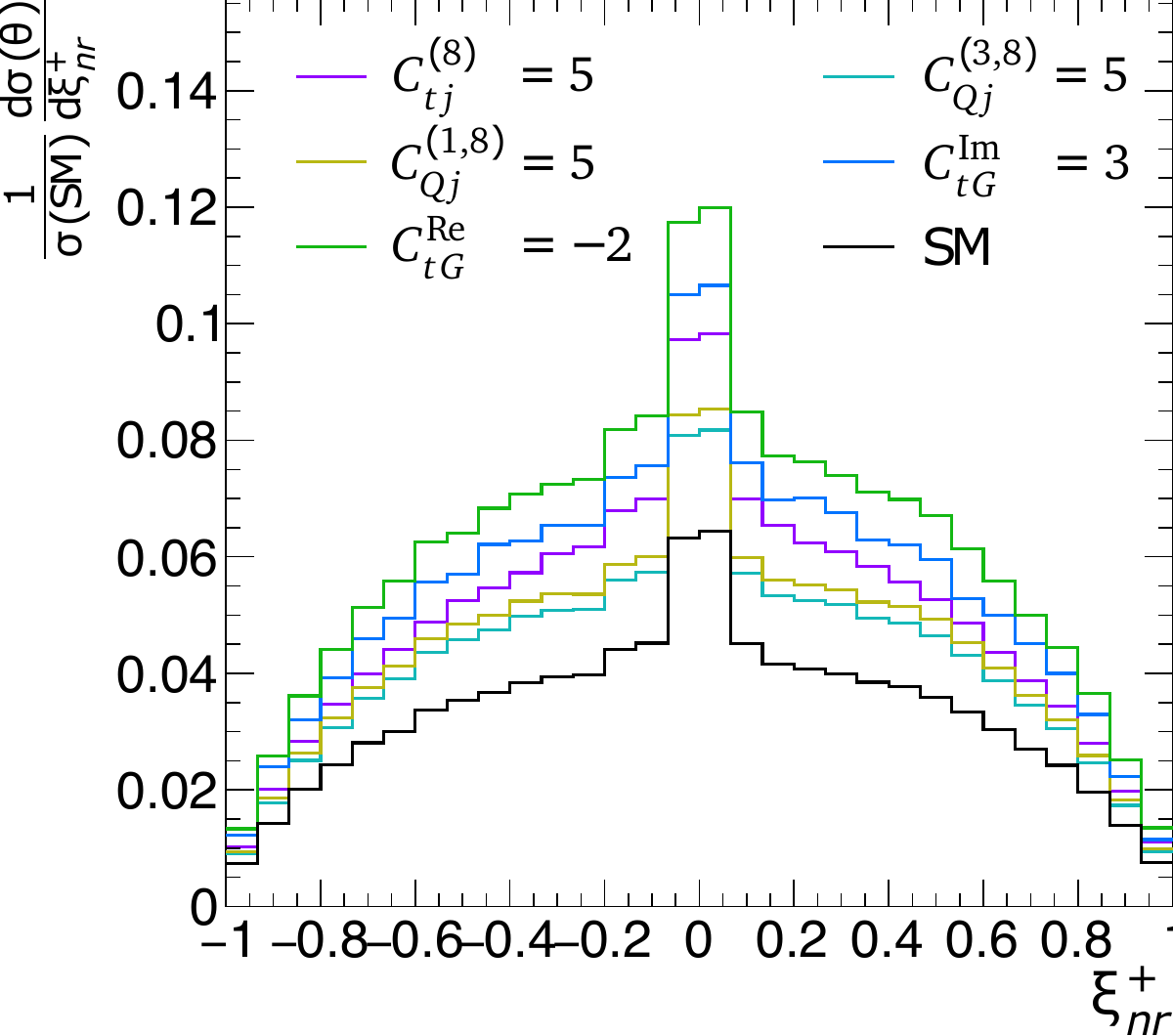}\hfill
\includegraphics[width=.32\textwidth]{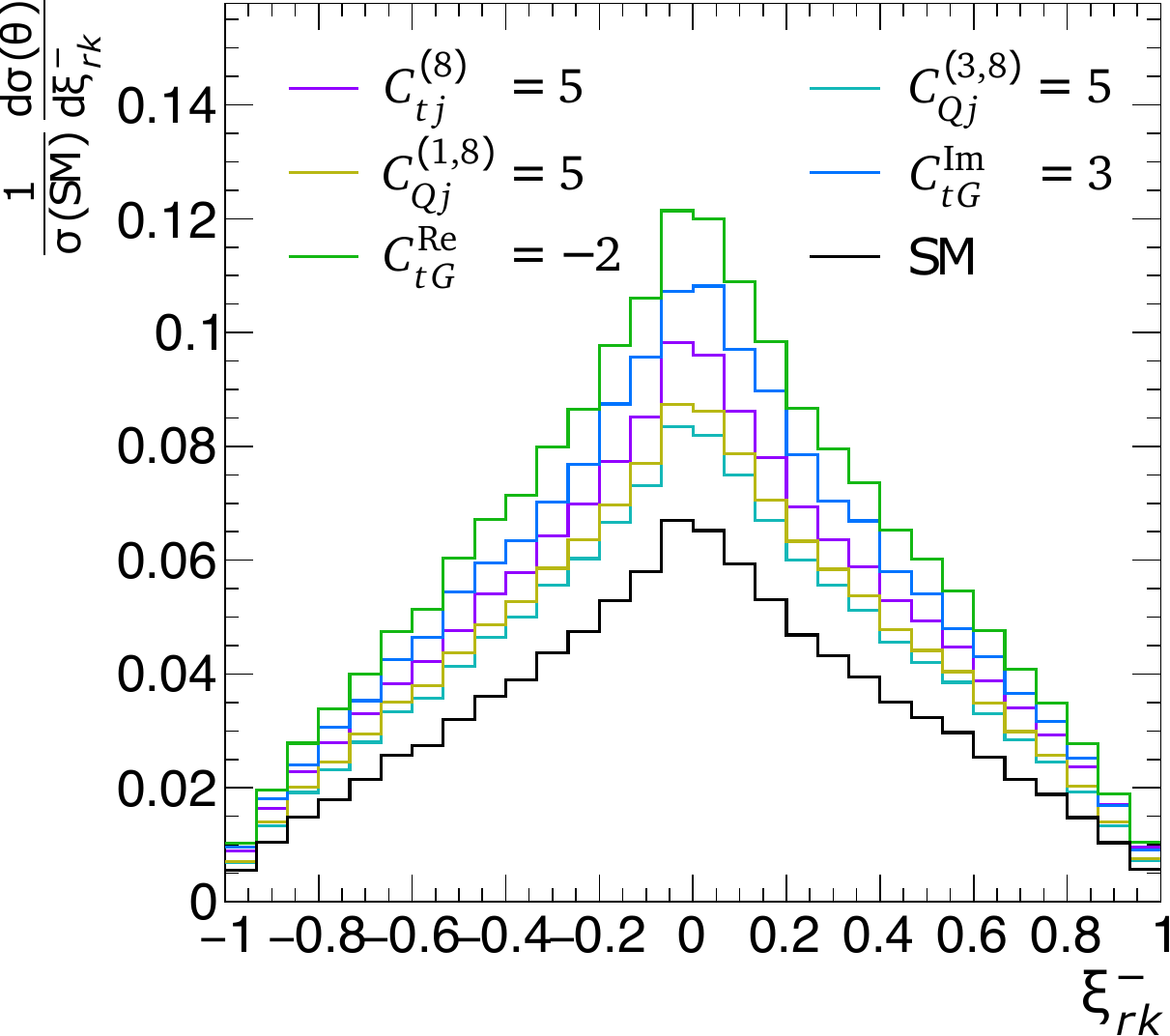}\hfill
\includegraphics[width=.32\textwidth]{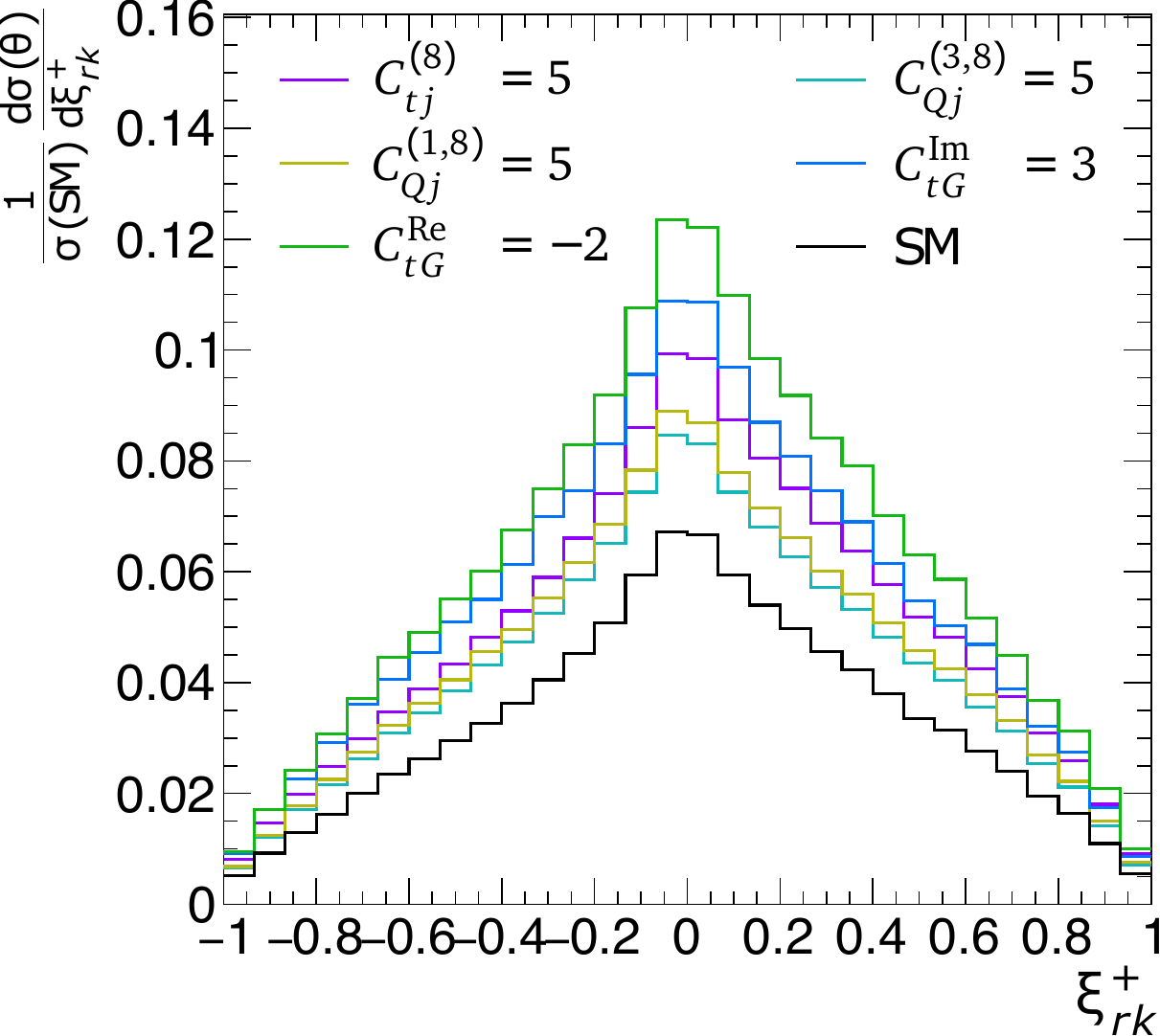}\\\vspace{.2cm}
\includegraphics[width=.32\textwidth]{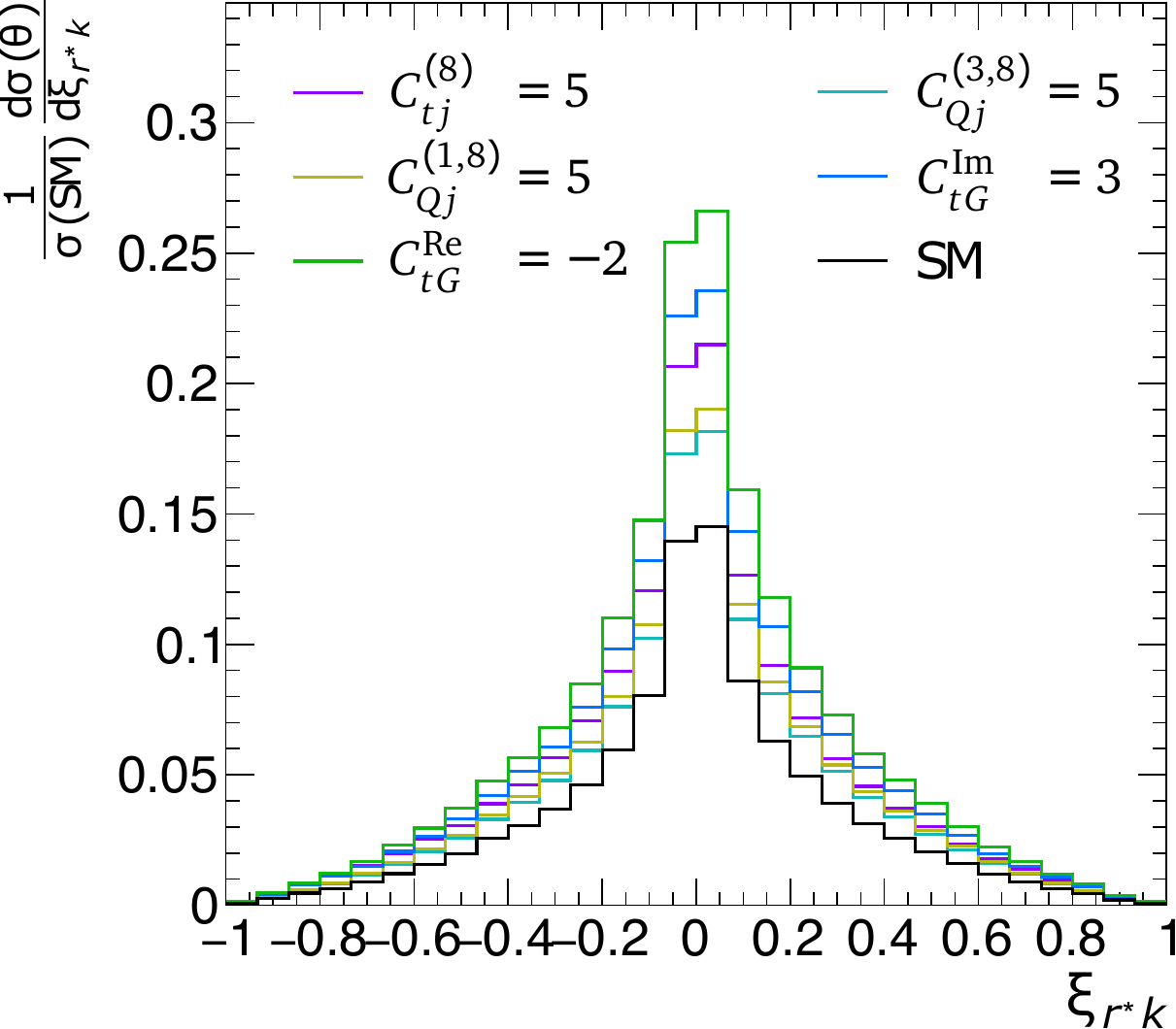}\hfill
\includegraphics[width=.32\textwidth]{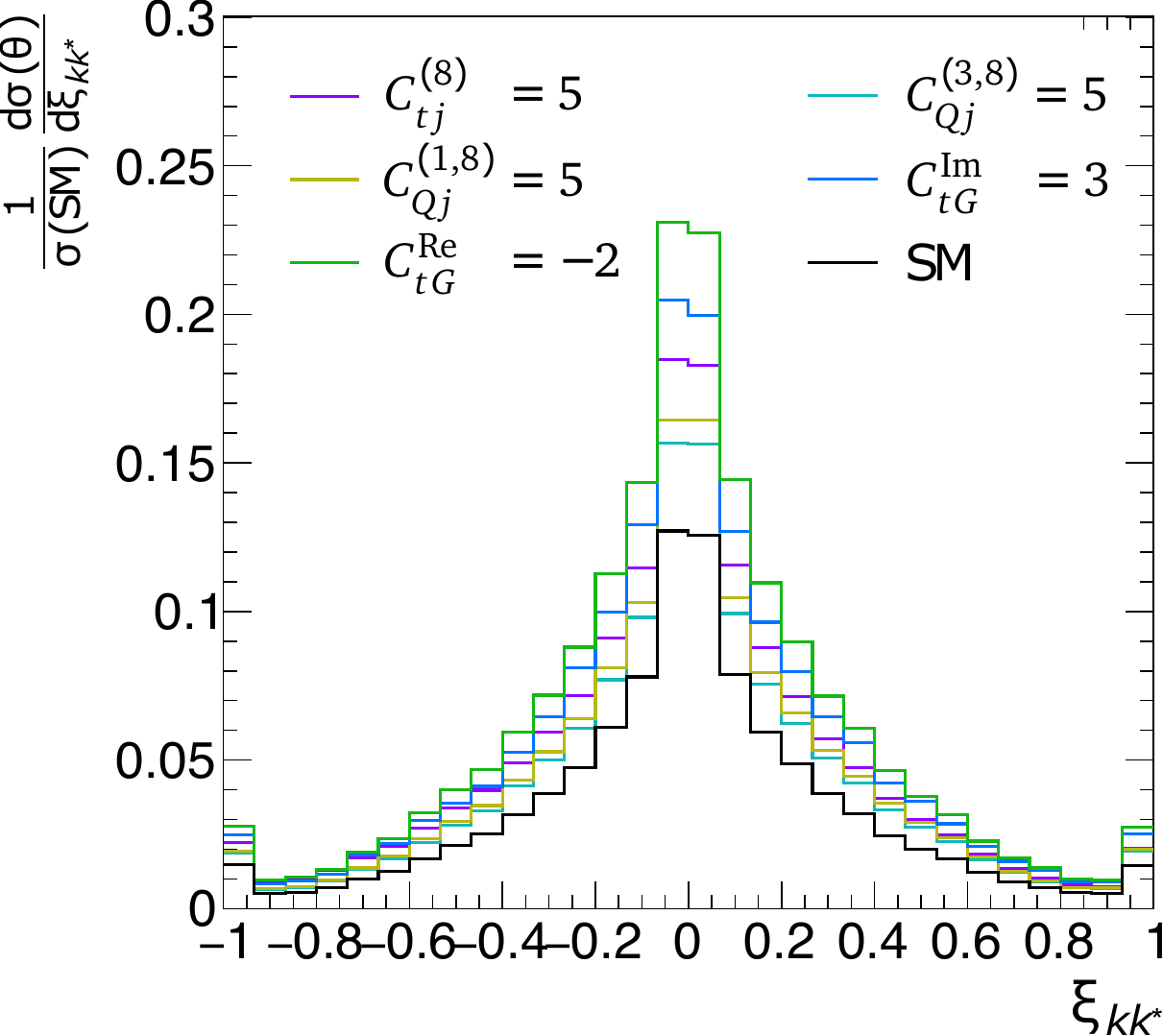}\hfill
\includegraphics[width=.32\textwidth]{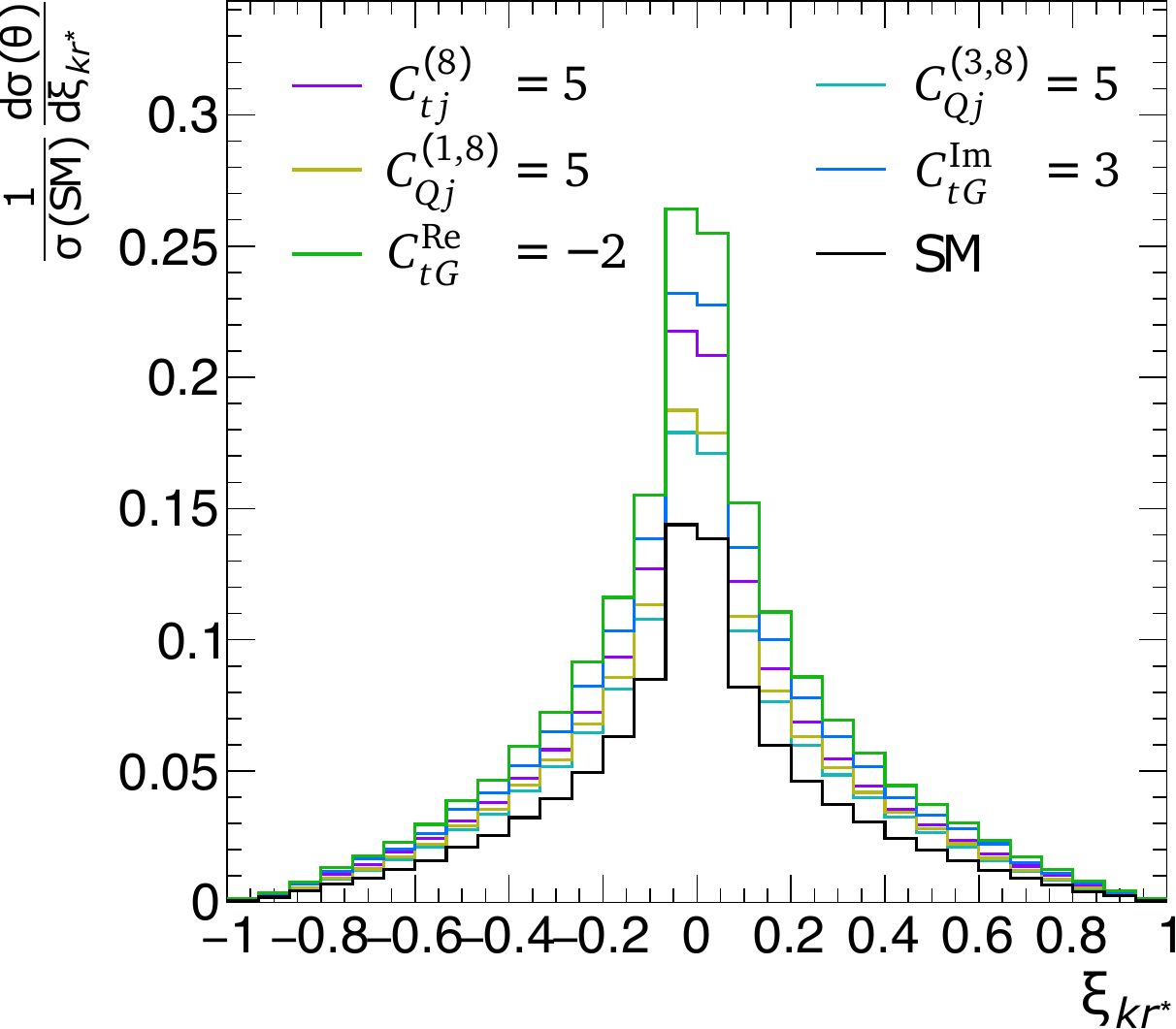}
  \caption{Same as Fig.~\ref{fig:ttbar-kinematics} for the distribution of the products of leptonic observables described in the text.}
  \label{fig:ttbar-xi}
\end{figure}

We briefly describe the angular observables introduced in Ref.~\cite{Bernreuther:2015yna}. A measurement of these quantities is performed in Ref.~\cite{CMS:2019nrx}. 
After reconstructing the top-quark momenta, the event is boosted into the \ttbar rest frame, and the following axes are defined. The axis $\hat k$ points toward the positively charged top quark. The axis $\hat r$ is orthogonal to $\hat k$ and must lie within the beam plane, spanned by the $\hat k$ and the momentum of the incoming parton in the \ttbar rest frame. The axis $\hat n$ is orthogonal to the beam plane, and $\{\hat r, \hat k, \hat  n\}$ must form a right-handed orthonormal basis. 
The lepton directions of flight, denoted by $\hat \ell^+$ and $\hat \ell^-$, are measured in the corresponding top quark center-of-mass frame, which is reached from the \ttbar frame by a rotation-free Lorentz transformation. Then, the quantities $\xi_{ab}=\cos\theta_a^+\cos\theta_b^-$ are defined where $\cos\theta_a^+=\hat\ell^+\cdot\hat a$ and  $\cos\theta_b^-=\hat\ell^-\cdot\hat a$ and the axis $a$ and $b$ can each be one of $\{\hat r, \hat k, \hat  n\}$. For $a\neq b$, sums and differences of these are considered, e.g., $\xi_{nr}^\pm=\xi_{nr}\pm\xi_{rn}$ and analogously for the other combinations. Two more axes $\hat r^\ast$ and $\hat k^\ast$ are defined by flipping the direction of $\hat r$ and $\hat k$ depending on the sign of the top quarks' rapidity difference in the laboratory frame while keeping the system orthonormal. The resulting 12 independent quantities characterize the spin-density matrix of the $\ttbar(2\ell)$ system. More details, including the behavior of these quantities under the discrete SM symmetries, are provided in Ref.~\cite{Bernreuther:2015yna}.
We show the distribution of the 12 quantities in Fig.~\ref{fig:ttbar-xi}.

\end{appendix}

\clearpage
\bibliographystyle{lucas_unsrt}

\bibliography{biblio.bib}

\providecommand{\href}[2]{#2}\begingroup\raggedright\begin{thebibliography}{10}%
\makeatletter
\providecommand{\hrefCMSnoop }[0]{\@secondoftwo}%
\makeatother
\providecommand{\doi}{\texttt{doi:}\begingroup \urlstyle{tt}\Url}

\bibitem{Buchmuller:1985jz}
\hrefCMSnoop {}{W.~Buchmuller and D.~Wyler, ``{{Effective Lagrangian Analysis
  of New Interactions and Flavor Conservation}}'',} \textit{ Nucl. Phys. B}
  \textbf{ 268} (1986) 621,
  \href{http://dx.doi.org/10.1016/0550-3213(86)90262-2}{\doi{10.1016/0550-3213(86)90262-2}}.

\bibitem{Leung:1984ni}
\hrefCMSnoop {}{C.~N. Leung, S.~T. Love, and S.~Rao, ``{Low-Energy
  Manifestations of a New Interaction Scale: Operator Analysis}'',} \textit{ Z.
  Phys. C} \textbf{ 31} (1986) 433,
  \href{http://dx.doi.org/10.1007/BF01588041}{\doi{10.1007/BF01588041}}.

\bibitem{Degrande:2012wf}
C.~Degrande\hrefCMSnoop {}{ {et~al.}, ``{Effective Field Theory: A Modern
  Approach to Anomalous Couplings}'',} \textit{ Annals Phys.} \textbf{ 335}
  (2013) 21,
  \href{http://dx.doi.org/10.1016/j.aop.2013.04.016}{\doi{10.1016/j.aop.2013.04.016}},
  \href{http://www.arXiv.org/abs/1205.4231}{\texttt{arXiv:1205.4231}}.

\bibitem{Brivio:2017vri}
\hrefCMSnoop {}{I.~Brivio and M.~Trott, ``{The Standard Model as an Effective
  Field Theory}'',} \textit{ Phys. Rept.} \textbf{ 793} (2019) 1,
  \href{http://dx.doi.org/10.1016/j.physrep.2018.11.002}{\doi{10.1016/j.physrep.2018.11.002}},
  \href{http://www.arXiv.org/abs/1706.08945}{\texttt{arXiv:1706.08945}}.

\bibitem{Isidori:2023pyp}
\hrefCMSnoop {}{G.~Isidori, F.~Wilsch, and D.~Wyler, ``{The standard model
  effective field theory at work}'',} \textit{ Rev. Mod. Phys.} \textbf{ 96}
  (2024) 015006,
  \href{http://dx.doi.org/10.1103/RevModPhys.96.015006}{\doi{10.1103/RevModPhys.96.015006}},
  \href{http://www.arXiv.org/abs/2303.16922}{\texttt{arXiv:2303.16922}}.

\bibitem{Grzadkowski:2010es}
\hrefCMSnoop {}{B.~Grzadkowski, M.~Iskrzynski, M.~Misiak, and J.~Rosiek,
  ``{{Dimension-Six Terms in the Standard Model Lagrangian}}'',} \textit{ JHEP}
  \textbf{ 10} (2010) 085,
  \href{http://dx.doi.org/10.1007/JHEP10(2010)085}{\doi{10.1007/JHEP10(2010)085}},
  \href{http://www.arXiv.org/abs/1008.4884}{\texttt{arXiv:1008.4884}}.

\bibitem{Belvedere:2024nzh}
\hrefCMSnoop {}{A.~Belvedere {et~al.}, ``{LHC EFT WG Note: SMEFT predictions,
  event reweighting, and simulation}'',}
  \href{http://www.arXiv.org/abs/2406.14620}{\texttt{arXiv:2406.14620}}.

\bibitem{GomezAmbrosio:2022mpm}
R.~Gomez~Ambrosio\hrefCMSnoop {}{ {et~al.}, ``Unbinned multivariate observables
  for global {SMEFT} analyses from machine learning'',} \textit{ JHEP} \textbf{
  03} (2023) 033,
  \href{http://dx.doi.org/10.1007/JHEP03(2023)033}{\doi{10.1007/JHEP03(2023)033}},
  \href{http://www.arXiv.org/abs/2211.02058}{\texttt{arXiv:2211.02058}}.

\bibitem{Chatterjee:2022oco}
\hrefCMSnoop {}{S.~Chatterjee, S.~Rohshap, R.~Sch\"ofbeck, and D.~Schwarz,
  ``Learning the {EFT} likelihood with tree boosting'',}
  \href{http://www.arXiv.org/abs/2205.12976}{\texttt{arXiv:2205.12976}}.

\bibitem{Chatterjee:2021nms}
S.~Chatterjee\hrefCMSnoop {}{ {et~al.}, ``Tree boosting for learning {EFT}
  parameters'',} \textit{ Comput. Phys. Commun.} \textbf{ 277} (2022) 108385,
  \href{http://dx.doi.org/10.1016/j.cpc.2022.108385}{\doi{10.1016/j.cpc.2022.108385}},
  \href{http://www.arXiv.org/abs/2107.10859}{\texttt{arXiv:2107.10859}}.

\bibitem{Chen:2020mev}
\hrefCMSnoop {}{S.~Chen, A.~Glioti, G.~Panico, and A.~Wulzer, ``Parametrized
  classifiers for optimal {EFT} sensitivity'',} \textit{ JHEP} \textbf{ 05}
  (2021) 247,
  \href{http://dx.doi.org/10.1007/JHEP05(2021)247}{\doi{10.1007/JHEP05(2021)247}},
  \href{http://www.arXiv.org/abs/2007.10356}{\texttt{arXiv:2007.10356}}.

\bibitem{Chen:2023ind}
\hrefCMSnoop {}{S.~Chen, A.~Glioti, G.~Panico, and A.~Wulzer, ``{Boosting
  likelihood learning with event reweighting}'',} \textit{ JHEP} \textbf{ 03}
  (2024) 117,
  \href{http://dx.doi.org/10.1007/JHEP03(2024)117}{\doi{10.1007/JHEP03(2024)117}},
  \href{http://www.arXiv.org/abs/2308.05704}{\texttt{arXiv:2308.05704}}.

\bibitem{Cranmer:2015bka}
\hrefCMSnoop {}{K.~Cranmer, J.~Pavez, and G.~Louppe, ``Approximating likelihood
  ratios with calibrated discriminative classifiers'',}
  \href{http://www.arXiv.org/abs/1506.02169}{\texttt{arXiv:1506.02169}}.

\bibitem{Brehmer:2018kdj}
\hrefCMSnoop {}{J.~Brehmer, K.~Cranmer, G.~Louppe, and J.~Pavez, ``Constraining
  effective field theories with machine learning'',} \textit{ Phys. Rev. Lett.}
  \textbf{ 121} (2018) 111801,
  \href{http://dx.doi.org/10.1103/PhysRevLett.121.111801}{\doi{10.1103/PhysRevLett.121.111801}},
  \href{http://www.arXiv.org/abs/1805.00013}{\texttt{arXiv:1805.00013}}.

\bibitem{Brehmer:2018eca}
\hrefCMSnoop {}{J.~Brehmer, K.~Cranmer, G.~Louppe, and J.~Pavez, ``A guide to
  constraining effective field theories with machine learning'',} \textit{
  Phys. Rev. D} \textbf{ 98} (2018) 052004,
  \href{http://dx.doi.org/10.1103/PhysRevD.98.052004}{\doi{10.1103/PhysRevD.98.052004}},
  \href{http://www.arXiv.org/abs/1805.00020}{\texttt{arXiv:1805.00020}}.

\bibitem{Brehmer:2018hga}
\hrefCMSnoop {}{J.~Brehmer, G.~Louppe, J.~Pavez, and K.~Cranmer, ``Mining gold
  from implicit models to improve likelihood-free inference'',} \textit{ Proc.
  Nat. Acad. Sci.} \textbf{ 117} (2020) 5242,
  \href{http://dx.doi.org/10.1073/pnas.1915980117}{\doi{10.1073/pnas.1915980117}},
  \href{http://www.arXiv.org/abs/1805.12244}{\texttt{arXiv:1805.12244}}.

\bibitem{Brehmer:2019xox}
\hrefCMSnoop {}{J.~Brehmer, F.~Kling, I.~Espejo, and K.~Cranmer, ``{MadMiner}:
  Machine learning-based inference for particle physics'',} \textit{ Comput.
  Softw. Big Sci.} \textbf{ 4} (2020) 3,
  \href{http://dx.doi.org/10.1007/s41781-020-0035-2}{\doi{10.1007/s41781-020-0035-2}},
  \href{http://www.arXiv.org/abs/1907.10621}{\texttt{arXiv:1907.10621}}.

\bibitem{Brehmer:2019gmn}
J.~Brehmer\hrefCMSnoop {}{ {et~al.}, ``Benchmarking simplified template cross
  sections in {WH} production'',} \textit{ JHEP} \textbf{ 11} (2019) 034,
  \href{http://dx.doi.org/10.1007/JHEP11(2019)034}{\doi{10.1007/JHEP11(2019)034}},
  \href{http://www.arXiv.org/abs/1908.06980}{\texttt{arXiv:1908.06980}}.

\bibitem{Butter:2021rvz}
\hrefCMSnoop {}{A.~Butter, T.~Plehn, N.~Soybelman, and J.~Brehmer, ``Back to
  the formula -- {LHC} edition'',} (2021).
  \href{http://www.arXiv.org/abs/2109.10414}{\texttt{arXiv:2109.10414}}.
  Submitted to \textit{SciPost Phys.}

\bibitem{Plehn:2022ftl}
T.~Plehn\hrefCMSnoop {}{ {et~al.}, ``{Modern Machine Learning for LHC
  Physicists}'',}
  \href{http://www.arXiv.org/abs/2211.01421}{\texttt{arXiv:2211.01421}}.

\bibitem{Cranmer:2014lly}
\hrefCMSnoop {}{K.~Cranmer, ``{Practical Statistics for the LHC}'',} in
  \textit{ {2011 European School of High-Energy Physics}}, p.~267.
\newblock 2014.
\newblock
  \href{http://www.arXiv.org/abs/1503.07622}{\texttt{arXiv:1503.07622}}.
\newblock
  \href{http://dx.doi.org/10.5170/CERN-2014-003.267}{\doi{10.5170/CERN-2014-003.267}}.

\bibitem{dAgnolo:2021aun}
R.~T. d'Agnolo\hrefCMSnoop {}{ {et~al.}, ``{Learning new physics from an
  imperfect machine}'',} \textit{ Eur. Phys. J. C} \textbf{ 82} (2022) 275,
  \href{http://dx.doi.org/10.1140/epjc/s10052-022-10226-y}{\doi{10.1140/epjc/s10052-022-10226-y}},
  \href{http://www.arXiv.org/abs/2111.13633}{\texttt{arXiv:2111.13633}}.

\bibitem{DeCastro:2018psv}
\hrefCMSnoop {}{P.~De~Castro and T.~Dorigo, ``{INFERNO: Inference-Aware Neural
  Optimisation}'',} \textit{ Comput. Phys. Commun.} \textbf{ 244} (2019) 170,
  \href{http://dx.doi.org/10.1016/j.cpc.2019.06.007}{\doi{10.1016/j.cpc.2019.06.007}},
  \href{http://www.arXiv.org/abs/1806.04743}{\texttt{arXiv:1806.04743}}.

\bibitem{Layer:2023lwi}
\hrefCMSnoop {}{L.~Layer, T.~Dorigo, and G.~Strong, ``{Application of Inferno
  to a Top Pair Cross Section Measurement with CMS Open Data}'',}
  \href{http://www.arXiv.org/abs/2301.10358}{\texttt{arXiv:2301.10358}}.

\bibitem{Neyman:1933wgr}
\hrefCMSnoop {}{J.~Neyman and E.~S. Pearson, ``{On the Problem of the Most
  Efficient Tests of Statistical Hypotheses}'',} \textit{ Phil. Trans. Roy.
  Soc. Lond. A} \textbf{ 231} (1933) 289,
  \href{http://dx.doi.org/10.1098/rsta.1933.0009}{\doi{10.1098/rsta.1933.0009}}.

\bibitem{Wilks:1938dza}
\hrefCMSnoop {}{S.~S. Wilks, ``{The Large-Sample Distribution of the Likelihood
  Ratio for Testing Composite Hypotheses}'',} \textit{ Annals Math. Statist.}
  \textbf{ 9} (1938) 60,
  \href{http://dx.doi.org/10.1214/aoms/1177732360}{\doi{10.1214/aoms/1177732360}}.

\bibitem{Bernlochner:2022oiw}
\hrefCMSnoop {}{F.~U. Bernlochner, D.~C. Fry, S.~B. Menary, and E.~Persson,
  ``{Cover your bases: asymptotic distributions of the profile likelihood ratio
  when constraining effective field theories in high-energy physics}'',}
  \textit{ SciPost Phys. Core} \textbf{ 6} (2023) 013,
  \href{http://dx.doi.org/10.21468/SciPostPhysCore.6.1.013}{\doi{10.21468/SciPostPhysCore.6.1.013}},
  \href{http://www.arXiv.org/abs/2207.01350}{\texttt{arXiv:2207.01350}}.

\bibitem{CMS:2021xjt}
\hrefCMSnoop {}{{CMS} Collaboration, ``{Precision luminosity measurement in
  proton-proton collisions at $\sqrt{s} =$ 13 TeV in 2015 and 2016 at CMS}'',}
  \textit{ Eur. Phys. J. C} \textbf{ 81} (2021) 800,
  \href{http://dx.doi.org/10.1140/epjc/s10052-021-09538-2}{\doi{10.1140/epjc/s10052-021-09538-2}},
  \href{http://www.arXiv.org/abs/2104.01927}{\texttt{arXiv:2104.01927}}.

\bibitem{ATLAS:2022hro}
\hrefCMSnoop {}{{ATLAS} Collaboration, ``{Luminosity determination in $pp$
  collisions at $\sqrt{s}=13$ TeV using the ATLAS detector at the LHC}'',}
  \textit{ Eur. Phys. J. C} \textbf{ 83} (2023) 982,
  \href{http://dx.doi.org/10.1140/epjc/s10052-023-11747-w}{\doi{10.1140/epjc/s10052-023-11747-w}},
  \href{http://www.arXiv.org/abs/2212.09379}{\texttt{arXiv:2212.09379}}.

\bibitem{Campbell:2022qmc}
\hrefCMSnoop {}{J.~M. Campbell {et~al.}, ``{Event Generators for High-Energy
  Physics Experiments}'',} \textit{ SciPost Phys.} \textbf{ 16} (2024) 130,
  \href{http://dx.doi.org/10.21468/SciPostPhys.16.5.130}{\doi{10.21468/SciPostPhys.16.5.130}},
  \href{http://www.arXiv.org/abs/2203.11110}{\texttt{arXiv:2203.11110}}.

\bibitem{Alwall:2014hca}
J.~Alwall\hrefCMSnoop {}{ {et~al.}, ``The automated computation of tree-level
  and next-to-leading order differential cross sections, and their matching to
  parton shower simulations'',} \textit{ JHEP} \textbf{ 07} (2014) 079,
  \href{http://dx.doi.org/10.1007/JHEP07(2014)079}{\doi{10.1007/JHEP07(2014)079}},
  \href{http://www.arXiv.org/abs/1405.0301}{\texttt{arXiv:1405.0301}}.

\bibitem{Frederix:2012ps}
\hrefCMSnoop {}{R.~Frederix and S.~Frixione, ``{Merging meets matching in
  MC@NLO}'',} \textit{ JHEP} \textbf{ 12} (2012) 061,
  \href{http://dx.doi.org/10.1007/JHEP12(2012)061}{\doi{10.1007/JHEP12(2012)061}},
  \href{http://www.arXiv.org/abs/1209.6215}{\texttt{arXiv:1209.6215}}.

\bibitem{Sherpa:2019gpd}
\hrefCMSnoop {}{{Sherpa} Collaboration, ``{Event Generation with Sherpa
  2.2}'',} \textit{ SciPost Phys.} \textbf{ 7} (2019) 034,
  \href{http://dx.doi.org/10.21468/SciPostPhys.7.3.034}{\doi{10.21468/SciPostPhys.7.3.034}},
  \href{http://www.arXiv.org/abs/1905.09127}{\texttt{arXiv:1905.09127}}.

\bibitem{Nason:2004rx}
\hrefCMSnoop {}{P.~Nason, ``{A New method for combining NLO QCD with shower
  Monte Carlo algorithms}'',} \textit{ JHEP} \textbf{ 11} (2004) 040,
  \href{http://dx.doi.org/10.1088/1126-6708/2004/11/040}{\doi{10.1088/1126-6708/2004/11/040}},
  \href{http://www.arXiv.org/abs/hep-ph/0409146}{\texttt{arXiv:hep-ph/0409146}}.

\bibitem{Frixione:2007vw}
\hrefCMSnoop {}{S.~Frixione, P.~Nason, and C.~Oleari, ``{Matching NLO QCD
  computations with Parton Shower simulations: the POWHEG method}'',} \textit{
  JHEP} \textbf{ 11} (2007) 070,
  \href{http://dx.doi.org/10.1088/1126-6708/2007/11/070}{\doi{10.1088/1126-6708/2007/11/070}},
  \href{http://www.arXiv.org/abs/0709.2092}{\texttt{arXiv:0709.2092}}.

\bibitem{Frixione:2007nw}
\hrefCMSnoop {}{S.~Frixione, P.~Nason, and G.~Ridolfi, ``{A Positive-weight
  next-to-leading-order Monte Carlo for heavy flavour hadroproduction}'',}
  \textit{ JHEP} \textbf{ 09} (2007) 126,
  \href{http://dx.doi.org/10.1088/1126-6708/2007/09/126}{\doi{10.1088/1126-6708/2007/09/126}},
  \href{http://www.arXiv.org/abs/0707.3088}{\texttt{arXiv:0707.3088}}.

\bibitem{Alioli:2010xd}
\hrefCMSnoop {}{S.~Alioli, P.~Nason, C.~Oleari, and E.~Re, ``{A general
  framework for implementing NLO calculations in shower Monte Carlo programs:
  the POWHEG BOX}'',} \textit{ JHEP} \textbf{ 06} (2010) 043,
  \href{http://dx.doi.org/10.1007/JHEP06(2010)043}{\doi{10.1007/JHEP06(2010)043}},
  \href{http://www.arXiv.org/abs/1002.2581}{\texttt{arXiv:1002.2581}}.

\bibitem{Campbell:2014kua}
\hrefCMSnoop {}{J.~M. Campbell, R.~K. Ellis, P.~Nason, and E.~Re, ``{Top-Pair
  Production and Decay at NLO Matched with Parton Showers}'',} \textit{ JHEP}
  \textbf{ 04} (2015) 114,
  \href{http://dx.doi.org/10.1007/JHEP04(2015)114}{\doi{10.1007/JHEP04(2015)114}},
  \href{http://www.arXiv.org/abs/1412.1828}{\texttt{arXiv:1412.1828}}.

\bibitem{Sjostrand:2014zea}
T.~Sj\"ostrand\hrefCMSnoop {}{ {et~al.}, ``An introduction to {PYTHIA} 8.2'',}
  \textit{ Comput. Phys. Commun.} \textbf{ 191} (2015) 159,
  \href{http://dx.doi.org/10.1016/j.cpc.2015.01.024}{\doi{10.1016/j.cpc.2015.01.024}},
  \href{http://www.arXiv.org/abs/1410.3012}{\texttt{arXiv:1410.3012}}.

\bibitem{Bellm:2015jjp}
\hrefCMSnoop {}{J.~Bellm {et~al.}, ``{Herwig 7.0/Herwig++ 3.0 release note}'',}
  \textit{ Eur. Phys. J. C} \textbf{ 76} (2016) 196,
  \href{http://dx.doi.org/10.1140/epjc/s10052-016-4018-8}{\doi{10.1140/epjc/s10052-016-4018-8}},
  \href{http://www.arXiv.org/abs/1512.01178}{\texttt{arXiv:1512.01178}}.

\bibitem{GEANT4:2002zbu}
\hrefCMSnoop {}{{GEANT4} Collaboration, ``{GEANT4--a simulation toolkit}'',}
  \textit{ Nucl. Instrum. Meth. A} \textbf{ 506} (2003) 250,
  \href{http://dx.doi.org/10.1016/S0168-9002(03)01368-8}{\doi{10.1016/S0168-9002(03)01368-8}}.

\bibitem{ATLAS:2017ghe}
\hrefCMSnoop {}{{ATLAS} Collaboration, ``{Jet reconstruction and performance
  using particle flow with the ATLAS Detector}'',} \textit{ Eur. Phys. J. C}
  \textbf{ 77} (2017) 466,
  \href{http://dx.doi.org/10.1140/epjc/s10052-017-5031-2}{\doi{10.1140/epjc/s10052-017-5031-2}},
  \href{http://www.arXiv.org/abs/1703.10485}{\texttt{arXiv:1703.10485}}.

\bibitem{CMS:2017yfk}
\hrefCMSnoop {}{{CMS} Collaboration, ``{Particle-flow reconstruction and global
  event description with the CMS detector}'',} \textit{ JINST} \textbf{ 12}
  (2017) P10003,
  \href{http://dx.doi.org/10.1088/1748-0221/12/10/P10003}{\doi{10.1088/1748-0221/12/10/P10003}},
  \href{http://www.arXiv.org/abs/1706.04965}{\texttt{arXiv:1706.04965}}.

\bibitem{deFavereau:2013fsa}
\hrefCMSnoop {}{{DELPHES 3} Collaboration, ``{DELPHES} 3, a modular framework
  for fast simulation of a generic collider experiment'',} \textit{ JHEP}
  \textbf{ 02} (2014) 057,
  \href{http://dx.doi.org/10.1007/JHEP02(2014)057}{\doi{10.1007/JHEP02(2014)057}},
  \href{http://www.arXiv.org/abs/1307.6346}{\texttt{arXiv:1307.6346}}.

\bibitem{Komiske:2018cqr}
\hrefCMSnoop {}{P.~T. Komiske, E.~M. Metodiev, and J.~Thaler, ``Energy flow
  networks: Deep sets for particle jets'',} \textit{ JHEP} \textbf{ 01} (2019)
  121,
  \href{http://dx.doi.org/10.1007/JHEP01(2019)121}{\doi{10.1007/JHEP01(2019)121}},
  \href{http://www.arXiv.org/abs/1810.05165}{\texttt{arXiv:1810.05165}}.

\bibitem{Chatterjee:2024pbp}
\hrefCMSnoop {}{S.~Chatterjee, S.~S. Cruz, R.~Sch\"ofbeck, and D.~Schwarz,
  ``{Rotation-equivariant graph neural network for learning hadronic SMEFT
  effects}'',} \textit{ Phys. Rev. D} \textbf{ 109} (2024) 076012,
  \href{http://dx.doi.org/10.1103/PhysRevD.109.076012}{\doi{10.1103/PhysRevD.109.076012}},
  \href{http://www.arXiv.org/abs/2401.10323}{\texttt{arXiv:2401.10323}}.

\bibitem{Buckley:2014ana}
A.~Buckley\hrefCMSnoop {}{ {et~al.}, ``{LHAPDF6: parton density access in the
  LHC precision era}'',} \textit{ Eur. Phys. J. C} \textbf{ 75} (2015) 132,
  \href{http://dx.doi.org/10.1140/epjc/s10052-015-3318-8}{\doi{10.1140/epjc/s10052-015-3318-8}},
  \href{http://www.arXiv.org/abs/1412.7420}{\texttt{arXiv:1412.7420}}.

\bibitem{Mattelaer:2016gcx}
\hrefCMSnoop {}{O.~Mattelaer, ``{On the maximal use of Monte Carlo samples:
  re-weighting events at NLO accuracy}'',} \textit{ Eur. Phys. J. C} \textbf{
  76} (2016) 674,
  \href{http://dx.doi.org/10.1140/epjc/s10052-016-4533-7}{\doi{10.1140/epjc/s10052-016-4533-7}},
  \href{http://www.arXiv.org/abs/1607.00763}{\texttt{arXiv:1607.00763}}.

\bibitem{Serkin:2021bbn}
{ATLAS, CMS} Collaboration, \hrefCMSnoop {}{L.~Serkin, ``{Treatment of
  top-quark backgrounds in extreme phase spaces: the ''top $p_{T}$
  reweighting'' and novel data-driven estimations in ATLAS and CMS}'',} in
  \textit{ {13th International Workshop on Top Quark Physics}}.
\newblock 5, 2021.
\newblock
  \href{http://www.arXiv.org/abs/2105.03977}{\texttt{arXiv:2105.03977}}.

\bibitem{Frixione:2002ik}
\hrefCMSnoop {}{S.~Frixione and B.~R. Webber, ``{Matching NLO QCD computations
  and parton shower simulations}'',} \textit{ JHEP} \textbf{ 06} (2002) 029,
  \href{http://dx.doi.org/10.1088/1126-6708/2002/06/029}{\doi{10.1088/1126-6708/2002/06/029}},
  \href{http://www.arXiv.org/abs/hep-ph/0204244}{\texttt{arXiv:hep-ph/0204244}}.

\bibitem{Hoeche:2005vzu}
S.~Hoeche\hrefCMSnoop {}{ {et~al.}, ``{Matching parton showers and matrix
  elements}'',} in \textit{ {HERA and the LHC: A Workshop on the Implications
  of HERA for LHC Physics: CERN - DESY Workshop 2004/2005 (Midterm Meeting,
  CERN, 11-13 October 2004; Final Meeting, DESY, 17-21 January 2005)}}, p.~288.
\newblock 2005.
\newblock
  \href{http://www.arXiv.org/abs/hep-ph/0602031}{\texttt{arXiv:hep-ph/0602031}}.
\newblock
  \href{http://dx.doi.org/10.5170/CERN-2005-014.288}{\doi{10.5170/CERN-2005-014.288}}.

\bibitem{Finke:2023ltw}
T.~Finke\hrefCMSnoop {}{ {et~al.}, ``{Tree-based algorithms for weakly
  supervised anomaly detection}'',} \textit{ Phys. Rev. D} \textbf{ 109} (2024)
  034033,
  \href{http://dx.doi.org/10.1103/PhysRevD.109.034033}{\doi{10.1103/PhysRevD.109.034033}},
  \href{http://www.arXiv.org/abs/2309.13111}{\texttt{arXiv:2309.13111}}.

\bibitem{Speckmayer:2010zz}
\hrefCMSnoop {}{P.~Speckmayer, A.~Hocker, J.~Stelzer, and H.~Voss, ``{The
  toolkit for multivariate data analysis, TMVA 4}'',} \textit{ J. Phys. Conf.
  Ser.} \textbf{ 219} (2010) 032057,
  \href{http://dx.doi.org/10.1088/1742-6596/219/3/032057}{\doi{10.1088/1742-6596/219/3/032057}}.

\bibitem{breiman1984classification}
L.~Breiman, J.~H. Friedman, R.~A. Olshen, and C.~J. Stone, ``Classification and
  regression trees''.
\newblock Wadsworth Publishing Company, 1984.

\bibitem{TAO-1}
\hrefCMSnoop {}{A.~Zharmagambetov, S.~S. Hada, M.~Gabidolla, and M.~A.
  Carreira-Perpi\~{n}\'{a}n, ``Non-greedy algorithms for decision tree
  optimization: An experimental comparison'',} in \textit{ 2021 International
  Joint Conference on Neural Networks (IJCNN)}, p.~1.
\newblock 2021.
\newblock
  \href{http://dx.doi.org/10.1109/IJCNN52387.2021.9533597}{\doi{10.1109/IJCNN52387.2021.9533597}}.

\bibitem{TAO-2}
\hrefCMSnoop {}{M.~Gabidolla and M.~A. Carreira-Perpi\~{n}\'{a}n, ``Pushing the
  envelope of gradient boosting forests via globally-optimized oblique
  trees'',} in \textit{ 2022 IEEE/CVF Conference on Computer Vision and Pattern
  Recognition (CVPR)}, p.~285.
\newblock 2022.
\newblock
  \href{http://dx.doi.org/10.1109/CVPR52688.2022.00038}{\doi{10.1109/CVPR52688.2022.00038}}.

\bibitem{TAO-3}
\hrefCMSnoop {}{A.~Zharmagambetov and M.~A. Carreira-Perpi\~{n}\'{a}n,
  ``Smaller, more accurate regression forests using tree alternating
  optimization''.}
  \url{https://proceedings.mlr.press/v119/zharmagambetov20a.html}, 2020.

\bibitem{TAO-4}
\hrefCMSnoop {}{A.~Zharmagambetov and M.~A. Carreira-Perpi\~{n}\'{a}n,
  ``Ensembles of bagged tao trees consistently improve over random forests,
  adaboost and gradient boosting'',} in \textit{ Proceedings of the 2020
  ACM-IMS on Foundations of Data Science Conference}, FODS '20, p.~35.
\newblock Association for Computing Machinery, New York, NY, USA, 2020.
\newblock
  \href{http://dx.doi.org/10.1145/3412815.3416882}{\doi{10.1145/3412815.3416882}}.

\bibitem{CMS:2024onh}
\hrefCMSnoop {}{{CMS} Collaboration, ``{The CMS statistical analysis and
  combination tool: COMBINE}'',}
  \href{http://www.arXiv.org/abs/2404.06614}{\texttt{arXiv:2404.06614}}.

\bibitem{Cranmer:2012sba}
\hrefCMSnoop {}{{ROOT} Collaboration, ``{HistFactory: A tool for creating
  statistical models for use with RooFit and RooStats}'',} technical report,
  New York U., 2012.
\newblock
  \href{http://dx.doi.org/10.17181/CERN-OPEN-2012-016}{\doi{10.17181/CERN-OPEN-2012-016}}.

\bibitem{Kassabov:2023hbm}
Z.~Kassabov\hrefCMSnoop {}{ {et~al.}, ``{The top quark legacy of the LHC Run II
  for PDF and SMEFT analyses}'',} \textit{ JHEP} \textbf{ 05} (2023) 205,
  \href{http://dx.doi.org/10.1007/JHEP05(2023)205}{\doi{10.1007/JHEP05(2023)205}},
  \href{http://www.arXiv.org/abs/2303.06159}{\texttt{arXiv:2303.06159}}.

\bibitem{CMS-Open-Data}
\hrefCMSnoop {}{{CMS Collaboration} Collaboration, ``{CMS Open Data Guide}'',}
  2024.
\newblock \url{https://cms-opendata-guide.web.cern.ch/}.

\bibitem{ATLAS-Open-Data}
\hrefCMSnoop {}{{ATLAS Collaboration} Collaboration, ``{ATLAS Open Data
  portal}'',} 2024.
\newblock \url{https://atlas.cern/Resources/Opendata}.

\bibitem{ATLAS:2023gsl}
\hrefCMSnoop {}{{ATLAS} Collaboration, ``{Inclusive and differential
  cross-sections for dilepton $ t\overline{t} $ production measured in $
  \sqrt{s} $ = 13 TeV pp collisions with the ATLAS detector}'',} \textit{ JHEP}
  \textbf{ 07} (2023) 141,
  \href{http://dx.doi.org/10.1007/JHEP07(2023)141}{\doi{10.1007/JHEP07(2023)141}},
  \href{http://www.arXiv.org/abs/2303.15340}{\texttt{arXiv:2303.15340}}.

\bibitem{CMS:2024ybg}
\hrefCMSnoop {}{{CMS} Collaboration, ``{Differential cross section measurements
  for the production of top quark pairs and of additional jets using dilepton
  events from pp collisions at $\sqrt{s}$ = 13 TeV}'',}
  \href{http://www.arXiv.org/abs/2402.08486}{\texttt{arXiv:2402.08486}}.

\bibitem{Ball_2017}
\hrefCMSnoop {}{{{NNPDF}} Collaboration, ``{Parton distributions from
  high-precision collider data}'',} \textit{ Eur. Phys. J. C} \textbf{ 77}
  (2017) 663,
  \href{http://dx.doi.org/10.1140/epjc/s10052-017-5199-5}{\doi{10.1140/epjc/s10052-017-5199-5}},
  \href{http://www.arXiv.org/abs/1706.00428}{\texttt{arXiv:1706.00428}}.

\bibitem{Brivio:2020onw}
\hrefCMSnoop {}{I.~Brivio, ``{SMEFTsim} 3.0 \textemdash{} a practical guide'',}
  \textit{ JHEP} \textbf{ 04} (2021) 073,
  \href{http://dx.doi.org/10.1007/JHEP04(2021)073}{\doi{10.1007/JHEP04(2021)073}},
  \href{http://www.arXiv.org/abs/2012.11343}{\texttt{arXiv:2012.11343}}.

\bibitem{Skands:2014pea}
\hrefCMSnoop {}{P.~Skands, S.~Carrazza, and J.~Rojo, ``{Tuning PYTHIA 8.1: the
  Monash 2013 Tune}'',} \textit{ Eur. Phys. J. C} \textbf{ 74} (2014) 3024,
  \href{http://dx.doi.org/10.1140/epjc/s10052-014-3024-y}{\doi{10.1140/epjc/s10052-014-3024-y}},
  \href{http://www.arXiv.org/abs/1404.5630}{\texttt{arXiv:1404.5630}}.

\bibitem{CMS:2015wcf}
\hrefCMSnoop {}{{CMS} Collaboration, ``{Event generator tunes obtained from
  underlying event and multiparton scattering measurements}'',} \textit{ Eur.
  Phys. J. C} \textbf{ 76} (2016) 155,
  \href{http://dx.doi.org/10.1140/epjc/s10052-016-3988-x}{\doi{10.1140/epjc/s10052-016-3988-x}},
  \href{http://www.arXiv.org/abs/1512.00815}{\texttt{arXiv:1512.00815}}.

\bibitem{CMS:2019csb}
\hrefCMSnoop {}{{CMS} Collaboration, ``{Extraction and validation of a new set
  of CMS PYTHIA8 tunes from underlying-event measurements}'',} \textit{ Eur.
  Phys. J. C} \textbf{ 80} (2020) 4,
  \href{http://dx.doi.org/10.1140/epjc/s10052-019-7499-4}{\doi{10.1140/epjc/s10052-019-7499-4}},
  \href{http://www.arXiv.org/abs/1903.12179}{\texttt{arXiv:1903.12179}}.

\bibitem{Alwall:2007fs}
\hrefCMSnoop {}{J.~Alwall {et~al.}, ``{Comparative study of various algorithms
  for the merging of parton showers and matrix elements in hadronic
  collisions}'',} \textit{ Eur. Phys. J. C} \textbf{ 53} (2008) 473,
  \href{http://dx.doi.org/10.1140/epjc/s10052-007-0490-5}{\doi{10.1140/epjc/s10052-007-0490-5}},
  \href{http://www.arXiv.org/abs/0706.2569}{\texttt{arXiv:0706.2569}}.

\bibitem{Cacciari:2008gp}
\hrefCMSnoop {}{M.~Cacciari, G.~P. Salam, and G.~Soyez, ``The anti-$k_\text{T}$
  jet clustering algorithm'',} \textit{ JHEP} \textbf{ 04} (2008) 063,
  \href{http://dx.doi.org/10.1088/1126-6708/2008/04/063}{\doi{10.1088/1126-6708/2008/04/063}},
  \href{http://www.arXiv.org/abs/0802.1189}{\texttt{arXiv:0802.1189}}.

\bibitem{Cacciari:2011ma}
\hrefCMSnoop {}{M.~Cacciari, G.~P. Salam, and G.~Soyez, ``{FastJet} user
  manual'',} \textit{ Eur. Phys. J. C} \textbf{ 72} (2012) 1896,
  \href{http://dx.doi.org/10.1140/epjc/s10052-012-1896-2}{\doi{10.1140/epjc/s10052-012-1896-2}},
  \href{http://www.arXiv.org/abs/1111.6097}{\texttt{arXiv:1111.6097}}.

\bibitem{Elmer:2023wtr}
\hrefCMSnoop {}{N.~Elmer, M.~Madigan, T.~Plehn, and N.~Schmal, ``{Staying on
  Top of SMEFT-Likelihood Analyses}'',}
  \href{http://www.arXiv.org/abs/2312.12502}{\texttt{arXiv:2312.12502}}.

\bibitem{CMS:2022ged}
\hrefCMSnoop {}{{CMS} Collaboration, ``{Measurement of the tt\textasciimacron{}
  charge asymmetry in events with highly Lorentz-boosted top quarks in pp
  collisions at s=13 TeV}'',} \textit{ Phys. Lett. B} \textbf{ 846} (2023)
  137703,
  \href{http://dx.doi.org/10.1016/j.physletb.2023.137703}{\doi{10.1016/j.physletb.2023.137703}},
  \href{http://www.arXiv.org/abs/2208.02751}{\texttt{arXiv:2208.02751}}.

\bibitem{ATLAS:2022waa}
\hrefCMSnoop {}{{ATLAS} Collaboration, ``{Evidence for the charge asymmetry in
  pp \textrightarrow{} $ t\overline{t} $ production at $ \sqrt{s} $ = 13 TeV
  with the ATLAS detector}'',} \textit{ JHEP} \textbf{ 08} (2023) 077,
  \href{http://dx.doi.org/10.1007/JHEP08(2023)077}{\doi{10.1007/JHEP08(2023)077}},
  \href{http://www.arXiv.org/abs/2208.12095}{\texttt{arXiv:2208.12095}}.

\bibitem{CMS:2019nrx}
\hrefCMSnoop {}{{CMS} Collaboration, ``{Measurement of the top quark
  polarization and $\mathrm{t\bar{t}}$ spin correlations using dilepton final
  states in proton-proton collisions at $\sqrt{s} =$ 13 TeV}'',} \textit{ Phys.
  Rev. D} \textbf{ 100} (2019) 072002,
  \href{http://dx.doi.org/10.1103/PhysRevD.100.072002}{\doi{10.1103/PhysRevD.100.072002}},
  \href{http://www.arXiv.org/abs/1907.03729}{\texttt{arXiv:1907.03729}}.

\bibitem{Bernreuther:2015yna}
\hrefCMSnoop {}{W.~Bernreuther, D.~Heisler, and Z.-G. Si, ``{A set of top quark
  spin correlation and polarization observables for the LHC: Standard Model
  predictions and new physics contributions}'',} \textit{ JHEP} \textbf{ 12}
  (2015) 026,
  \href{http://dx.doi.org/10.1007/JHEP12(2015)026}{\doi{10.1007/JHEP12(2015)026}},
  \href{http://www.arXiv.org/abs/1508.05271}{\texttt{arXiv:1508.05271}}.

\bibitem{Butterworth:2015oua}
\hrefCMSnoop {}{J.~Butterworth {et~al.}, ``{PDF4LHC recommendations for LHC Run
  II}'',} \textit{ J. Phys. G} \textbf{ 43} (2016) 023001,
  \href{http://dx.doi.org/10.1088/0954-3899/43/2/023001}{\doi{10.1088/0954-3899/43/2/023001}},
  \href{http://www.arXiv.org/abs/1510.03865}{\texttt{arXiv:1510.03865}}.

\bibitem{open-data-ws-1}
\hrefCMSnoop {}{{CMS Collaboration} Collaboration, ``{CMS Open Data Workshop
  2021}'',} 2021.
\newblock
  \url{https://cms-opendata-workshop.github.io/2021-07-19-cms-open-data-workshop/}.

\bibitem{open-data-ws-2}
\hrefCMSnoop {}{{CMS Collaboration} Collaboration, ``{CMS Open Data Workshop
  2022}'',} 2022.
\newblock
  \url{https://cms-opendata-workshop.github.io/2022-08-01-cms-open-data-workshop/}.

\bibitem{open-data-ws-3}
\hrefCMSnoop {}{{CMS Collaboration} Collaboration, ``{CMS Open Data Workshop
  2023}'',} 2023.
\newblock
  \url{https://cms-opendata-workshop.github.io/2023-07-11-cms-open-data-workshop/}.

\bibitem{osti_826696}
\hrefCMSnoop {}{J.~H. Friedman, ``{On multivariate goodness of fit and two
  sample testing}'',} \textit{ eConf} \textbf{ C030908} (2003) THPD002.

\bibitem{lopezpaz2018revisiting}
\hrefCMSnoop {}{D.~Lopez-Paz and M.~Oquab, ``{Revisiting Classifier Two-Sample
  Tests}'',}
  \href{http://www.arXiv.org/abs/1610.06545}{\texttt{arXiv:1610.06545}}.

\bibitem{Cowan:2010js}
\hrefCMSnoop {}{G.~Cowan, K.~Cranmer, E.~Gross, and O.~Vitells, ``{Asymptotic
  formulae for likelihood-based tests of new physics}'',} \textit{ Eur. Phys.
  J. C} \textbf{ 71} (2011) 1554,
  \href{http://dx.doi.org/10.1140/epjc/s10052-011-1554-0}{\doi{10.1140/epjc/s10052-011-1554-0}},
  \href{http://www.arXiv.org/abs/1007.1727}{\texttt{arXiv:1007.1727}}.
  [Erratum: Eur.Phys.J.C 73, 2501 (2013)].

\bibitem{Wald}
\hrefCMSnoop {}{A.~Wald, ``Tests of statistical hypotheses concerning several
  parameters when the number of observations is large'',} \textit{ Transactions
  of the American Mathematical Society} \textbf{ 54} (1943) 426.

\bibitem{iminuit}
\hrefCMSnoop {}{H.~Dembinski and P.~O. et~al., ``scikit-hep/iminuit'',}
  \href{http://dx.doi.org/10.5281/zenodo.3949207}{\doi{10.5281/zenodo.3949207}}.

\bibitem{harris2020array}
C.~R. Harris\hrefCMSnoop {}{ {et~al.}, ``Array programming with {NumPy}'',}
  \textit{ Nature} \textbf{ 585} (September, 2020) 357,
  \href{http://dx.doi.org/10.1038/s41586-020-2649-2}{\doi{10.1038/s41586-020-2649-2}}.

\bibitem{code}
\hrefCMSnoop {}{{Github repository}, ``{Boosted Parametric Tree}'',} 2024.
\newblock \url{https://github.com/HephyAnalysisSW/BPT}.

\end{thebibliography}\endgroup

\end{document}